\documentclass[useAMS,usenatbib]{mnras}

\usepackage{graphicx}
\usepackage{graphics}
\usepackage{color}
\usepackage{url}
\usepackage[caption=false]{subfig}
\usepackage{verbatim}
\usepackage{array}
\usepackage{amssymb}
\usepackage{amsmath}
\usepackage{afterpage}
\usepackage{alphalph}

\usepackage{lscape}
\usepackage{pbox}
\usepackage{supertabular}

\newcommand{\swift}{{\it Swift}}

\pubyear{2017}

\begin{document}

\title[ASAS-SN Bright SN Catalog 2016]{The ASAS-SN Bright Supernova Catalog -- III. 2016}

\author[T.~W.-S.~Holoien et al.]{T.~W.-S.~Holoien$^{1,2,3}$\thanks{tholoien@astronomy.ohio-state.edu}, J.~S.~Brown$^{1}$, K.~Z.~Stanek$^{1,2}$, C.~S.~Kochanek$^{1,2}$,  
\newauthor
B.~J.~Shappee$^{4,5}$, J.~L.~Prieto$^{6,7}$, Subo~Dong$^{8}$, J.~Brimacombe$^{9}$, D.~W.~Bishop$^{10}$,     
\newauthor
S.~Bose$^{8}$, J.~F.~Beacom$^{1,2,11}$, D.~Bersier$^{12}$, Ping~Chen$^{8}$, L.~Chomiuk$^{13}$, 
\newauthor 
E.~Falco$^{14}$, D.~Godoy-Rivera$^{1}$, N.~Morrell$^{15}$, G.~Pojmanski$^{16}$, J.~V.~Shields$^{1}$, 
\newauthor
J.~Strader$^{13}$, M.~D.~Stritzinger$^{17}$, Todd~A.~Thompson$^{1,2}$, P.~R.~Wo\'zniak$^{18}$,
\newauthor
G.~Bock$^{19}$, P.~Cacella$^{20}$, E.~Conseil$^{21}$, I.~Cruz$^{22}$, J.~M.~Fernandez$^{23}$, S.~Kiyota$^{24}$, 
\newauthor
R.~A.~Koff$^{25}$, G. Krannich$^{26}$, P.~Marples$^{27}$, G.~Masi$^{28}$, L.~A.~G.~Monard$^{29}$, 
\newauthor
B.~Nicholls$^{30}$, J.~Nicolas$^{31}$, R.~S.~Post$^{32}$, G.~Stone$^{33}$, and W.~S.~Wiethoff$^{34}$\\ \\
  $^{1}$ Department of Astronomy, The Ohio State University, 140 West 18th Avenue, Columbus, OH 43210, USA \\
  $^{2}$ Center for Cosmology and AstroParticle Physics (CCAPP), The Ohio State University, 191 W. Woodruff Ave., \\
             \hspace{0.6cm}Columbus, OH 43210, USA \\
  $^{3}$ US Department of Energy Computational Science Graduate Fellow \\
  $^{4}$ Carnegie Observatories, 813 Santa Barbara Street, Pasadena, CA 91101, USA \\
  $^{5}$ Hubble and Carnegie-Princeton Fellow\\
  $^{6}$ N\'ucleo de Astronom\'ia de la Facultad de Ingenier\'ia y Ciencias, Universidad Diego Portales, Av. Ej\'ercito 441, Santiago, Chile \\
  $^{7}$ Millennium Institute of Astrophysics, Santiago, Chile \\
  $^{8}$ Kavli Institute for Astronomy and Astrophysics, Peking University, Yi He Yuan Road 5, Hai Dian District, \\
               \hspace{0.6cm}Beijing 100871, China \\
  $^{9}$ Coral Towers Observatory, Cairns, Queensland 4870, Australia \\
  $^{10}$ Rochester Academy of Science, 1194 West Avenue, Hilton, NY 14468, USA \\
  $^{11}$ Department of Physics, The Ohio State University, 191 W. Woodruff Ave., Columbus, OH 43210, USA \\
  $^{12}$ Astrophysics Research Institute, Liverpool John Moores University, 146 Brownlow Hill, Liverpool L3 5RF, UK \\
  $^{13}$ Department of Physics and Astronomy, Michigan State University, East Lansing, MI 48824, USA \\
  $^{14}$ Harvard-Smithsonian Center for Astrophysics, 60 Garden St., Cambridge, MA 02138, USA \\
  $^{15}$ Las Campanas Observatory, Carnegie Observatories, Casilla 601, La Serena, Chile \\
  $^{16}$ Warsaw University Astronomical Observatory, Al. Ujazdowskie 4, 00-478 Warsaw, Poland \\
  $^{17}$ Department of Physics and Astronomy, Aarhus University, Ny Munkegade 120, DK-8000 Aarhus C, Denmark \\
  $^{18}$ Los Alamos National Laboratory, Mail Stop B244, Los Alamos, NM 87545, USA \\
  $^{19}$ Runaway Bay Observatory, 1 Lee Road, Runaway Bay, Queensland 4216, Australia \\
  $^{20}$ DogsHeaven Observatory, SMPW Q25 CJ1 LT10B, Brasilia, DF 71745-501, Brazil \\
  $^{21}$ Association Francaise des Observateurs d'Etoiles Variables (AFOEV), Observatoire de Strasbourg, 11 Rue de l'Universite, \\
               \hspace{0.6cm}67000 Strasbourg, France \\
  $^{22}$ Cruz Observatory, 1971 Haverton Drive, Reynoldsburg, OH 43068, USA \\
  $^{23}$ Observatory Inmaculada del Molino, Hernando de Esturmio 46, Osuna, 41640 Sevilla, Spain \\
  $^{24}$ Variable Star Observers League in Japan, 7-1 Kitahatsutomi, Kamagaya, Chiba 273-0126, Japan \\
  $^{25}$ Antelope Hills Observatory, 980 Antelope Drive West, Bennett, CO 80102, USA \\
  $^{26}$ Roof Observatory Kaufering, Lessingstr. 16, D-86916 Kaufering, Germany \\
  $^{27}$ Leyburn \& Loganholme Observatories, 45 Kiewa Drive, Loganholme, Queensland 4129, Australia \\
  $^{28}$ Virtual Telescope Project, Via Madonna de Loco, 47-03023 Ceccano (FR), Italy \\
  $^{29}$ Kleinkaroo Observatory, Calitzdorp, St. Helena 1B, P.O. Box 281, 6660 Calitzdorp, Western Cape, South Africa \\
  $^{30}$ Mount Vernon Observatory, 6 Mount Vernon Place, Nelson, New Zealand \\
  $^{31}$ Groupe SNAude France, 364 Chemin de Notre Dame, 06220 Vallauris, France \\
  $^{32}$ Post Observatory, Lexington, MA 02421, USA \\
  $^{33}$ Sierra Remote Observatories, 44325 Alder Heights Road, Auberry, CA 93602, USA \\
  $^{34}$ Department of Earth and Environmental Sciences, University of Minnesota, 230 Heller Hall, 1114 Kirby Drive, \\
               \hspace{0.6cm}Duluth, MN. 55812, USA
  }
\maketitle
\begin{abstract}
This catalog summarizes information for all supernovae discovered by the All-Sky Automated Survey for SuperNovae (ASAS-SN) and all other bright ($m_{peak}\leq17$), spectroscopically confirmed supernovae discovered in 2016. We then gather the near-IR through UV magnitudes of all host galaxies and the offsets of the supernovae from the centers of their hosts from public databases. We illustrate the results using a sample that now totals 668 supernovae discovered since 2014 May 1, including the supernovae from our previous catalogs, with type distributions closely matching those of the ideal magnitude limited sample from \citet{li11}. This is the third of a series of yearly papers on bright supernovae and their hosts from the ASAS-SN team. 
\end{abstract}

\raggedbottom

%%%%%%%%%%%%%%%%%
% Section: Introduction
%%%%%%%%%%%%%%%%%

\section{Introduction}
\label{sec:intro}

The last two decades have seen the proliferation of large, systematic surveys that search some or all of the sky for supernovae (SNe) and other transient phenomena. Significant examples include the Lick Observatory Supernova Search \citep[LOSS;][]{li00}, the Panoramic Survey Telescope \& Rapid Response System \citep[Pan-STARRS;][]{kaiser02}, the Texas Supernova Search \citep{quimby06}, the Sloan Digital Sky Survey (SDSS) Supernova Survey \citep[][]{frieman08}, the Catalina Real-Time Transient Survey \citep[CRTS;][]{drake09}, the CHilean Automatic Supernova sEarch \citep[CHASE;][]{pignata09}, the Palomar Transient Factory \citep[PTF;][]{law09}, the Gaia transient survey \citep{hodgkin13}, the La Silla-QUEST (LSQ) Low Redshift Supernova Survey \citep{baltay13}, the Mobile Astronomical System of TElescope Robots \citep[MASTER;][]{gorbovskoy13} survey, the Optical Gravitational Lensing Experiment-IV \citep[OGLE-IV;][]{wyrzykowski14}, and the Asteroid Terrestrial-impact Last Alert System \citep[ATLAS;][]{tonry11}.

Despite the number of transient survey projects, there has been no rapid-cadence optical survey scanning the entire visible sky to find the bright and nearby transients that can be observed in the greatest detail. Such events can provide the detailed observational data needed to have the greatest physical impact. 

This is the goal of the All-Sky Automated Survey for SuperNovae (ASAS-SN\footnote{\url{http://www.astronomy.ohio-state.edu/~assassin/}}; \citealt{shappee14}). ASAS-SN is a long-term project designed to find bright transients, and has been highly successful since the beginning of survey operations, finding many interesting and nearby supernovae \citep[e.g.,][]{dong16,holoien16c,shappee16,godoy-rivera17}, tidal disruption events \citep{holoien14b,brown16a,brown16b,holoien16b,holoien16a,prieto16,romero16}, active galactic nucleus flares \citep{shappee14}, stellar outbursts \citep{holoien14a,schmidt14,herczeg16,schmidt16}, and cataclysmic variable stars \citep{kato13,kato14,kato15,kato16}. 

During 2016, ASAS-SN comprised eight 14-cm telescopes with standard $V$-band filters, each with a $4.5\times4.5$ degree field-of-view and a limiting magnitude of $m_V\sim17$ (see \citet{shappee14} for further technical details). These telescopes are divided into two units each with four telescopes on a common mount hosted by the Las Cumbres Observatory \citep{brown13}. Brutus, our northern unit, is housed at the Las Cumbres Observatory site on Mount Haleakala in Hawaii, and Cassius, our southern unit, is hosted at the Las Cumbres Observatory site at Cerro Tololo, Chile. The two units combined give ASAS-SN roughly 20000 square degrees of coverage per clear night, allowing us to cover the entire observable sky (roughly 30000 square degrees at any given time) with a $2-3$ day cadence. In 2017, ASAS-SN will expand to five units (20 telescopes) at four sites (Hawaii, McDonald Observatory in Texas, Sutherland, South Africa, and two in Chile), allowing nightly coverage of the visible sky with little sensitivity to local weather. For a more detailed history of the ASAS-SN project, see the introduction of \citet{holoien16d} and \citet{shappee14}.

ASAS-SN data are processed and searched in real-time and all ASAS-SN discoveries are announced publicly upon confirmation, allowing for rapid discovery and response by both the ASAS-SN team and others. Our untargeted survey approach and complete spectroscopic identification makes our sample less biased than those of many other supernova searches. This makes it ideal for population studies of nearby supernovae and their hosts. 

This manuscript is the third of a series of yearly catalogs provided by the ASAS-SN team and presents collected information on supernovae discovered by ASAS-SN in 2016 and their host galaxies. As in our previous catalogs \citep{holoien16d,holoien17a}, we also provide the same information for bright supernovae (those with $m_{peak}\leq17$) that were discovered by other professional surveys and amateur astronomers in 2016 to construct a complete sample of bright supernovae discovered in 2016. This includes whether ASAS-SN independently found the supernovae after the initial announcement.

The analyses and information presented in this paper supersede the information presented in discovery and classification Astronomer's Telegrams (ATels), which we cite in this manuscript, and the information publicly available on ASAS-SN webpages and the Transient Name Server (TNS\footnote{\url{https://wis-tns.weizmann.ac.il/}}). 

In \S\ref{sec:sample} we describe the sources of the information presented in this manuscript and list ASAS-SN supernovae with updated classifications or redshift measurements. In \S\ref{sec:analysis}, we give statistics on the supernova and host galaxy populations in our full cumulative sample, including the discoveries listed in \citet{holoien16d,holoien17a}, and discuss overall trends in the sample. Throughout our analyses, we assume a standard $\Lambda$CDM cosmology with $H_0=69.3$~km~s$^{-1}$~Mpc$^{-1}$, $\Omega_M=0.29$, and $\Omega_{\Lambda}=0.71$ for converting host redshifts into distances. In \S\ref{sec:disc}, we conclude with our overall findings and discuss how the upcoming expansion to ASAS-SN will impact our future discoveries.

%%%%%%%%%%%%%%%%%
% Section: Sample
%%%%%%%%%%%%%%%%%

\section{Data Samples}
\label{sec:sample}

Below we outline the sources of the data collected in our supernova and host galaxy samples. These data are presented in Tables~\ref{table:asassn_sne}, \ref{table:other_sne}, \ref{table:asassn_hosts}, and \ref{table:other_hosts}.

%%%%%%%%%%%%%%%%%
% Subsection: ASAS-SN Sample
%%%%%%%%%%%%%%%%%

\subsection{The ASAS-SN Supernova Sample}
\label{sec:asassn_sample}

Table~\ref{table:asassn_sne} includes information for all supernovae discovered by ASAS-SN between 2016 January 1 and 2016 December 31. As in \citet{holoien16d,holoien17a}, all names, discovery dates, and host names are taken from our discovery ATels, all of which are cited in Table~\ref{table:asassn_sne}. We also include the supernova names designated by TNS, the official IAU mechanism for reporting new astronomical transients. As is noted in our ATels, the ASAS-SN team is participating in the TNS system to minimize potential confusion, but we use the ASAS-SN designations as our primary nomenclature and encourage others to do the same in order to preserve the origin of the transient in future literature.

ASAS-SN supernova redshifts were spectroscopically measured from classification spectra. For those cases where a supernova host had a previously measured redshift and the host redshift is consistent with the transient redshift, we list the redshift of the host taken from the NASA/IPAC Extragalactic Database (NED)\footnote{\url{https://ned.ipac.caltech.edu/}}. For all other cases we typically report the redshifts given in the classification telegrams, excepting those that have been updated in this work (see below).

ASAS-SN supernova classifications are taken from classification telegrams, which we have cited in  Table~\ref{table:asassn_sne}, when available. In some cases a classification was only reported on TNS and was not reported in an ATel; for these cases we list ``TNS'' in the ``Classification Telegram'' column. When available in the classification ATel or on TNS we also give the approximate age at discovery measured in days relative to peak. Classifications were typically obtained using either the Supernova Identification code \citep[SNID;][]{blondin07} or the Generic Classification Tool (GELATO\footnote{\url{gelato.tng.iac.es}}; \citealt{harutyunyan08}), which both compare observed input spectra to template spectra in order to estimate the supernova age and type.

Using archival classification and late-time spectra of the ASAS-SN supernova discoveries taken from TNS and Weizmann Interactive Supernova data REPository \citep[WISEREP;][]{yaron12}, we also update a number of redshifts and classifications that differ from what was reported in the discovery and classification telegrams. ASASSN-16ah, ASASSN-16bm, ASASSN-16bq, ASASSN-16bv, ASASSN-16cr, ASASSN-16cs, ASASSN-16es, ASASSN-16fa, ASASSN-16fc, ASASSN-16fx, ASASSN-16gz, ASASSN-16hr, ASASSN-16hw, ASASSM-16ip, ASASSN-16je, ASASSN-16jj, ASASSN-16la, ASASSN-16lc, ASASSN-16ll, ASASSN-16mj, ASASSN-16ns, ASASSN-16oj, ASASSN-16ok, ASASSN-16ol, ASASSN-16oy, ASASSN-16pd, ASASSN-16pj, and ASASSN-16pk have updated redshifts based on archival spectra. ASASSN-16dx has been reclassified from archival spectra, and ASASSN-16fp has been classified as a Ib/Ic-BL \citep{yamanaka17}. All updated redshifts and classifications are reported in Table~\ref{table:asassn_sne}.

Using the astrometry.net \citep{barron08,lang10} package we solved the astrometry in follow-up images for all ASAS-SN supernovae and measured a centroid position for the supernova using {\sc Iraf}. This approach typically yields errors of $<$1\farcs{0} in position, which is significantly more accurate than measuring the supernova position directly in ASAS-SN images, which have a 7\farcs{0} pixel scale. The images used to measure astrometry were obtained using the Las Cumbres Observatory 1-m telescopes \citep{brown13}, OSMOS mounted on the MDM Hiltner 2.4-m telescope, or from amateur collaborators working with the ASAS-SN team. For most cases, the coordinates measured from follow-up images were reported in our discovery ATels, but we report new, more accurate coordinates in Table~\ref{table:asassn_sne} for those cases where the supernovae were announced with coordinates measured in ASAS-SN data. The offset from the host galaxy nucleus is also reported, and was calculated using the coordinates measured from follow-up images and host coordinates available in NED.

$V$-band, host-subtracted discovery and peak magnitudes were re-measured from ASAS-SN data for all ASAS-SN supernova discoveries, and these magnitudes are reported in Table~\ref{table:asassn_sne}. This has resulted in differences between the magnitudes reported in this work and the magnitudes reported in the original discovery ATels for some cases, where re-reduction of the data has led to improvements in our photometry. We define the ``discovery magnitude'' as the magnitude of the supernova on the announced discovery date. For cases with enough detections in the light curve, we also perform a parabolic fit to the light curve and estimate a peak magnitude based on the fit. The ``peak magnitude'' reported in Table~\ref{table:asassn_sne} is the brighter value between the brightest measured magnitude and the peak of the parabolic fit.

We include all supernovae discovered by ASAS-SN in 2016 in this catalog and in Table~\ref{table:asassn_sne}, including those that peaked at magnitudes fainter than $m_V=17$. In the comparison analyses presented in \S\ref{sec:analysis}, however, we only include those ASAS-SN supernovae with $m_{V,peak}\leq17$ so that our sample is consistent with the non-ASAS-SN sample.

%%%%%%%%%%%%%%%%%
% Subsection: Other SN Sample
%%%%%%%%%%%%%%%%%

\subsection{The Non-ASAS-SN Supernova Sample}
\label{sec:other_sample}

In Table~\ref{table:other_sne} we give information for all spectroscopically confirmed supernovae with peak magnitudes of $m_{peak}\leq17$ that were discovered by other professional and amateur supernova searches between 2016 January 1 and 2016 December 31.

%%%%%%%%%%%%%%%%%
% Figure: Pie Charts
%%%%%%%%%%%%%%%%%

\begin{figure*}
\begin{minipage}{\textwidth}
\centering
\subfloat{{\includegraphics[width=0.31\textwidth]{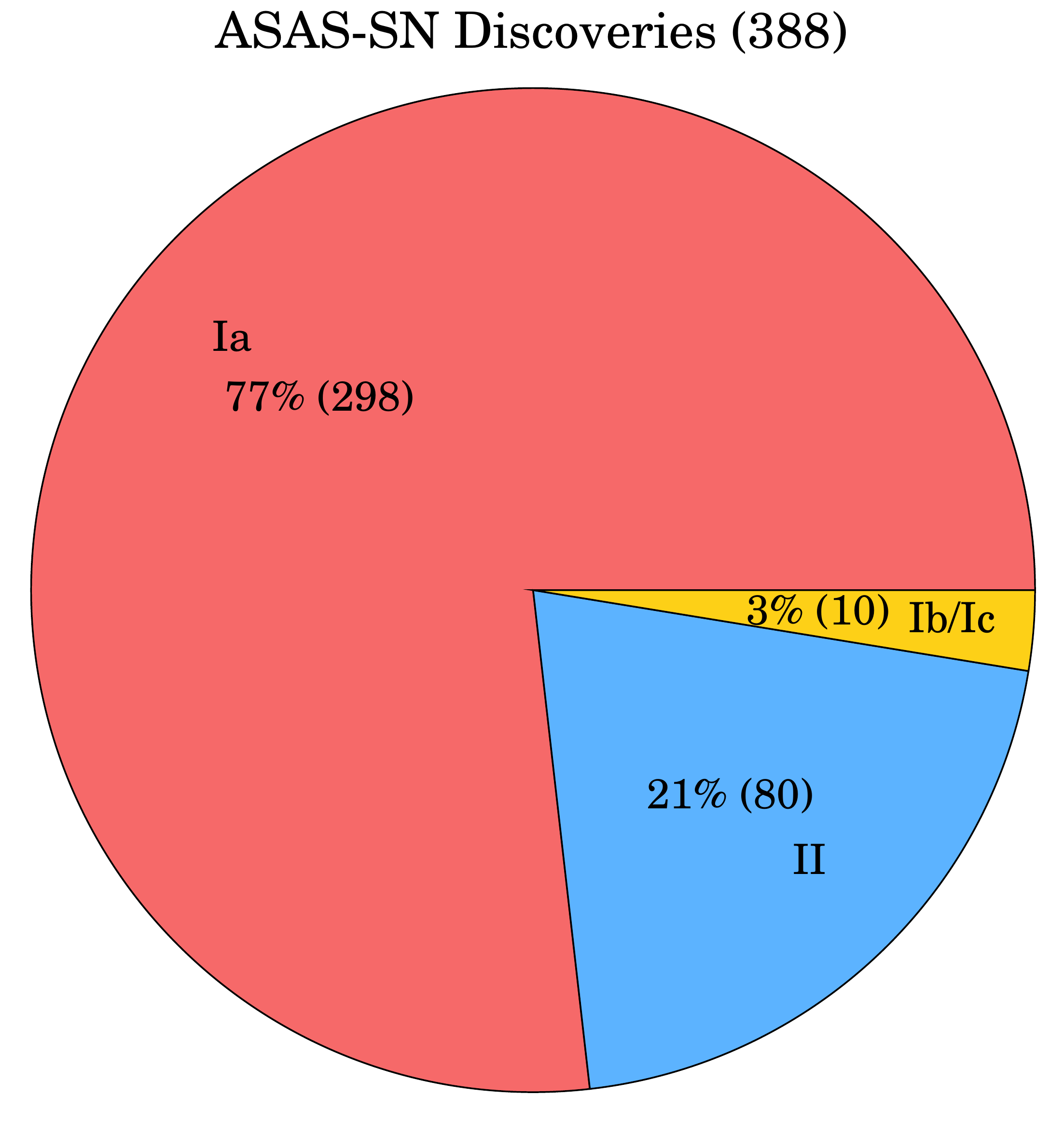}}}
\subfloat{{\includegraphics[width=0.31\textwidth]{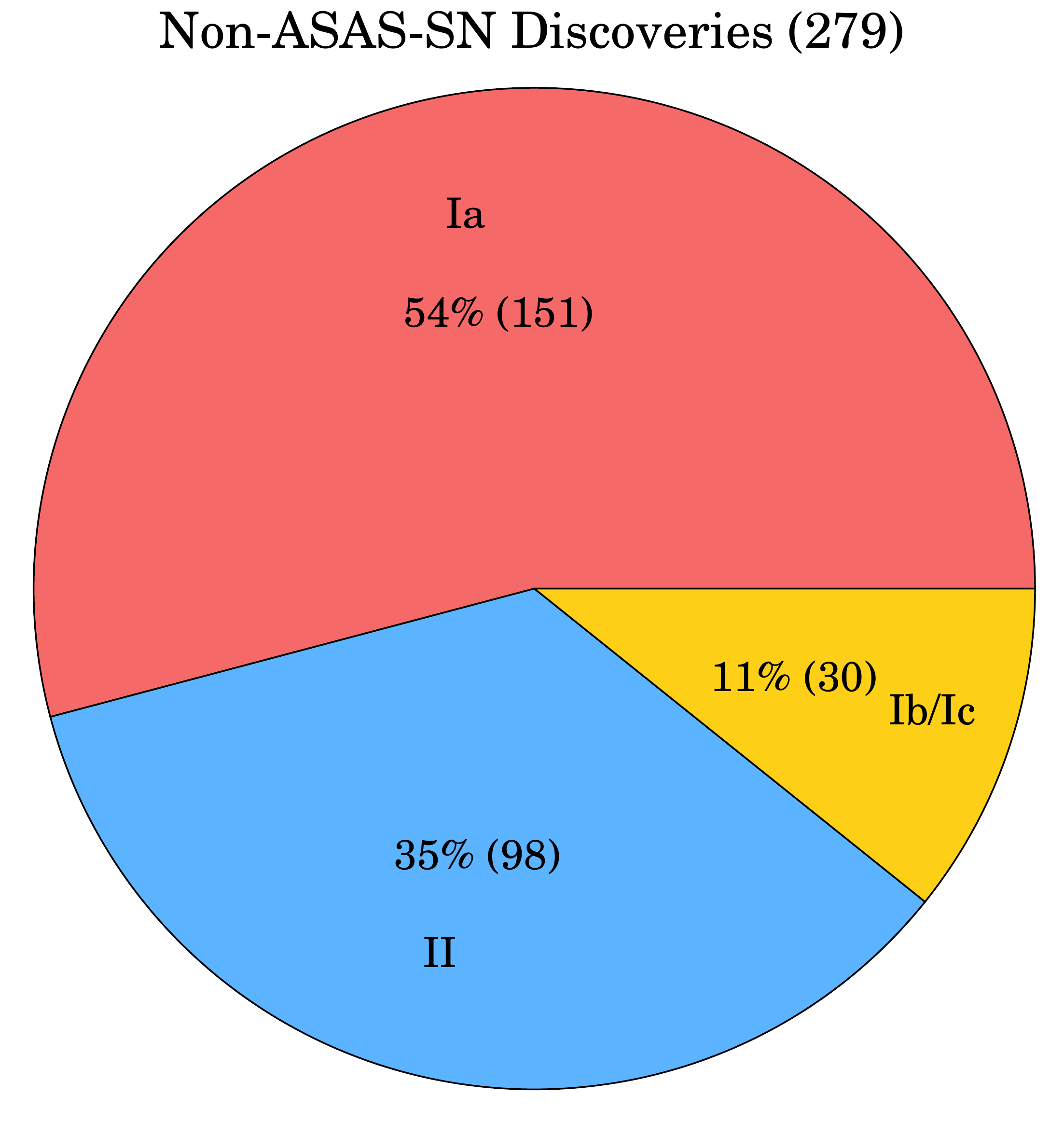}}}
\subfloat{{\includegraphics[width=0.31\textwidth]{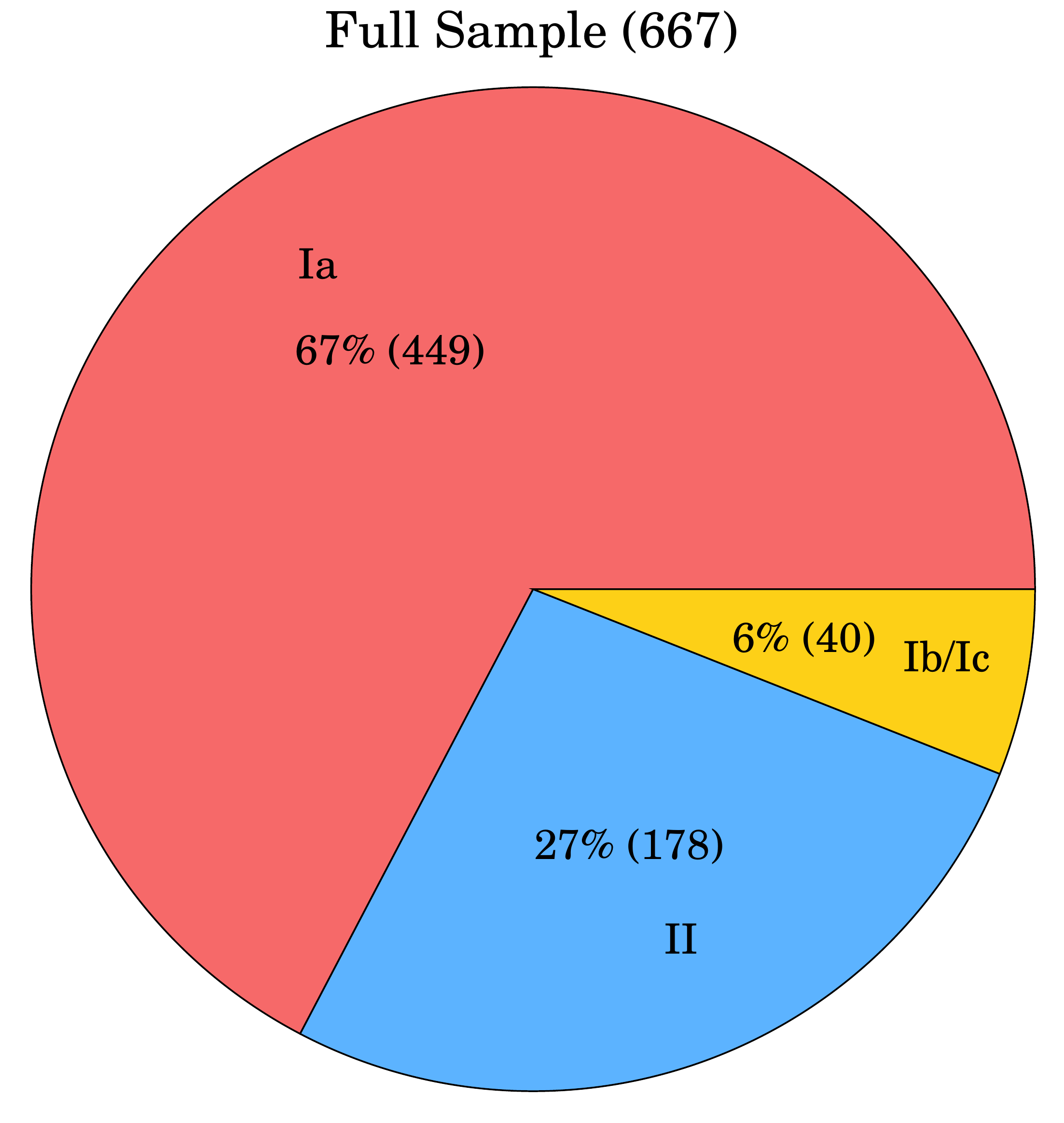}}}
\caption{\emph{Left Panel}: Breakdown by type of the supernovae discovered by ASAS-SN between 2014 May 01 and 2016 December 31. The proportions of each type is very similar to that of an ideal magnitude-limited sample \citep{li11}. \emph{Center Panel}: The same chart for the non-ASAS-SN sample in the same time period. \emph{Right Panel}: The same breakdown of types for the combined supernova sample. For the purposes of this analysis, we exclude superluminous supernovae and include Type IIb supernovae in the ``Type II'' sample.} 
\label{fig:piechart}
\end{minipage}
\end{figure*}

We compiled data for the non-ASAS-SN discoveries from the ``latest supernovae'' website\footnote{\url{http://www.rochesterastronomy.org/snimages/}} designed and maintained by D.~W.~Bishop \citep{galyam13}. This site compiles discoveries reported from different channels and links objects reported by different sources at different times, making it an ideal source for collecting information on supernovae discovered by different search groups. While we did use TNS for verification of the data from the latest supernovae website, we did not use it as the primary source of information on non-ASAS-SN discoveries, as some supernova searches do not participate in the TNS system. 

Names, IAU names, discovery dates, coordinates, host names, host offsets, peak magnitudes, spectral types, and discovery sources for each supernova in the non-ASAS-SN sample were taken from the latest supernovae website when possible. Host galaxy redshifts were collected from NED when available and were taken from the latest supernovae website otherwise. For cases where a host name or host offset was not listed on the website for a supernova, the primary name and offset were taken from NED. In these cases, we define the offset as the distance between the reported coordinates of the supernova and the galaxy coordinates in NED. In some cases, no catalogued galaxy was listed at the position of the host in NED, but a host galaxy was clearly visible in archival Pan-STARRS data \citep{chambers16} . In these cases, we measured the centroid position of the host nucleus using {\sc Iraf} and calculated the offset using those coordinates. For all supernovae in both samples we use the primary name of the host galaxy listed in NED, which sometimes differs from the name listed on the ASAS-SN supernova page or the latest supernova website.

We update the redshifts and classifications of several supernovae discovered by non-ASAS-SN sources that have missing or incorrect information on the latest supernovae website. Using archival classification and late-time spectra from TNS and WISeREP we have updated the classifications of SN 2016ajf, SN 2016bau, SN 2016gxp, KAIT-16az, Gaia16cbd, and SN 2016aqt, which were previously mis-typed. CSS160708:151956+052419/SN 2016ehy has a redshift of $z=0.045$ (I. Shivvers, private communication) and Gaia16alq/SN 2016dxv has a redshift of $z=0.023$ (A. Piascik, private communication), measured from supernova lines in the spectra in both cases. Finally, based on an examination of available spectra, PS16dtm/SN 2016ezh, which was previously reported as a Type-II superluminous supernova \citep[SLSN-II,][]{ps16dtm_atel1,ps16dtm_atel2}, appears to be a highly unique object, possibly consistent with a SLSN-II, a tidal disruption event, or even rare AGN activity \citep[e.g.,][]{blanchard17}. For this reason, we include it in the catalog, but exclude it from the analyses that follow. All updated types and redshifts are reported in Table~\ref{table:other_sne}.

The name of the discovery group is listed for all supernovae discovered by other professional surveys. For those supernovae discovered by non-professional astronomers, we list ``Amateurs'' as the discovery source to differentiate these from those discovered by ASAS-SN and other professional astronomers and surveys. As in previous years, amateurs account for the largest number of bright supernova discoveries in 2016 after ASAS-SN.

As in our previous catalogs, we note in Table~\ref{table:other_sne} if the ASAS-SN team independently recovered these supernovae while scanning our data. This is done to help quantify the impact ASAS-SN has on the discovery of bright supernovae in the absence of other supernovae searches.

%%%%%%%%%%%%%%%%%
% Subsection: Host Galaxy Sample
%%%%%%%%%%%%%%%%%

\subsection{The Host Galaxy Samples}
\label{sec:host_sample}

For the host galaxies of both supernova samples, we collected Galactic extinction estimates for the direction to the host and host magnitudes spanning from the near-ultraviolet (NUV) to the infrared (IR) wavelengths. We present these data in Tables~\ref{table:asassn_hosts} and \ref{table:other_hosts} for ASAS-SN hosts and non-ASAS-SN hosts, respectively. Galactic $A_V$ from \citet{schlafly11} at the positions of the supernovae were gathered from NED. NUV magnitudes are taken from the Galaxy Evolution Explorer \citep[GALEX;][]{morrissey07} All Sky Imaging Survey (AIS), optical $ugriz$ magnitudes are gathered from the Sloan Digital Sky Survey Data Release 13 \citep[SDSS DR13;][]{albareti16}, NIR $JHK_S$ magnitudes are gathered from the Two-Micron All Sky Survey \citep[2MASS;][]{skrutskie06}, and IR $W1$ and $W2$ magnitudes are gathered from the Wide-field Infrared Survey Explorer \citep[WISE;][]{wright10} AllWISE source catalog. 

%%%%%%%%%%%%%%%%%
% Figure: Offset-Mag
%%%%%%%%%%%%%%%%%

\begin{figure*}
\begin{minipage}{\textwidth}
\centering
\subfloat{{\includegraphics[width=0.85\textwidth]{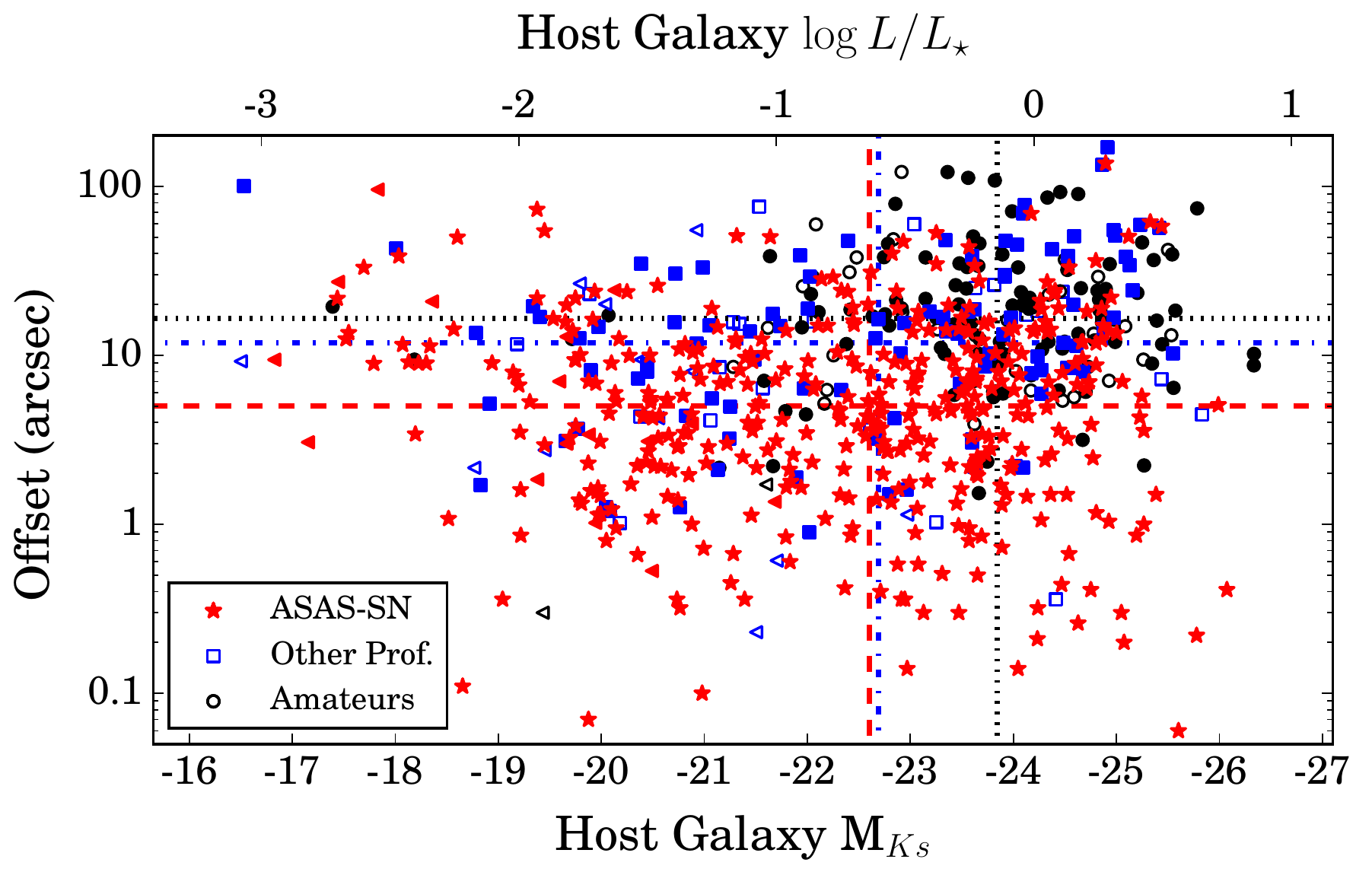}}}

\centering
\subfloat{{\includegraphics[width=0.85\textwidth]{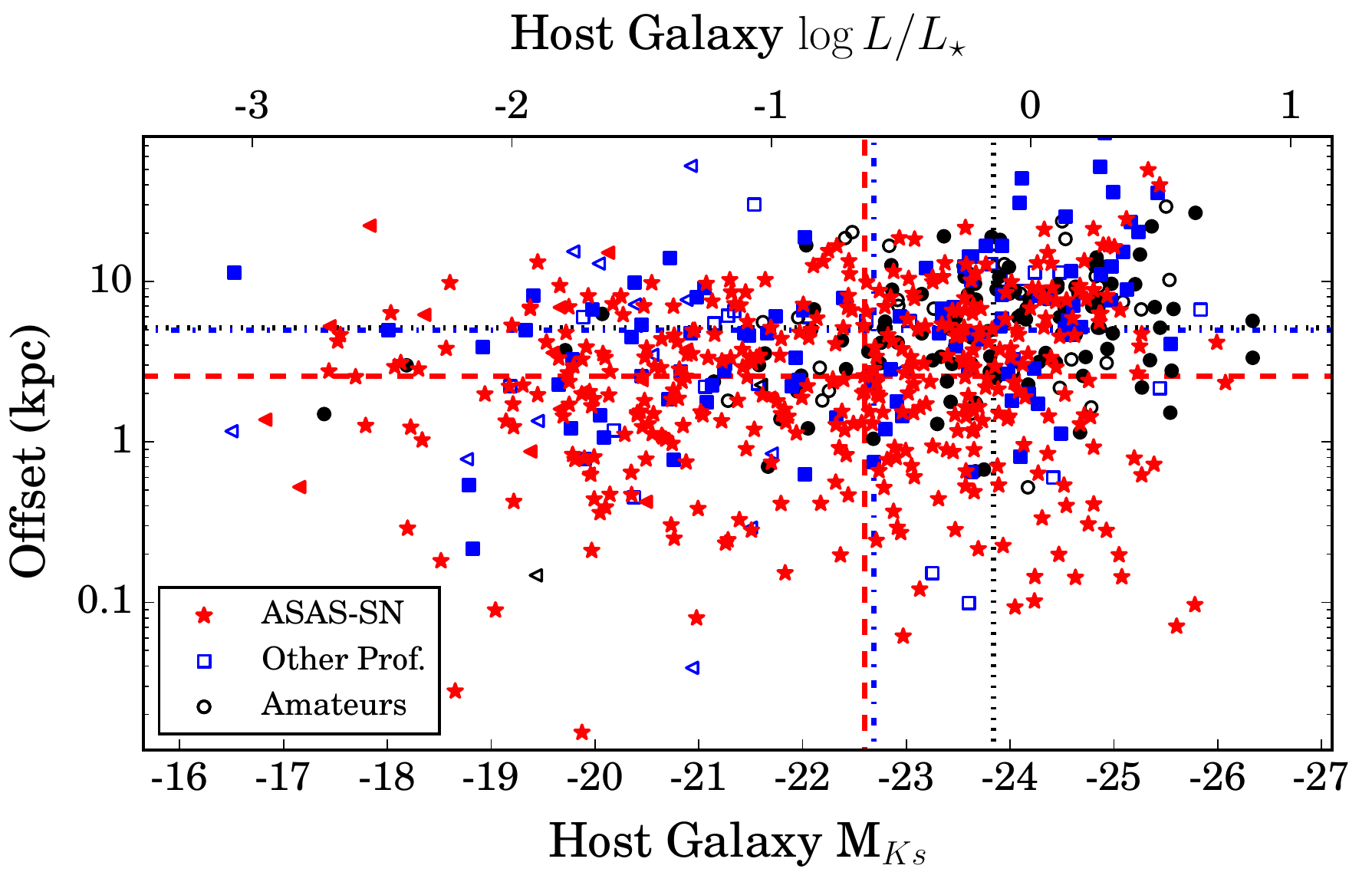}}}
\caption{\emph{Upper Panel}: Offset from the host nucleus in arcseconds compared to the absolute $K_S$-band host magnitude for all supernovae in our combined sample discovered between 2014 May 1 and 2016 December 31. The top axis shows $\log{(L/L_\star)}$ values corresponding to the magnitude range shown on the bottom scale assuming $M_{\star,K_S}=-24.2$ \citep{kochanek01}. ASAS-SN supernova discoveries are shown as red stars, amateur discoveries are shown as black circles, and discoveries by other professional searches are shown as blue squares. Triangles indicate upper limits on the host galaxy magnitudes for hosts that were not detected in 2MASS or WISE. Points are filled for supernovae that were independently recovered by ASAS-SN. We indicate the median offsets and host magnitudes for ASAS-SN discoveries, amateur discoveries, and other professional discoveries using dashed, dotted, and dash-dotted lines, respectively, in colors that match the data points. \emph{Lower Panel}: As above, but with the offset measured in kiloparsecs.}
\label{fig:offmag}
\end{minipage}
\end{figure*}

When a host galaxy was not detected in 2MASS, we adopted an upper limit corresponding to the faintest 2MASS host magnitudes in our sample for the $J$ and $H$ bands ($m_J>16.5$, $m_H>15.7$). For hosts that are not detected in 2MASS but are detected in the WISE $W1$ band, we estimated a host magnitude by adding the mean $K_S-W1$ offset from the sample to the WISE $W1$ data. This offset was calculated by averaging the offsets for all hosts that are detected in both the $K_S$ and $W1$ bands from both supernova samples from 2014 May 1 through 2016 December 31. The average offset is equal to $-0.51$ magnitudes with a scatter of $0.04$ magnitudes and a standard error of $0.002$ magnitudes. If a host was not detected in either 2MASS or WISE, we adopted an upper limit of $m_{K_S}>15.6$, corresponding to the faintest detected host in our sample.

%%%%%%%%%%%%%%%%%
% Figure: Offset-Mag Distributions
%%%%%%%%%%%%%%%%%

\begin{figure*}
\begin{minipage}{\textwidth}
\centering
\subfloat{{\includegraphics[width=0.75\textwidth]{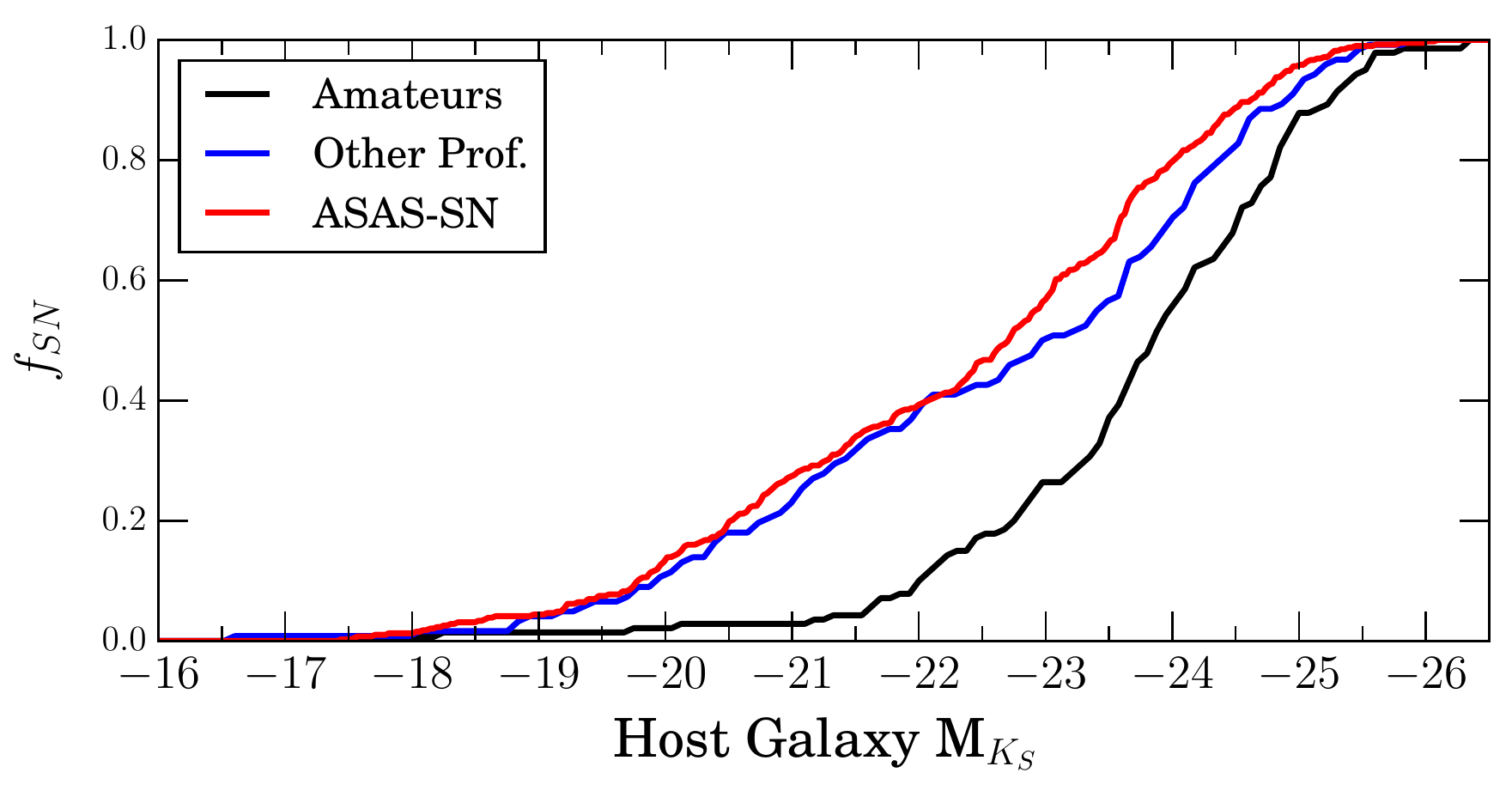}}}

\centering
\subfloat{{\includegraphics[width=0.75\textwidth]{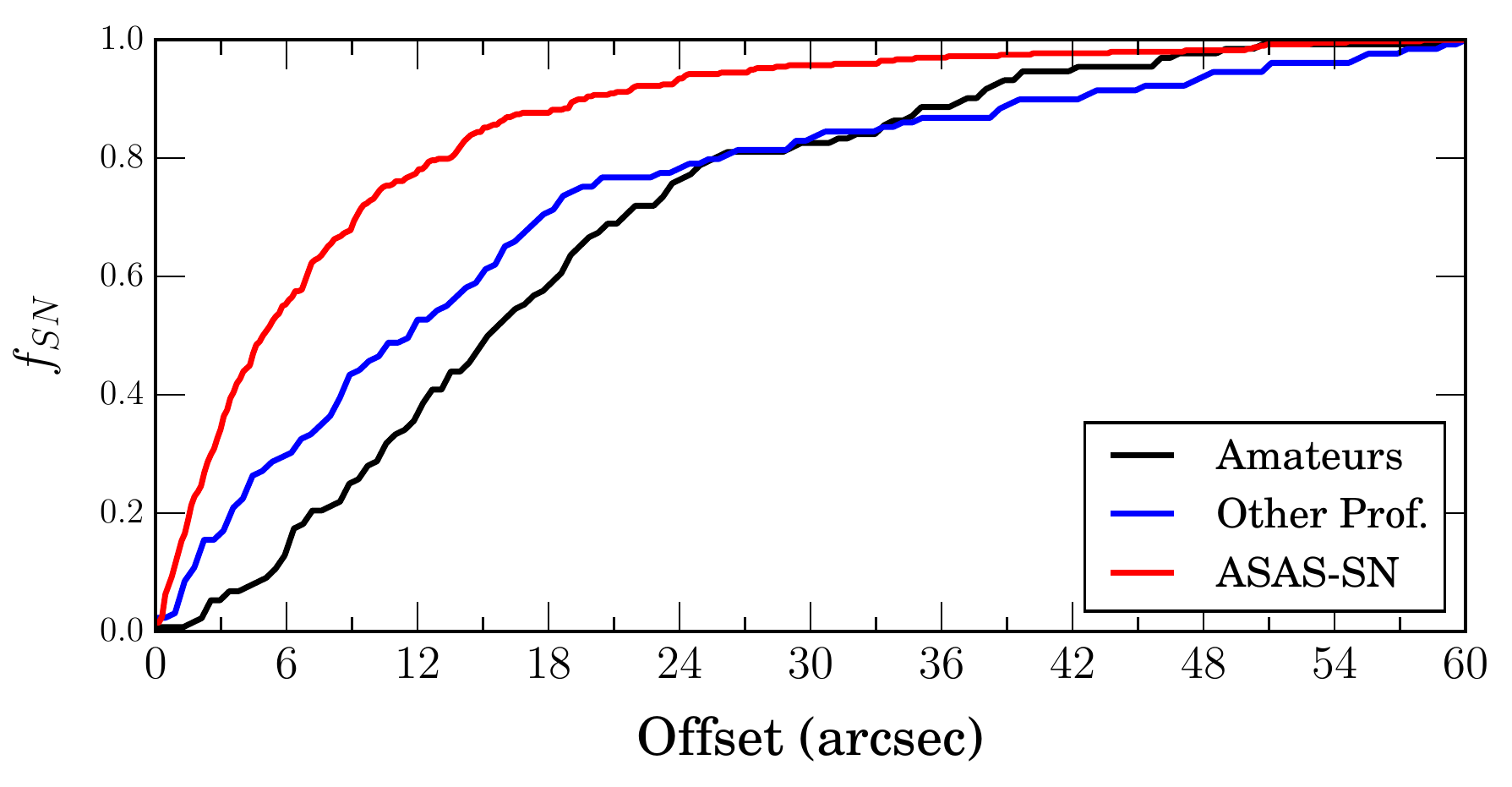}}}

\centering
\subfloat{{\includegraphics[width=0.75\textwidth]{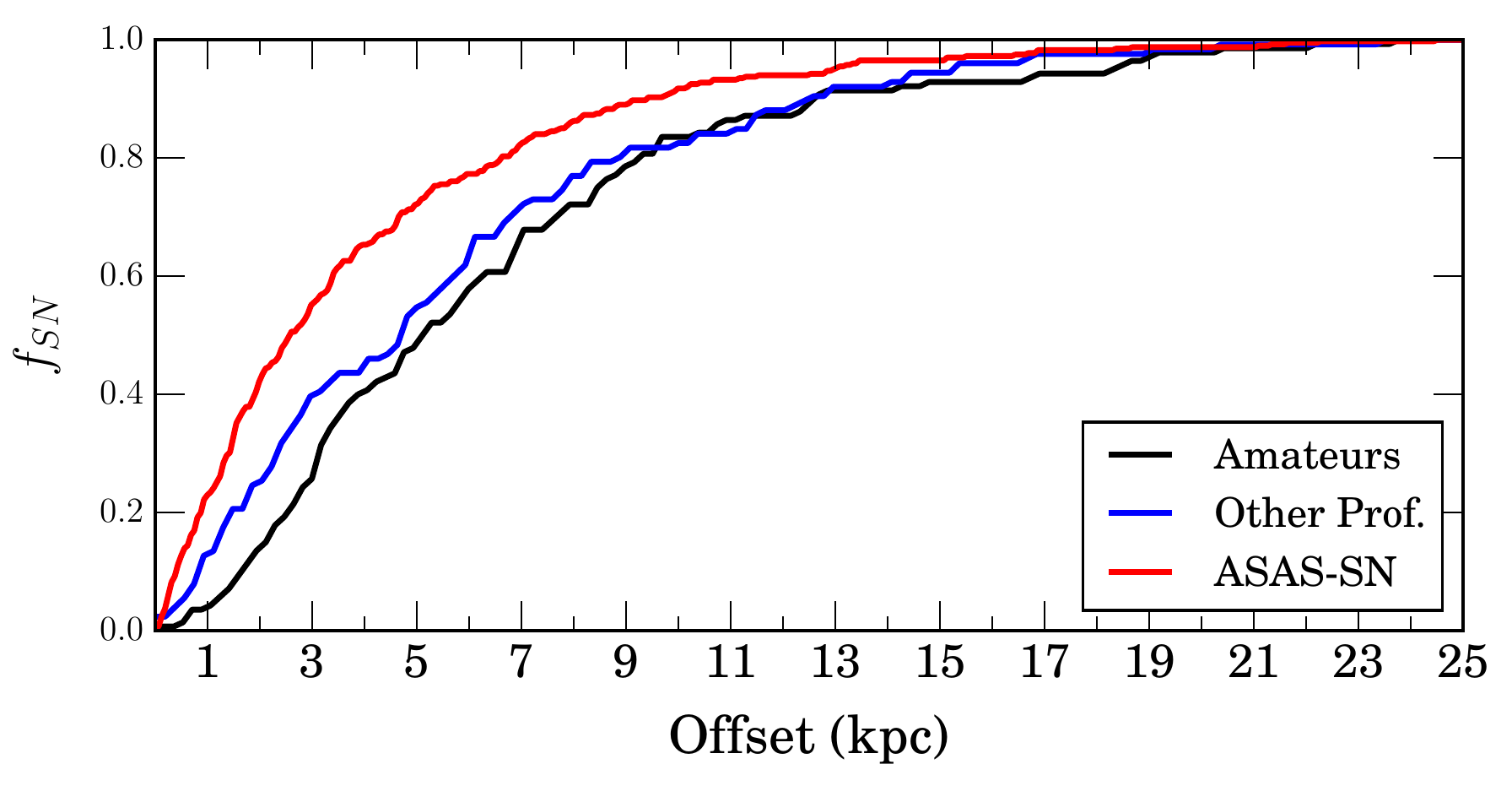}}}
\caption{Cumulative, normalized distributions of host galaxy absolute magnitude (upper panel), offset from host nucleus in arcseconds (center panel), and offset from host nucleus in kpc (bottom panel) for the ASAS-SN supernova sample (red), the other professional sample (blue), and the amateur sample (black). These figures further illustrate the trends from Figure~\ref{fig:offmag}: Amateur discoveries are clearly more biased towards more luminous hosts than professional surveys (including ASAS-SN), while ASAS-SN finds supernovae at smaller offsets, regardless of whether offset is measured in arcseconds or kpc.}
\label{fig:offmag_dist}
\end{minipage}
\end{figure*}

%%%%%%%%%%%%%%%%%
% Section: Analysis
%%%%%%%%%%%%%%%%%

\section{Analysis of the Sample}
\label{sec:analysis}

Combining all the bright supernovae discovered between 2014 May 01, when ASAS-SN became operational in both hemispheres, and 2016 December 31 provides a sample of 668 supernovae once we exclude ASAS-SN discoveries with $m_{peak}>17.0$ \citep{holoien16d,holoien17a}. Of these, 58\% (389) were discovered by ASAS-SN, 21\% (137) were discovered by other professional surveys, and 21\% (142) were discovered by amateur astronomers. 449 were Type Ia supernovae, 178 were Type II supernovae, 40 were Type Ib/Ic supernovae, and 1 was a superluminous supernova. As in our previous catalogs, we consider Type IIb supernovae as part of the Type II sample to allow for more direct comparison with the results of \citet{li11}. ASASSN-15lh is excluded in analyses that follow looking at trends by type, as all available evidence points to it being an extremely luminous Type I SLSN \citep{dong16,godoy-rivera17}, though it has also been classified as a tidal disruption event around a Kerr black hole \citep{leloudas16}. ASAS-SN discoveries account for 66\% of the Type Ia supernovae, 45\% of the Type II supernovae, and 25\% of the Type Ib/Ic supernovae. Amateur discoveries account for 16\%, 30\%, and 48\% of the Type Ia, Type II, and Type Ib/Ic supernovae in the sample, respectively, and discoveries from other professional surveys account for the remaining 18\%, 25\%, and 28\% of each type.

Figure~\ref{fig:piechart} shows pie charts breaking down the type distributions of supernovae in the ASAS-SN, non-ASAS-SN, and combined samples. Type Ia supernovae represent the largest fraction of supernovae in all three samples, as expected for a magnitude-limited sample \citep[e.g.,][]{li11}. Comparing to the ``ideal magnitude-limited sample'' breakdown predicted from the LOSS sample in \citet{li11}, where there are 79\% Type Ia, 17\% Type II, and 4\% Type Ib/Ic, the ASAS-SN sample matches the LOSS prediction almost exactly. The non-ASAS-SN sample and the combined sample have higher fractions of core-collapse supernovae, as was the case in our previous catalogs \citep{holoien16d,holoien17a}.

ASAS-SN continues to be the dominant source of bright supernova discoveries, and we often discover supernovae shortly after explosion due to our rapid cadence: of the 336 ASAS-SN discoveries with approximate discovery ages, 69\% (232) were discovered prior to reaching their peak brightness. As was seen in \citet{holoien17a}, ASAS-SN is less affected by host galaxy selection effects than other bright supernova searches. For example, 25\% (96) of the ASAS-SN bright supernovae were found in catalogued hosts that did not have previous redshift measurements available in NED, and an additional 4\% (14) were discovered in uncatalogued hosts or have no apparent host galaxy. Conversely, only 16\% (44) of non-ASAS-SN discoveries were found in catalogued hosts without redshift measurements, and only 3\% (8) were in uncatalogued galaxies or were hostless.

%%%%%%%%%%%%%%%%%
% Figure: 2016 Histogram
%%%%%%%%%%%%%%%%%

\begin{figure*}
\begin{minipage}{\textwidth}
\centering
\subfloat{{\includegraphics[width=0.95\linewidth]{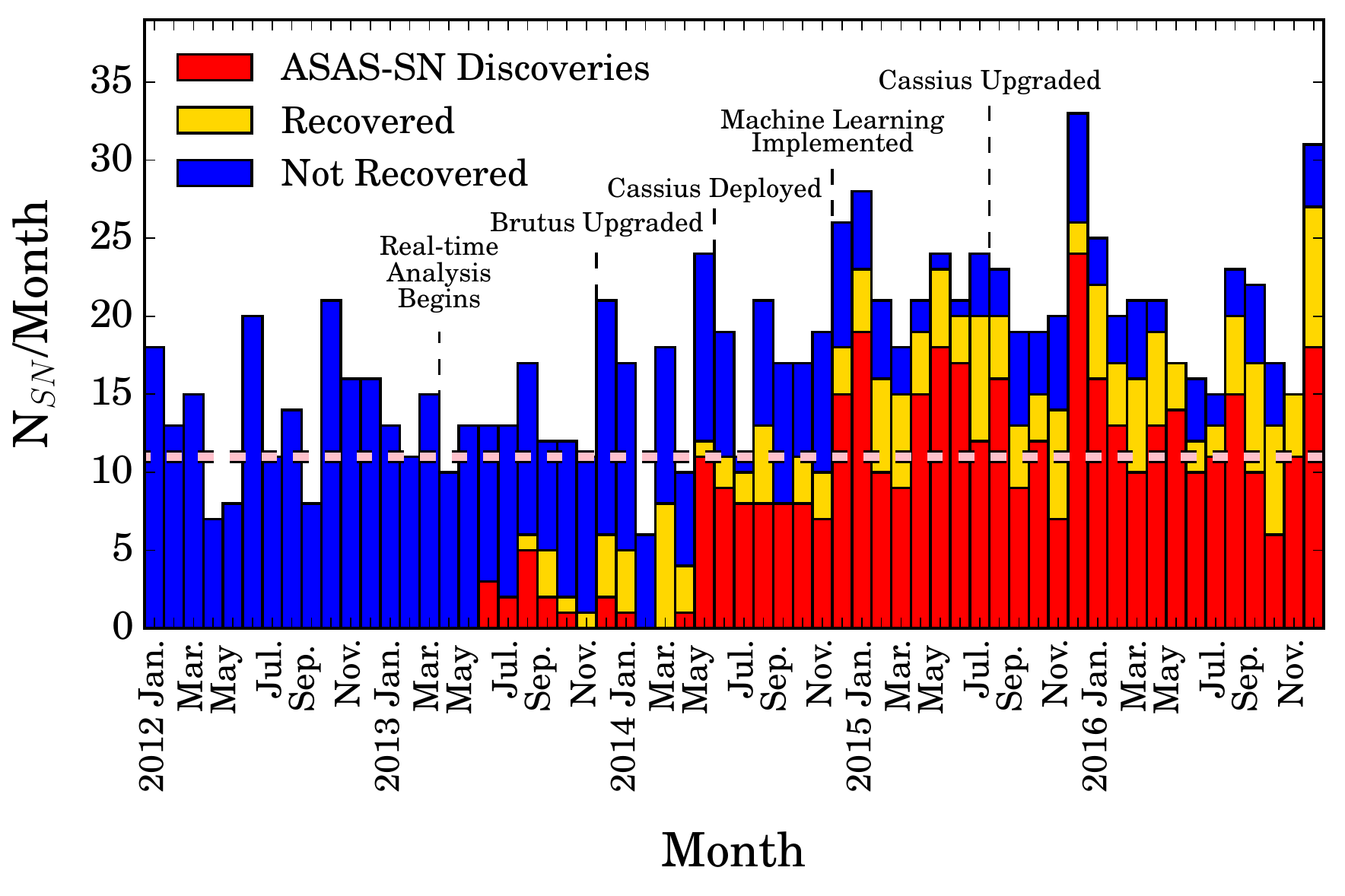}}}
\caption{Histogram of bright supernova discoveries in each month from 2012 through 2016. ASAS-SN discoveries are shown in red, supernovae discovered by other sources and recovered in ASAS-SN data are shown in yellow, and supernovae that were not recovered by ASAS-SN are shown in blue. Significant milestones in the ASAS-SN timeline are also shown. The dashed pink line shows the median number of supernovae discovered in each month from 2010 through 2012. The number of bright supernova discoveries has exceeded this previous median in every month since ASAS-SN became operational in both hemispheres in 2014 May, and ASAS-SN discoveries account for at least half of all bright supernova discoveries in every month since 2014 April.}
\label{fig:histogram}
\end{minipage}
\end{figure*}

As we showed in our previous catalogs, ASAS-SN discoveries have a smaller average offset from their host galaxy nuclei than bright supernovae discovered by other searches. The host galaxy $K_S$-band absolute magnitudes and the offsets of the supernovae from the host centers for all supernovae in our sample are shown in Figure~\ref{fig:offmag}. The median offsets and magnitudes are shown with horizontal and vertical lines for each supernova source (ASAS-SN, amateurs, or other professionals). A luminosity scale corresponding to the magnitude scale is given on the upper axis of the figure to help put the magnitude scale in perspective, assuming that a typical $L_\star$ galaxy has $M_{\star,K_S}=-24.2$ \citep{kochanek01}.

Amateur supernova searches tend to observe bright, nearby galaxies and use less sophisticated detection techniques than professional surveys, resulting in discoveries that are significantly biased towards more luminous hosts and larger offsets from the host nucleus. As we found previously \citep{holoien17a}, other professional surveys continue to discover supernovae with smaller angular separations than amateurs (median value of 11\farcs{8} vs. 16\farcs{5}), but show a similar median offset in terms of physical separation (5.0 kpc for professionals, 5.2 kpc for amateurs). ASAS-SN continues to be less biased against discoveries close to the host nucleus than either comparison group, as ASAS-SN discoveries show median offsets of 5\farcs{0} and 2.6 kpc. 

These trends are more easily visible when looking at the cumulative distributions of the host galaxy magnitudes and offsets from host nuclei, as shown in Figure~\ref{fig:offmag_dist}. The distributions clearly show that the ASAS-SN and other professional samples stand out from the amateur sample in host galaxy luminosity, and that supernovae discovered by ASAS-SN are more concentrated towards the centers of their hosts than those discovered by either amateurs or other professionals. While the majority of non-ASAS-SN professional discoveries continue to be made by professional searches that do not use difference imaging (e.g., MASTER, Gaia, CRTS), a larger fraction of other professional discoveries were made by surveys that do use difference imaging in 2016 than in previous years due to the start of the ATLAS survey. ASAS-SN continues to find sources with smaller median offsets than its competitors despite this fact, implying that the avoidance of the central regions of galaxies is still fairly common in surveys other than ASAS-SN, regardless of survey strategy and techniques.

The median host magnitudes are $M_{K_S}\simeq -22.6$, $M_{K_S}\simeq -22.8$, and $M_{K_S}\simeq -23.8$ for ASAS-SN discoveries, other professional discoveries, and amateur discoveries, respectively. There remains a clear distinction between professional surveys (including ASAS-SN) and amateurs in terms of host luminosity, and ASAS-SN discoveries now have a fainter median than those of other professional surveys.

As we have shown previously, one way the impact of ASAS-SN on the discovery of bright supernovae can be seen is by looking at the number of bright supernovae discovered per month in recent years \citep[e.g.,][]{holoien17a}. In Figure~\ref{fig:histogram} we show the number of supernovae with $m_{peak}\leq17$ per month in each month from 2012 through 2016. Milestones in the ASAS-SN timeline, such as the deployment of our southern unit Cassius and software improvements, are shown on the figure to help visualize the impact of these hardware and software improvements.

In its first year of operation, ASAS-SN had little effect on the number of bright supernovae being discovered per month: the average number of bright supernovae discovered per month from 2012 January through 2013 May was 13 with a scatter of 4 supernovae per month, and from 2013 June through 2014 May the average was 15 with a scatter of 5 supernovae per month. However, the addition of our southern unit Cassius and improvements to our pipeline dramatically impacted our detection efficiency and survey cadence, resulting in a significant increase in the number of supernovae discovered per month: since ASAS-SN became operational in both hemispheres, the average number of bright supernova discoveries has increased to 20 with a scatter of 5 supernovae per month. This indicates that ASAS-SN has increased the rate of bright supernovae discovered per month since becoming operational in the southern hemisphere, from $\sim13\pm2$ supernovae per month to $\sim20\pm2$ supernovae per month, and has continued to maintain this increased rate for the last 2.5 years---the addition of the 2016 supernovae has only decreased the average number of discoveries by 1 supernova per month from the previous 2014$+$2015 sample \citep{holoien17a}. ASAS-SN is discovering supernovae that otherwise would not be found, allowing us to construct a more complete sample of bright, nearby supernovae than was previously possible.

\begin{figure}
\centering
\subfloat{{\includegraphics[width=0.95\linewidth]{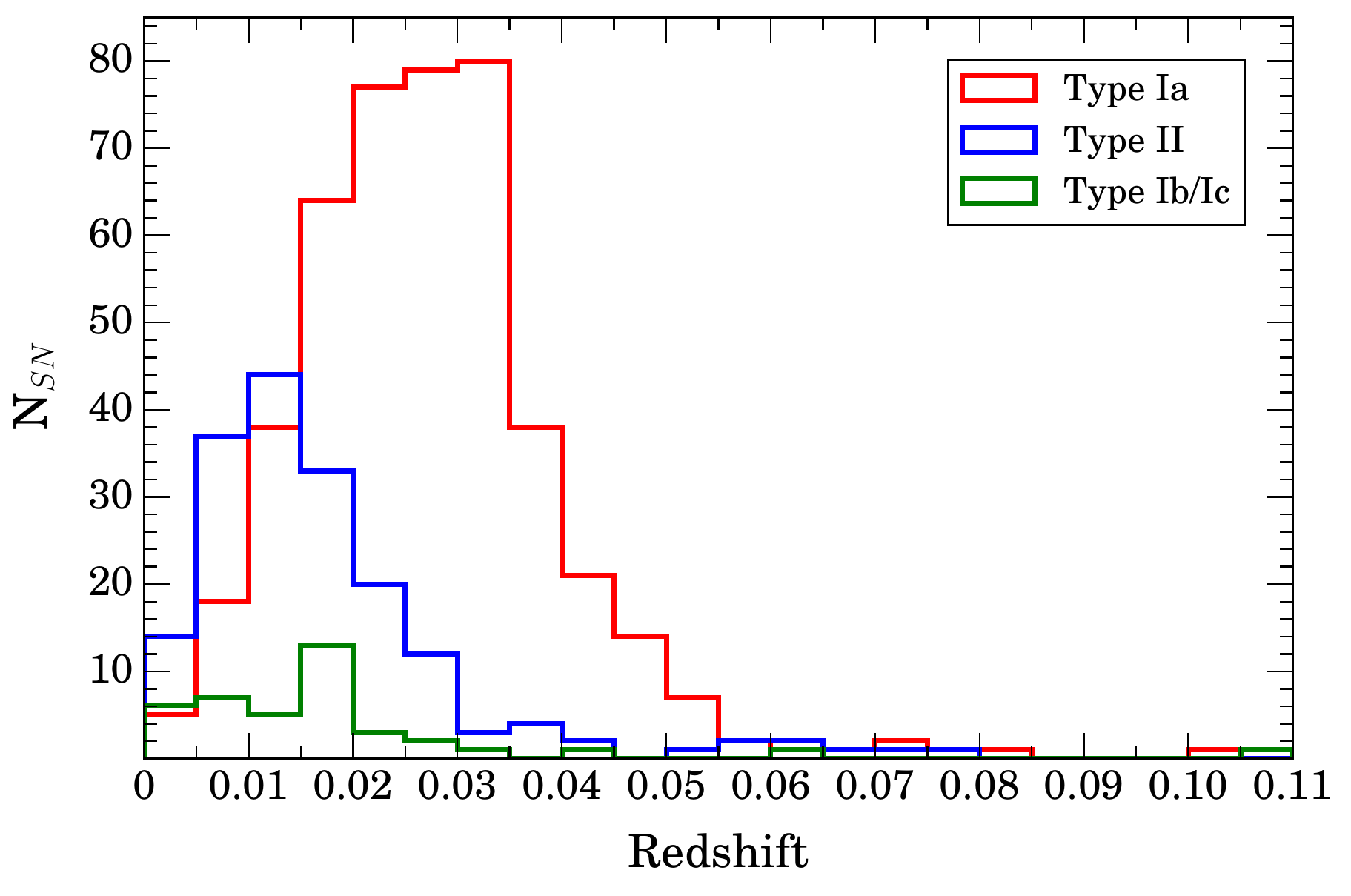}}}
\caption{Histograms of supernova redshifts from our complete sample with a bin width of $z=0.005$. Distributions for Type Ia (red line), Type II (blue line), and Type Ib/Ic (green line) supernovae are shown separately, with subtypes (such as SN 1991T-like and SN 1991bg-like Type Ia supernovae) included as part of their parent groups. As expected due to their larger intrinsic brightness, Type Ia supernovae are predominantly found at higher redshifts, while less luminous core-collapse supernovae are found at comparatively lower redshifts.}
\label{fig:redshift}
\end{figure}

Figure~\ref{fig:redshift} shows the redshift distribution of our full sample, divided by type. There is a clear distinction between the three types shown, with the Type Ia distribution peaking between $z=0.03$ and $z=0.035$, the Type II distribution peaking between $z=0.01$ and $z=0.015$, and the Type Ib/Ic distribution peaking between $z=0.015$ and $z=0.02$. Type Ia supernovae have a more luminous mean peak luminosity than core-collapse supernovae, so this distribution is expected for our magnitude-limited sample, and is similar to what we have seen in our previous catalogs \citep{holoien16d,holoien17a}.

Finally, we show a cumulative histogram of supernova peak magnitudes with $13.5<m_{peak}<17.0$ in Figure~\ref{fig:mag_dist}. As in our previous catalogs, the Figure shows ASAS-SN discoveries, ASAS-SN discoveries and supernovae recovered by ASAS-SN, and all supernovae from our sample separately. While amateur observers still account for a large number of the brightest discoveries (those with $m_{peak}\lesssim14.5$; \citealt{holoien17a}), ASAS-SN has discovered a significant fraction of these very bright supernovae in 2015 and 2016, accounting for roughly half of such discoveries in our complete sample. ASAS-SN recovers the vast majority of such very bright cases that it does not discover, showing that it is competitive with amateurs who observe the small number of very low-redshift galaxies with high cadence. ASAS-SN discovered or recovered every supernova with $m_{peak}<14.3$ in 2016 and accounts for a large fraction of the brightest supernovae overall.

Figure~\ref{fig:mag_dist} also illustrates an estimate of the completeness of our sample. We fit a broken power-law (shown with a green dashed line in the Figure) to the magnitudes of the observable supernovae brighter than $m_{peak}=17.01$, assuming a Euclidean slope below the break magnitude and a variable slope for higher magnitudes. We derived the parameters of the fit using Markov Chain Monte Carlo (MCMC) methods. For only the supernovae discovered by ASAS-SN in our complete sample, the number counts are consistent with the Euclidean slope down to $m=16.34\pm0.07$, similar to what we found in the 2015 sample \citep{holoien17a}. We find break magnitudes of $m=16.26\pm0.06$ and $m=16.19\pm0.09$ for the sample of supernovae discovered and recovered by ASAS-SN and the sample of all bright supernovae, respectively, again similar to the 2015 results.

We find that the integral completenesses of the three samples relative to Euclidean predictions are $0.97\pm0.02$ ($0.68\pm0.03$), $0.94\pm0.02$ ($0.65\pm0.03$), and $0.93\pm0.03$ ($0.71\pm0.03$) at 16.5 (17.0)~mag for the ASAS-SN discovered sample, the ASAS-SN discovered $+$ recovered sample, and the total sample, respectively. The differential completenesses relative to Euclidean predictions are $0.71\pm0.10$ ($0.22\pm0.04$), $0.62\pm0.06$ ($0.22\pm0.04$), and $0.67\pm0.05$ ($0.36\pm0.04$) at 16.5 (17.0)~mag, respectively. These results imply that roughly 70\% of the supernovae brighter than $m_{peak}=17$ are being found, and that 20$-$30\% of the $m_{peak}=17$ supernovae are being found, relative to the Euclidean expectation extrapolated from brighter supernovae, an improvement over the 15$-$20\% seen in the 2015 sample. The Euclidean approximation used here does not take into account deviations from Euclidean geometry, the effects of time dilation on supernova rates, or K-corrections, and thus likely modestly underestimates the true completeness for faint supernovae. These higher order corrections will be included when we carry out a full analysis of nearby supernova rates.

%%%%%%%%%%%%%%%%%
% Figure: Redshift Distribution
%%%%%%%%%%%%%%%%%

%%%%%%%%%%%%%%%%%
% Section: Conclusions
%%%%%%%%%%%%%%%%%

\section{Conclusions}
\label{sec:disc}

This paper represents a comprehensive catalog of spectroscopically confirmed bright supernovae and their hosts from the ASAS-SN team, comprising 248 supernovae discovered by ASAS-SN, other professional surveys, and amateur observers in 2016. Our total combined bright supernova sample now includes 668 supernovae, 387 discovered by ASAS-SN. The combined sample remains similar to that of an ideal magnitude-limited sample from \citet{li11} with a smaller proportion of Type Ia supernova relative to core-collapse supernovae than expected.

ASAS-SN is the only professional survey that provides a complete, rapid-cadence, all-sky survey of the nearby transient Universe, and continues to have a major impact on the discovery and follow-up of bright supernovae. Even with the advent of recent professional surveys, amateur astronomers, who focus on bright and nearby galaxies for their supernova searches, remain the primary competition to ASAS-SN for new discoveries. Our analyses show that ASAS-SN continues to find supernovae that would not be found otherwise (e.g., Figure~\ref{fig:histogram}) and that it finds supernovae closer to galactic nuclei and in less luminous hosts than its competitors (Figure~\ref{fig:offmag}). In 2016 ASAS-SN recovered the majority of bright supernovae that it did not discover, as was the case in 2015, and discovered or recovered all but one of the very bright ($m_{peak}\leq15$) supernovae that were discovered in 2016.

%%%%%%%%%%%%%%%%%
% Figure: Mag Distribution
%%%%%%%%%%%%%%%%%

\begin{figure}
\centering
\subfloat{{\includegraphics[width=0.95\linewidth]{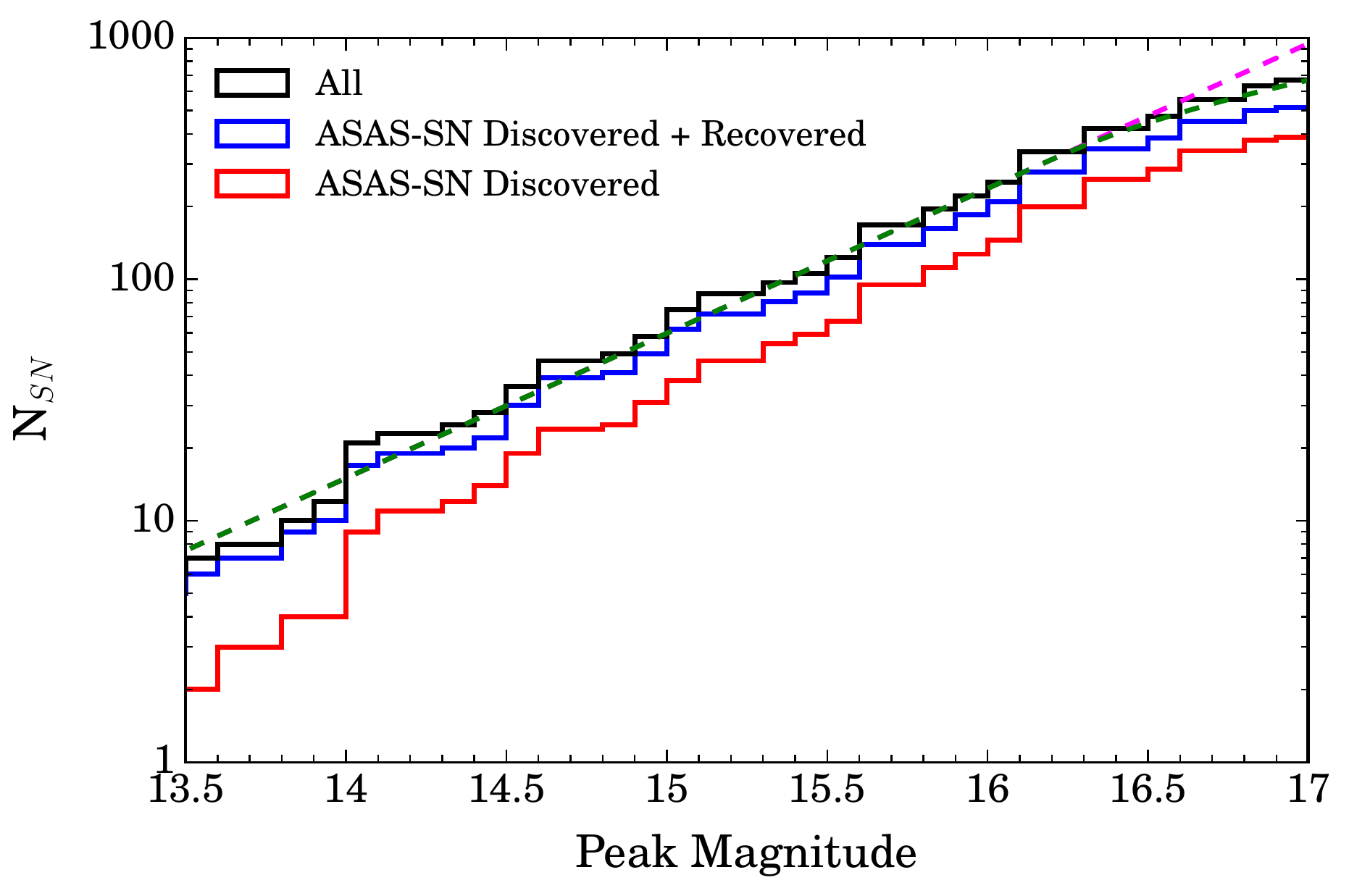}}}
\caption{Cumulative histogram of supernova peak magnitudes using a 0.1 magnitude bin width. The distributions for only ASAS-SN discoveries (red line), ASAS-SN discoveries and supernovae recovered independently by ASAS-SN (blue line), and all supernovae in the sample (black line) are shown separately. The green dashed line shows a broken power-law fit that has been normalized to the complete sample with a Euclidean slope below the break magnitude and a variable slope for fainter sources, and the lavender dashed line shows an extrapolation of the Euclidean slope to $m=17$. The sample is roughly 70\% complete for $m_{peak}<17$.}
\label{fig:mag_dist}
\end{figure}

Our sample completeness is comparable to what it was at the end of 2015. Figure~\ref{fig:mag_dist} shows that the magnitude distribution of supernovae discovered between 2014 May 1 and 2016 December 31 is roughly complete to a peak magnitude of $m_{peak}=16.2$, slightly worse than in 2015, but that it is roughly 70\% complete for $m_{peak}\leq17.0$, a slight improvement over 2015. This analysis serves as a precursor to rate calculations which will be presented in Holoien et al. (\emph{in prep.}). These rate calculations may have a significant impact on a number of fields, including the nearby core-collapse rate \citep[e.g.,][]{horiuchi11,horiuchi13} and multi-messenger studies ranging from gravitational waves \citep[e.g.,][]{ando13,nakamura16}, to MeV gamma rays from Type Ia supernovae \citep[e.g.,][]{horiuchi10,diehl14,churazov15} to GeV--TeV gamma rays and neutrinos from rare types of core-collapse supernovae \citep[e.g.,][]{ando05,murase11,abbasi12}. Such joint measurements would be a great increase in the scientific reach of ASAS-SN discoveries.

This is the third of a yearly series of bright supernova catalogs provided by the ASAS-SN team, and it is our hope that these catalogs will provide convenient and useful repositories of bright supernovae and their host galaxies that can be used for new and interesting population studies. ASAS-SN continues to discover many of the best and brightest transients in the sky, and these catalogs are one way in which we can use our unbiased sample to impact supernova research now and in the future. 

\section*{Acknowledgments}

The authors thank A. Piascik and I. Shivvers for providing redshift measurements.

We thank Las Cumbres Observatory and its staff for their continued support of ASAS-SN.  

ASAS-SN is supported by the Gordon and Betty Moore Foundation through grant GBMF5490 to the Ohio State University and NSF grant AST-1515927. Development of ASAS-SN has been supported by NSF grant AST-0908816, the Center for Cosmology and AstroParticle Physics at the Ohio State University, the Mt. Cuba Astronomical Foundation, the Chinese Academy of Sciences South America Center for Astronomy (CASSACA), and by George Skestos.

TW-SH is supported by the DOE Computational Science Graduate Fellowship, grant number DE-FG02-97ER25308. JSB, KZS, and CSK are supported by NSF grant AST-1515927. KZS and CSK are also supported by NSF grant AST-1515876. BJS is supported by NASA through Hubble Fellowship grant HST-HF-51348.001 awarded by the Space Telescope Science Institute, which is operated by the Association of Universities for Research in Astronomy, Inc., for NASA, under contract NAS 5-26555. Support for JLP is in part provided by FONDECYT through the grant 1151445 and by the Ministry of Economy, Development, and Tourism's Millennium Science Initiative through grant IC120009, awarded to The Millennium Institute of Astrophysics, MAS. SD and PC are supported by ?the Strategic Priority Research Program-The Emergence of Cosmological Structures? of the Chinese Academy of Sciences (Grant No. XDB09000000). SD, SB and PC are also supported by Project 11573003 supported by NSFC. JFB is supported by NSF grant PHY-1404311. JS acknowledges support from the Packard Foundation. MDS is supported by generous grants provided by the Danish Agency for Science and Technology and Innovation realized through a Sapere Aude Level 2 grant and the Villum foundation. PRW acknowledges support from the US Department of Energy as part of the Laboratory Directed Research and Development program at LANL.

This research uses data obtained through the Telescope Access Program (TAP), which has been funded by ?the Strategic Priority Research Program-The Emergence of Cosmological Structures? of the Chinese Academy of Sciences (Grant No.11 XDB09000000) and the Special Fund for Astronomy from the Ministry of Finance.

This research has made use of the XRT Data Analysis Software (XRTDAS) developed under the responsibility of the ASI Science Data Center (ASDC), Italy. At Penn State the NASA {\swift} program is support through contract NAS5-00136.

This research was made possible through the use of the AAVSO Photometric All-Sky Survey (APASS), funded by the Robert Martin Ayers Sciences Fund.

This research has made use of data provided by Astrometry.net \citep{barron08,lang10}.

This paper uses data products produced by the OIR Telescope Data Center, supported by the Smithsonian Astrophysical Observatory.

Observations made with the NASA Galaxy Evolution Explorer (GALEX) were used in the analyses presented in this manuscript. Some of the data presented in this paper were obtained from the Mikulski Archive for Space Telescopes (MAST). STScI is operated by the Association of Universities for Research in Astronomy, Inc., under NASA contract NAS5-26555. Support for MAST for non-HST data is provided by the NASA Office of Space Science via grant NNX13AC07G and by other grants and contracts.

Funding for SDSS-III has been provided by the Alfred P. Sloan Foundation, the Participating Institutions, the National Science Foundation, and the U.S. Department of Energy Office of Science. The SDSS-III web site is http://www.sdss3.org/.

This publication makes use of data products from the Two Micron All Sky Survey, which is a joint project of the University of Massachusetts and the Infrared Processing and Analysis Center/California Institute of Technology, funded by NASA and the National Science Foundation.

This publication makes use of data products from the Wide-field Infrared Survey Explorer, which is a joint project of the University of California, Los Angeles, and the Jet Propulsion Laboratory/California Institute of Technology, funded by NASA.

This research is based in part on observations obtained at the Southern Astrophysical Research (SOAR) telescope, which is a joint project of the Minist\'{e}rio da Ci\^{e}ncia, Tecnologia, e Inova\c{c}\~{a}o (MCTI) da Rep\'{u}blica Federativa do Brasil, the U.S. National Optical Astronomy Observatory (NOAO), the University of North Carolina at Chapel Hill (UNC), and Michigan State University (MSU). 

This research has made use of the NASA/IPAC Extragalactic Database (NED), which is operated by the Jet Propulsion Laboratory, California Institute of Technology, under contract with NASA.

\bibliographystyle{mnras2}
\bibliography{../../bibliography/bibliography}

\begin{thebibliography}{}
\makeatletter
\relax
\def\mn@urlcharsother{\let\do\@makeother \do\$\do\&\do\#\do\^\do\_\do\%\do\~}
\def\mn@doi{\begingroup\mn@urlcharsother \@ifnextchar [ {\mn@doi@}
  {\mn@doi@[]}}
\def\mn@doi@[#1]#2{\def\@tempa{#1}\ifx\@tempa\@empty \href
  {http://dx.doi.org/#2} {doi:#2}\else \href {http://dx.doi.org/#2} {#1}\fi
  \endgroup}
\def\mn@eprint#1#2{\mn@eprint@#1:#2::\@nil}
\def\mn@eprint@arXiv#1{\href {http://arxiv.org/abs/#1} {{\tt arXiv:#1}}}
\def\mn@eprint@dblp#1{\href {http://dblp.uni-trier.de/rec/bibtex/#1.xml}
  {dblp:#1}}
\def\mn@eprint@#1:#2:#3:#4\@nil{\def\@tempa {#1}\def\@tempb {#2}\def\@tempc
  {#3}\ifx \@tempc \@empty \let \@tempc \@tempb \let \@tempb \@tempa \fi \ifx
  \@tempb \@empty \def\@tempb {arXiv}\fi \@ifundefined
  {mn@eprint@\@tempb}{\@tempb:\@tempc}{\expandafter \expandafter \csname
  mn@eprint@\@tempb\endcsname \expandafter{\@tempc}}}

\bibitem[\protect\citeauthoryear{{Abbasi} et~al.,}{{Abbasi}
  et~al.}{2012}]{abbasi12}
{Abbasi} R.,  et~al., 2012, \mn@doi [\aap] {10.1051/0004-6361/201118071}, \href
  {http://adsabs.harvard.edu/abs/2012A%26A...539A..60A} {539, A60}

\bibitem[\protect\citeauthoryear{{Ando} \& {Beacom}}{{Ando} \&
  {Beacom}}{2005}]{ando05}
{Ando} S.,  {Beacom} J.~F.,  2005, \mn@doi [Physical Review Letters]
  {10.1103/PhysRevLett.95.061103}, \href
  {http://adsabs.harvard.edu/abs/2005PhRvL..95f1103A} {95, 061103}

\bibitem[\protect\citeauthoryear{{Ando} et~al.,}{{Ando} et~al.}{2013}]{ando13}
{Ando} S.,  et~al., 2013, \mn@doi [Reviews of Modern Physics]
  {10.1103/RevModPhys.85.1401}, \href
  {http://adsabs.harvard.edu/abs/2013RvMP...85.1401A} {85, 1401}

\bibitem[\protect\citeauthoryear{{Balam} \& {Graham}}{{Balam} \&
  {Graham}}{2016a}]{asassn16cs_spec_atel}
{Balam} D.~D.,  {Graham} M.~L.,  2016a, The Astronomer's Telegram, \href
  {http://adsabs.harvard.edu/abs/2016ATel.8788....1B} {8788}

\bibitem[\protect\citeauthoryear{{Balam} \& {Graham}}{{Balam} \&
  {Graham}}{2016b}]{asassn16es_spec_atel}
{Balam} D.~D.,  {Graham} M.~L.,  2016b, The Astronomer's Telegram, \href
  {http://adsabs.harvard.edu/abs/2016ATel.9016....1B} {9016}

\bibitem[\protect\citeauthoryear{{Balam} \& {Graham}}{{Balam} \&
  {Graham}}{2016c}]{asassn16fa_spec_atel}
{Balam} D.~D.,  {Graham} M.~L.,  2016c, The Astronomer's Telegram, \href
  {http://adsabs.harvard.edu/abs/2016ATel.9047....1B} {9047}

\bibitem[\protect\citeauthoryear{{Baltay} et~al.,}{{Baltay}
  et~al.}{2013}]{baltay13}
{Baltay} C.,  et~al., 2013, \mn@doi [\pasp] {10.1086/671198}, \href
  {http://adsabs.harvard.edu/abs/2013PASP..125..683B} {125, 683}

\bibitem[\protect\citeauthoryear{{Barron}, {Stumm}, {Hogg}, {Lang}  \&
  {Roweis}}{{Barron} et~al.}{2008}]{barron08}
{Barron} J.~T.,  {Stumm} C.,  {Hogg} D.~W.,  {Lang} D.,   {Roweis} S.,  2008,
  \mn@doi [\aj] {10.1088/0004-6256/135/1/414}, \href
  {http://adsabs.harvard.edu/abs/2008AJ....135..414B} {135, 414}

\bibitem[\protect\citeauthoryear{{Bersier}}{{Bersier}}{2016a}]{asassn16hd_spec_atel}
{Bersier} D.,  2016a, The Astronomer's Telegram, \href
  {http://adsabs.harvard.edu/abs/2016ATel.9253....1B} {9253}

\bibitem[\protect\citeauthoryear{{Bersier}}{{Bersier}}{2016b}]{asassn16hh_spec_atel}
{Bersier} D.,  2016b, The Astronomer's Telegram, \href
  {http://adsabs.harvard.edu/abs/2016ATel.9258....1B} {9258}

\bibitem[\protect\citeauthoryear{{Bersier}}{{Bersier}}{2016c}]{asassn16hr_spec_atel}
{Bersier} D.,  2016c, The Astronomer's Telegram, \href
  {http://adsabs.harvard.edu/abs/2016ATel.9273....1B} {9273}

\bibitem[\protect\citeauthoryear{{Bersier} et~al.,}{{Bersier}
  et~al.}{2016}]{asassn16mj_atel}
{Bersier} D.,  et~al., 2016, The Astronomer's Telegram, \href
  {http://adsabs.harvard.edu/abs/2016ATel.9697....1B} {9697}

\bibitem[\protect\citeauthoryear{{Blanchard} et~al.,}{{Blanchard}
  et~al.}{2017}]{blanchard17}
{Blanchard} P.~K.,  et~al., 2017, preprint, \href
  {http://adsabs.harvard.edu/abs/2017arXiv170307816B} {} (\mn@eprint {arXiv}
  {1703.07816})

\bibitem[\protect\citeauthoryear{{Blondin} \& {Tonry}}{{Blondin} \&
  {Tonry}}{2007}]{blondin07}
{Blondin} S.,  {Tonry} J.~L.,  2007, \mn@doi [\apj] {10.1086/520494}, \href
  {http://adsabs.harvard.edu/abs/2007ApJ...666.1024B} {666, 1024}

\bibitem[\protect\citeauthoryear{{Bock} et~al.,}{{Bock}
  et~al.}{2016a}]{asassn16at_atel}
{Bock} G.,  et~al., 2016a, The Astronomer's Telegram, \href
  {http://adsabs.harvard.edu/abs/2016ATel.8566....1B} {8566}

\bibitem[\protect\citeauthoryear{{Bock} et~al.,}{{Bock}
  et~al.}{2016b}]{asassn16ec_atel}
{Bock} G.,  et~al., 2016b, The Astronomer's Telegram, \href
  {http://adsabs.harvard.edu/abs/2016ATel.8915....1B} {8915}

\bibitem[\protect\citeauthoryear{{Bock} et~al.,}{{Bock}
  et~al.}{2016c}]{asassn16fq_atel}
{Bock} G.,  et~al., 2016c, The Astronomer's Telegram, \href
  {http://adsabs.harvard.edu/abs/2016ATel.9091....1B} {9091}

\bibitem[\protect\citeauthoryear{{Bock} et~al.,}{{Bock}
  et~al.}{2016d}]{asassn16lz_atel}
{Bock} G.,  et~al., 2016d, The Astronomer's Telegram, \href
  {http://adsabs.harvard.edu/abs/2016ATel.9668....1B} {9668}

\bibitem[\protect\citeauthoryear{{Bose}, {Thorstensen}, {Klusmeyer}, {Stanek},
  {Prieto}  \& {Dong}}{{Bose} et~al.}{2016}]{asassn16jf_spec_atel}
{Bose} S.,  {Thorstensen} J.,  {Klusmeyer} J.,  {Stanek} K.,  {Prieto} J.,
  {Dong} S.,  2016, The Astronomer's Telegram, \href
  {http://adsabs.harvard.edu/abs/2016ATel.9405....1B} {9405}

\bibitem[\protect\citeauthoryear{{Brimacombe} et~al.,}{{Brimacombe}
  et~al.}{2016a}]{asassn16ah_atel}
{Brimacombe} J.,  et~al., 2016a, The Astronomer's Telegram, \href
  {http://adsabs.harvard.edu/abs/2016ATel.8539....1B} {8539}

\bibitem[\protect\citeauthoryear{{Brimacombe} et~al.,}{{Brimacombe}
  et~al.}{2016b}]{asassn16aj_atel}
{Brimacombe} J.,  et~al., 2016b, The Astronomer's Telegram, \href
  {http://adsabs.harvard.edu/abs/2016ATel.8542....1B} {8542}

\bibitem[\protect\citeauthoryear{{Brimacombe} et~al.,}{{Brimacombe}
  et~al.}{2016c}]{asassn16bl_atel}
{Brimacombe} J.,  et~al., 2016c, The Astronomer's Telegram, \href
  {http://adsabs.harvard.edu/abs/2016ATel.8652....1B} {8652}

\bibitem[\protect\citeauthoryear{{Brimacombe} et~al.,}{{Brimacombe}
  et~al.}{2016d}]{asassn16bq_atel}
{Brimacombe} J.,  et~al., 2016d, The Astronomer's Telegram, \href
  {http://adsabs.harvard.edu/abs/2016ATel.8685....1B} {8685}

\bibitem[\protect\citeauthoryear{{Brimacombe} et~al.,}{{Brimacombe}
  et~al.}{2016e}]{asassn16bv_atel}
{Brimacombe} J.,  et~al., 2016e, The Astronomer's Telegram, \href
  {http://adsabs.harvard.edu/abs/2016ATel.8703....1B} {8703}

\bibitem[\protect\citeauthoryear{{Brimacombe} et~al.,}{{Brimacombe}
  et~al.}{2016f}]{asassn16bx_atel}
{Brimacombe} J.,  et~al., 2016f, The Astronomer's Telegram, \href
  {http://adsabs.harvard.edu/abs/2016ATel.8712....1B} {8712}

\bibitem[\protect\citeauthoryear{{Brimacombe} et~al.,}{{Brimacombe}
  et~al.}{2016g}]{asassn16dw_atel}
{Brimacombe} J.,  et~al., 2016g, The Astronomer's Telegram, \href
  {http://adsabs.harvard.edu/abs/2016ATel.8897....1B} {8897}

\bibitem[\protect\citeauthoryear{{Brimacombe} et~al.,}{{Brimacombe}
  et~al.}{2016h}]{asassn16dx_atel}
{Brimacombe} J.,  et~al., 2016h, The Astronomer's Telegram, \href
  {http://adsabs.harvard.edu/abs/2016ATel.8898....1B} {8898}

\bibitem[\protect\citeauthoryear{{Brimacombe} et~al.,}{{Brimacombe}
  et~al.}{2016i}]{asassn16eq_atel}
{Brimacombe} J.,  et~al., 2016i, The Astronomer's Telegram, \href
  {http://adsabs.harvard.edu/abs/2016ATel.8979....1B} {8979}

\bibitem[\protect\citeauthoryear{{Brimacombe} et~al.,}{{Brimacombe}
  et~al.}{2016j}]{asassn16et_atel}
{Brimacombe} J.,  et~al., 2016j, The Astronomer's Telegram, \href
  {http://adsabs.harvard.edu/abs/2016ATel.9005....1B} {9005}

\bibitem[\protect\citeauthoryear{{Brimacombe} et~al.,}{{Brimacombe}
  et~al.}{2016k}]{asassn16fc_atel}
{Brimacombe} J.,  et~al., 2016k, The Astronomer's Telegram, \href
  {http://adsabs.harvard.edu/abs/2016ATel.9057....1B} {9057}

\bibitem[\protect\citeauthoryear{{Brimacombe} et~al.,}{{Brimacombe}
  et~al.}{2016l}]{asassn16ff_atel}
{Brimacombe} J.,  et~al., 2016l, The Astronomer's Telegram, \href
  {http://adsabs.harvard.edu/abs/2016ATel.9058....1B} {9058}

\bibitem[\protect\citeauthoryear{{Brimacombe} et~al.,}{{Brimacombe}
  et~al.}{2016m}]{asassn16ft_atel}
{Brimacombe} J.,  et~al., 2016m, The Astronomer's Telegram, \href
  {http://adsabs.harvard.edu/abs/2016ATel.9117....1B} {9117}

\bibitem[\protect\citeauthoryear{{Brimacombe} et~al.,}{{Brimacombe}
  et~al.}{2016n}]{asassn16fv_atel}
{Brimacombe} J.,  et~al., 2016n, The Astronomer's Telegram, \href
  {http://adsabs.harvard.edu/abs/2016ATel.9123....1B} {9123}

\bibitem[\protect\citeauthoryear{{Brimacombe} et~al.,}{{Brimacombe}
  et~al.}{2016o}]{asassn16gm_atel}
{Brimacombe} J.,  et~al., 2016o, The Astronomer's Telegram, \href
  {http://adsabs.harvard.edu/abs/2016ATel.9188....1B} {9188}

\bibitem[\protect\citeauthoryear{{Brimacombe} et~al.,}{{Brimacombe}
  et~al.}{2016p}]{asassn16gp_atel}
{Brimacombe} J.,  et~al., 2016p, The Astronomer's Telegram, \href
  {http://adsabs.harvard.edu/abs/2016ATel.9199....1B} {9199}

\bibitem[\protect\citeauthoryear{{Brimacombe} et~al.,}{{Brimacombe}
  et~al.}{2016q}]{asassn16gv_atel}
{Brimacombe} J.,  et~al., 2016q, The Astronomer's Telegram, \href
  {http://adsabs.harvard.edu/abs/2016ATel.9212....1B} {9212}

\bibitem[\protect\citeauthoryear{{Brimacombe} et~al.,}{{Brimacombe}
  et~al.}{2016r}]{asassn16hn_atel}
{Brimacombe} J.,  et~al., 2016r, The Astronomer's Telegram, \href
  {http://adsabs.harvard.edu/abs/2016ATel.9262....1B} {9262}

\bibitem[\protect\citeauthoryear{{Brimacombe} et~al.,}{{Brimacombe}
  et~al.}{2016s}]{asassn16hr_atel}
{Brimacombe} J.,  et~al., 2016s, The Astronomer's Telegram, \href
  {http://adsabs.harvard.edu/abs/2016ATel.9270....1B} {9270}

\bibitem[\protect\citeauthoryear{{Brimacombe} et~al.,}{{Brimacombe}
  et~al.}{2016t}]{asassn16hy_atel}
{Brimacombe} J.,  et~al., 2016t, The Astronomer's Telegram, \href
  {http://adsabs.harvard.edu/abs/2016ATel.9305....1B} {9305}

\bibitem[\protect\citeauthoryear{{Brimacombe} et~al.,}{{Brimacombe}
  et~al.}{2016u}]{asassn16hz_atel}
{Brimacombe} J.,  et~al., 2016u, The Astronomer's Telegram, \href
  {http://adsabs.harvard.edu/abs/2016ATel.9332....1B} {9332}

\bibitem[\protect\citeauthoryear{{Brimacombe} et~al.,}{{Brimacombe}
  et~al.}{2016v}]{asassn16if_atel}
{Brimacombe} J.,  et~al., 2016v, The Astronomer's Telegram, \href
  {http://adsabs.harvard.edu/abs/2016ATel.9338....1B} {9338}

\bibitem[\protect\citeauthoryear{{Brimacombe} et~al.,}{{Brimacombe}
  et~al.}{2016w}]{asassn16in_atel}
{Brimacombe} J.,  et~al., 2016w, The Astronomer's Telegram, \href
  {http://adsabs.harvard.edu/abs/2016ATel.9344....1B} {9344}

\bibitem[\protect\citeauthoryear{{Brimacombe} et~al.,}{{Brimacombe}
  et~al.}{2016x}]{asassn16jq_atel}
{Brimacombe} J.,  et~al., 2016x, The Astronomer's Telegram, \href
  {http://adsabs.harvard.edu/abs/2016ATel.9430....1B} {9430}

\bibitem[\protect\citeauthoryear{{Brimacombe} et~al.,}{{Brimacombe}
  et~al.}{2016y}]{asassn16jt_atel}
{Brimacombe} J.,  et~al., 2016y, The Astronomer's Telegram, \href
  {http://adsabs.harvard.edu/abs/2016ATel.9439....1B} {9439}

\bibitem[\protect\citeauthoryear{{Brimacombe} et~al.,}{{Brimacombe}
  et~al.}{2016z}]{asassn16jw_atel}
{Brimacombe} J.,  et~al., 2016z, The Astronomer's Telegram, \href
  {http://adsabs.harvard.edu/abs/2016ATel.9457....1B} {9457}

\bibitem[\protect\citeauthoryear{{Brimacombe} et~al.,}{{Brimacombe}
  et~al.}{2016aa}]{asassn16ke_atel}
{Brimacombe} J.,  et~al., 2016aa, The Astronomer's Telegram, \href
  {http://adsabs.harvard.edu/abs/2016ATel.9474....1B} {9474}

\bibitem[\protect\citeauthoryear{{Brimacombe} et~al.,}{{Brimacombe}
  et~al.}{2016ab}]{asassn16kv_atel}
{Brimacombe} J.,  et~al., 2016ab, The Astronomer's Telegram, \href
  {http://adsabs.harvard.edu/abs/2016ATel.9560....1B} {9560}

\bibitem[\protect\citeauthoryear{{Brimacombe} et~al.,}{{Brimacombe}
  et~al.}{2016ac}]{asassn16mt_atel}
{Brimacombe} J.,  et~al., 2016ac, The Astronomer's Telegram, \href
  {http://adsabs.harvard.edu/abs/2016ATel.9714....1B} {9714}

\bibitem[\protect\citeauthoryear{{Brimacombe} et~al.,}{{Brimacombe}
  et~al.}{2016ad}]{asassn16mz_atel}
{Brimacombe} J.,  et~al., 2016ad, The Astronomer's Telegram, \href
  {http://adsabs.harvard.edu/abs/2016ATel.9724....1B} {9724}

\bibitem[\protect\citeauthoryear{{Brimacombe} et~al.,}{{Brimacombe}
  et~al.}{2016ae}]{asassn16na_atel}
{Brimacombe} J.,  et~al., 2016ae, The Astronomer's Telegram, \href
  {http://adsabs.harvard.edu/abs/2016ATel.9727....1B} {9727}

\bibitem[\protect\citeauthoryear{{Brimacombe} et~al.,}{{Brimacombe}
  et~al.}{2016af}]{asassn16nm_atel}
{Brimacombe} J.,  et~al., 2016af, The Astronomer's Telegram, \href
  {http://adsabs.harvard.edu/abs/2016ATel.9785....1B} {9785}

\bibitem[\protect\citeauthoryear{{Brimacombe} et~al.,}{{Brimacombe}
  et~al.}{2016ag}]{asassn16no_atel}
{Brimacombe} J.,  et~al., 2016ag, The Astronomer's Telegram, \href
  {http://adsabs.harvard.edu/abs/2016ATel.9796....1B} {9796}

\bibitem[\protect\citeauthoryear{{Brimacombe} et~al.,}{{Brimacombe}
  et~al.}{2016ah}]{asassn16np_atel}
{Brimacombe} J.,  et~al., 2016ah, The Astronomer's Telegram, \href
  {http://adsabs.harvard.edu/abs/2016ATel.9798....1B} {9798}

\bibitem[\protect\citeauthoryear{{Brimacombe} et~al.,}{{Brimacombe}
  et~al.}{2016ai}]{asassn16nv_atel}
{Brimacombe} J.,  et~al., 2016ai, The Astronomer's Telegram, \href
  {http://adsabs.harvard.edu/abs/2016ATel.9804....1B} {9804}

\bibitem[\protect\citeauthoryear{{Brimacombe} et~al.,}{{Brimacombe}
  et~al.}{2016aj}]{asassn16ns_atel}
{Brimacombe} J.,  et~al., 2016aj, The Astronomer's Telegram, \href
  {http://adsabs.harvard.edu/abs/2016ATel.9812....1B} {9812}

\bibitem[\protect\citeauthoryear{{Brimacombe} et~al.,}{{Brimacombe}
  et~al.}{2016ak}]{asassn16nt_atel}
{Brimacombe} J.,  et~al., 2016ak, The Astronomer's Telegram, \href
  {http://adsabs.harvard.edu/abs/2016ATel.9822....1B} {9822}

\bibitem[\protect\citeauthoryear{{Brimacombe} et~al.,}{{Brimacombe}
  et~al.}{2016al}]{asassn16pa_atel}
{Brimacombe} J.,  et~al., 2016al, The Astronomer's Telegram, \href
  {http://adsabs.harvard.edu/abs/2016ATel.9893....1B} {9893}

\bibitem[\protect\citeauthoryear{{Brimacombe} et~al.,}{{Brimacombe}
  et~al.}{2017}]{asassn16po_atel}
{Brimacombe} J.,  et~al., 2017, The Astronomer's Telegram, \href
  {http://adsabs.harvard.edu/abs/2017ATel.9927....1B} {9927}

\bibitem[\protect\citeauthoryear{{Brown} et~al.,}{{Brown}
  et~al.}{2013}]{brown13}
{Brown} T.~M.,  et~al., 2013, \mn@doi [\pasp] {10.1086/673168}, \href
  {http://adsabs.harvard.edu/abs/2013PASP..125.1031B} {125, 1031}

\bibitem[\protect\citeauthoryear{{Brown}, {W.-S Holoien}, {Auchettl}, {Stanek},
  {Kochanek}, {Shappee}, {Prieto}  \& {Grupe}}{{Brown}
  et~al.}{2016a}]{brown16b}
{Brown} J.~S.,  {W.-S Holoien} T.,  {Auchettl} K.,  {Stanek} K.~Z.,  {Kochanek}
  C.~S.,  {Shappee} B.~J.,  {Prieto} J.~L.,   {Grupe} D.,  2016a, preprint,
  \href {http://adsabs.harvard.edu/abs/2016arXiv160904403B} {} (\mn@eprint
  {arXiv} {1609.04403})

\bibitem[\protect\citeauthoryear{{Brown}, {Shappee}, {Holoien}, {Stanek},
  {Kochanek}  \& {Prieto}}{{Brown} et~al.}{2016b}]{brown16a}
{Brown} J.~S.,  {Shappee} B.~J.,  {Holoien} T.~W.-S.,  {Stanek} K.~Z.,
  {Kochanek} C.~S.,   {Prieto} J.~L.,  2016b, \mn@doi [\mnras]
  {10.1093/mnras/stw1928}, \href
  {http://adsabs.harvard.edu/abs/2016MNRAS.462.3993B} {462, 3993}

\bibitem[\protect\citeauthoryear{{Brown} et~al.,}{{Brown}
  et~al.}{2016c}]{asassn16aa_atel}
{Brown} J.~S.,  et~al., 2016c, The Astronomer's Telegram, \href
  {http://adsabs.harvard.edu/abs/2016ATel.8496....1B} {8496}

\bibitem[\protect\citeauthoryear{{Brown} et~al.,}{{Brown}
  et~al.}{2016d}]{asassn16ai_atel}
{Brown} J.~S.,  et~al., 2016d, The Astronomer's Telegram, \href
  {http://adsabs.harvard.edu/abs/2016ATel.8537....1B} {8537}

\bibitem[\protect\citeauthoryear{{Brown} et~al.,}{{Brown}
  et~al.}{2016e}]{asassn16bg_atel}
{Brown} J.~S.,  et~al., 2016e, The Astronomer's Telegram, \href
  {http://adsabs.harvard.edu/abs/2016ATel.8647....1B} {8647}

\bibitem[\protect\citeauthoryear{{Brown} et~al.,}{{Brown}
  et~al.}{2016f}]{asassn16cc_atel}
{Brown} J.~S.,  et~al., 2016f, The Astronomer's Telegram, \href
  {http://adsabs.harvard.edu/abs/2016ATel.8736....1B} {8736}

\bibitem[\protect\citeauthoryear{{Brown} et~al.,}{{Brown}
  et~al.}{2016g}]{asassn16cm_atel}
{Brown} J.~S.,  et~al., 2016g, The Astronomer's Telegram, \href
  {http://adsabs.harvard.edu/abs/2016ATel.8775....1B} {8775}

\bibitem[\protect\citeauthoryear{{Brown} et~al.,}{{Brown}
  et~al.}{2016h}]{asassn16dn_atel}
{Brown} J.~S.,  et~al., 2016h, The Astronomer's Telegram, \href
  {http://adsabs.harvard.edu/abs/2016ATel.8885....1B} {8885}

\bibitem[\protect\citeauthoryear{{Brown} et~al.,}{{Brown}
  et~al.}{2016i}]{asassn16eh_atel}
{Brown} J.~S.,  et~al., 2016i, The Astronomer's Telegram, \href
  {http://adsabs.harvard.edu/abs/2016ATel.8930....1B} {8930}

\bibitem[\protect\citeauthoryear{{Brown} et~al.,}{{Brown}
  et~al.}{2016j}]{asassn16el_atel}
{Brown} J.~S.,  et~al., 2016j, The Astronomer's Telegram, \href
  {http://adsabs.harvard.edu/abs/2016ATel.8947....1B} {8947}

\bibitem[\protect\citeauthoryear{{Brown} et~al.,}{{Brown}
  et~al.}{2016k}]{asassn16es_atel}
{Brown} J.~S.,  et~al., 2016k, The Astronomer's Telegram, \href
  {http://adsabs.harvard.edu/abs/2016ATel.9001....1B} {9001}

\bibitem[\protect\citeauthoryear{{Brown} et~al.,}{{Brown}
  et~al.}{2016l}]{asassn16fx_atel}
{Brown} J.~S.,  et~al., 2016l, The Astronomer's Telegram, \href
  {http://adsabs.harvard.edu/abs/2016ATel.9127....1B} {9127}

\bibitem[\protect\citeauthoryear{{Brown} et~al.,}{{Brown}
  et~al.}{2016m}]{asassn16gz_atel}
{Brown} J.~S.,  et~al., 2016m, The Astronomer's Telegram, \href
  {http://adsabs.harvard.edu/abs/2016ATel.9222....1B} {9222}

\bibitem[\protect\citeauthoryear{{Brown} et~al.,}{{Brown}
  et~al.}{2016n}]{asassn16hd_atel}
{Brown} J.~S.,  et~al., 2016n, The Astronomer's Telegram, \href
  {http://adsabs.harvard.edu/abs/2016ATel.9227....1B} {9227}

\bibitem[\protect\citeauthoryear{{Brown} et~al.,}{{Brown}
  et~al.}{2016o}]{asassn16hw_atel}
{Brown} J.~S.,  et~al., 2016o, The Astronomer's Telegram, \href
  {http://adsabs.harvard.edu/abs/2016ATel.9278....1B} {9278}

\bibitem[\protect\citeauthoryear{{Brown} et~al.,}{{Brown}
  et~al.}{2016p}]{asassn16ie_atel}
{Brown} J.~S.,  et~al., 2016p, The Astronomer's Telegram, \href
  {http://adsabs.harvard.edu/abs/2016ATel.9341....1B} {9341}

\bibitem[\protect\citeauthoryear{{Brown} et~al.,}{{Brown}
  et~al.}{2016q}]{asassn16iq_atel}
{Brown} J.~S.,  et~al., 2016q, The Astronomer's Telegram, \href
  {http://adsabs.harvard.edu/abs/2016ATel.9354....1B} {9354}

\bibitem[\protect\citeauthoryear{{Brown} et~al.,}{{Brown}
  et~al.}{2016r}]{asassn16je_atel}
{Brown} J.~S.,  et~al., 2016r, The Astronomer's Telegram, \href
  {http://adsabs.harvard.edu/abs/2016ATel.9395....1B} {9395}

\bibitem[\protect\citeauthoryear{{Brown} et~al.,}{{Brown}
  et~al.}{2016s}]{asassn16jt_spec_atel}
{Brown} J.~S.,  et~al., 2016s, The Astronomer's Telegram, \href
  {http://adsabs.harvard.edu/abs/2016ATel.9445....1B} {9445}

\bibitem[\protect\citeauthoryear{{Brown} et~al.,}{{Brown}
  et~al.}{2016t}]{asassn16kk_atel}
{Brown} J.~S.,  et~al., 2016t, The Astronomer's Telegram, \href
  {http://adsabs.harvard.edu/abs/2016ATel.9484....1B} {9484}

\bibitem[\protect\citeauthoryear{{Brown} et~al.,}{{Brown}
  et~al.}{2016u}]{asassn16km_atel}
{Brown} J.~S.,  et~al., 2016u, The Astronomer's Telegram, \href
  {http://adsabs.harvard.edu/abs/2016ATel.9494....1B} {9494}

\bibitem[\protect\citeauthoryear{{Brown} et~al.,}{{Brown}
  et~al.}{2016v}]{asassn16la_atel}
{Brown} J.~S.,  et~al., 2016v, The Astronomer's Telegram, \href
  {http://adsabs.harvard.edu/abs/2016ATel.9571....1B} {9571}

\bibitem[\protect\citeauthoryear{{Brown} et~al.,}{{Brown}
  et~al.}{2016w}]{asassn16lc_atel}
{Brown} J.~S.,  et~al., 2016w, The Astronomer's Telegram, \href
  {http://adsabs.harvard.edu/abs/2016ATel.9584....1B} {9584}

\bibitem[\protect\citeauthoryear{{Brown} et~al.,}{{Brown}
  et~al.}{2016x}]{asassn16lg_atel}
{Brown} J.~S.,  et~al., 2016x, The Astronomer's Telegram, \href
  {http://adsabs.harvard.edu/abs/2016ATel.9601....1B} {9601}

\bibitem[\protect\citeauthoryear{{Brown} et~al.,}{{Brown}
  et~al.}{2016y}]{asassn16ll_atel}
{Brown} J.~S.,  et~al., 2016y, The Astronomer's Telegram, \href
  {http://adsabs.harvard.edu/abs/2016ATel.9602....1B} {9602}

\bibitem[\protect\citeauthoryear{{Brown} et~al.,}{{Brown}
  et~al.}{2016z}]{asassn16lx_atel}
{Brown} J.~S.,  et~al., 2016z, The Astronomer's Telegram, \href
  {http://adsabs.harvard.edu/abs/2016ATel.9666....1B} {9666}

\bibitem[\protect\citeauthoryear{{Brown} et~al.,}{{Brown}
  et~al.}{2016aa}]{asassn16pd_atel}
{Brown} J.~S.,  et~al., 2016aa, The Astronomer's Telegram, \href
  {http://adsabs.harvard.edu/abs/2016ATel.9898....1B} {9898}

\bibitem[\protect\citeauthoryear{{Chambers} et~al.,}{{Chambers}
  et~al.}{2016}]{chambers16}
{Chambers} K.~C.,  et~al., 2016, preprint, \href
  {http://adsabs.harvard.edu/abs/2016arXiv161205560C} {} (\mn@eprint {arXiv}
  {1612.05560})

\bibitem[\protect\citeauthoryear{{Chen} et~al.,}{{Chen}
  et~al.}{2016a}]{asassn16jf_atel}
{Chen} P.,  et~al., 2016a, The Astronomer's Telegram, \href
  {http://adsabs.harvard.edu/abs/2016ATel.9398....1C} {9398}

\bibitem[\protect\citeauthoryear{{Chen} et~al.,}{{Chen}
  et~al.}{2016b}]{asassn16jh_atel}
{Chen} P.,  et~al., 2016b, The Astronomer's Telegram, \href
  {http://adsabs.harvard.edu/abs/2016ATel.9407....1C} {9407}

\bibitem[\protect\citeauthoryear{{Chornock} et~al.,}{{Chornock}
  et~al.}{2016}]{asassn16ci_atel}
{Chornock} R.,  et~al., 2016, The Astronomer's Telegram, \href
  {http://adsabs.harvard.edu/abs/2016ATel.8765....1C} {8765}

\bibitem[\protect\citeauthoryear{{Churazov} et~al.,}{{Churazov}
  et~al.}{2015}]{churazov15}
{Churazov} E.,  et~al., 2015, \mn@doi [\apj] {10.1088/0004-637X/812/1/62},
  \href {http://adsabs.harvard.edu/abs/2015ApJ...812...62C} {812, 62}

\bibitem[\protect\citeauthoryear{{Cikota} et~al.,}{{Cikota}
  et~al.}{2016a}]{asassn16aw_spec_atel}
{Cikota} A.,  et~al., 2016a, The Astronomer's Telegram, \href
  {http://adsabs.harvard.edu/abs/2016ATel.8613....1C} {8613}

\bibitem[\protect\citeauthoryear{{Cikota} et~al.,}{{Cikota}
  et~al.}{2016b}]{asassn16ov_spec_atel}
{Cikota} A.,  et~al., 2016b, The Astronomer's Telegram, \href
  {http://adsabs.harvard.edu/abs/2016ATel.9878....1C} {9878}

\bibitem[\protect\citeauthoryear{{Cikota} et~al.,}{{Cikota}
  et~al.}{2016c}]{asassn16oz_spec_atel}
{Cikota} A.,  et~al., 2016c, The Astronomer's Telegram, \href
  {http://adsabs.harvard.edu/abs/2016ATel.9889....1C} {9889}

\bibitem[\protect\citeauthoryear{{Cruz} et~al.,}{{Cruz}
  et~al.}{2016a}]{asassn16bp_atel}
{Cruz} I.,  et~al., 2016a, The Astronomer's Telegram, \href
  {http://adsabs.harvard.edu/abs/2016ATel.8666....1C} {8666}

\bibitem[\protect\citeauthoryear{{Cruz} et~al.,}{{Cruz}
  et~al.}{2016b}]{asassn16cr_atel}
{Cruz} I.,  et~al., 2016b, The Astronomer's Telegram, \href
  {http://adsabs.harvard.edu/abs/2016ATel.8784....1C} {8784}

\bibitem[\protect\citeauthoryear{{Cruz} et~al.,}{{Cruz}
  et~al.}{2016c}]{asassn16ek_atel}
{Cruz} I.,  et~al., 2016c, The Astronomer's Telegram, \href
  {http://adsabs.harvard.edu/abs/2016ATel.8944....1C} {8944}

\bibitem[\protect\citeauthoryear{{Cruz} et~al.,}{{Cruz}
  et~al.}{2016d}]{asassn16em_atel}
{Cruz} I.,  et~al., 2016d, The Astronomer's Telegram, \href
  {http://adsabs.harvard.edu/abs/2016ATel.8953....1C} {8953}

\bibitem[\protect\citeauthoryear{{Cruz} et~al.,}{{Cruz}
  et~al.}{2016e}]{asassn16fj_atel}
{Cruz} I.,  et~al., 2016e, The Astronomer's Telegram, \href
  {http://adsabs.harvard.edu/abs/2016ATel.9073....1C} {9073}

\bibitem[\protect\citeauthoryear{{Diehl} et~al.,}{{Diehl}
  et~al.}{2014}]{diehl14}
{Diehl} R.,  et~al., 2014, \mn@doi [Science] {10.1126/science.1254738}, \href
  {http://adsabs.harvard.edu/abs/2014Sci...345.1162D} {345, 1162}

\bibitem[\protect\citeauthoryear{{Dimitriadis} et~al.,}{{Dimitriadis}
  et~al.}{2016}]{asassn16al_spec_atel}
{Dimitriadis} G.,  et~al., 2016, The Astronomer's Telegram, \href
  {http://adsabs.harvard.edu/abs/2016ATel.8555....1D} {8555}

\bibitem[\protect\citeauthoryear{{Dong} et~al.,}{{Dong} et~al.}{2016a}]{dong16}
{Dong} S.,  et~al., 2016a, Science, \href
  {http://adsabs.harvard.edu/abs/2015arXiv150703010D} {351, 257}

\bibitem[\protect\citeauthoryear{{Dong} et~al.,}{{Dong}
  et~al.}{2016b}]{asassn16ad_atel}
{Dong} S.,  et~al., 2016b, The Astronomer's Telegram, \href
  {http://adsabs.harvard.edu/abs/2016ATel.8521....1D} {8521}

\bibitem[\protect\citeauthoryear{{Dong} et~al.,}{{Dong}
  et~al.}{2016c}]{asassn16dp_atel}
{Dong} S.,  et~al., 2016c, The Astronomer's Telegram, \href
  {http://adsabs.harvard.edu/abs/2016ATel.8892....1D} {8892}

\bibitem[\protect\citeauthoryear{{Dong} et~al.,}{{Dong}
  et~al.}{2016d}]{ps16dtm_atel2}
{Dong} S.,  et~al., 2016d, The Astronomer's Telegram, \href
  {http://adsabs.harvard.edu/abs/2016ATel.9843....1D} {9843}

\bibitem[\protect\citeauthoryear{{Drake} et~al.,}{{Drake}
  et~al.}{2009}]{drake09}
{Drake} A.~J.,  et~al., 2009, \mn@doi [\apj] {10.1088/0004-637X/696/1/870},
  \href {http://adsabs.harvard.edu/abs/2009ApJ...696..870D} {696, 870}

\bibitem[\protect\citeauthoryear{{Elias-Rosa} et~al.,}{{Elias-Rosa}
  et~al.}{2016a}]{asassn16bw_spec_atel}
{Elias-Rosa} N.,  et~al., 2016a, The Astronomer's Telegram, \href
  {http://adsabs.harvard.edu/abs/2016ATel.8727....1E} {8727}

\bibitem[\protect\citeauthoryear{{Elias-Rosa} et~al.,}{{Elias-Rosa}
  et~al.}{2016b}]{asassn16fp_spec_atel}
{Elias-Rosa} N.,  et~al., 2016b, The Astronomer's Telegram, \href
  {http://adsabs.harvard.edu/abs/2016ATel.9090....1E} {9090}

\bibitem[\protect\citeauthoryear{{Falco}, {Macri}, {Stringer}, {Prieto},
  {Challis}  \& {Kirshner}}{{Falco} et~al.}{2016a}]{asassn16ax_spec_atel}
{Falco} E.,  {Macri} L.,  {Stringer} K.,  {Prieto} J.~L.,  {Challis} P.,
  {Kirshner} R.,  2016a, The Astronomer's Telegram, \href
  {http://adsabs.harvard.edu/abs/2016ATel.8636....1F} {8636}

\bibitem[\protect\citeauthoryear{{Falco}, {Challis}, {Kirshner}, {Berlind},
  {Calkins}, {Prieto}  \& {Stanek}}{{Falco}
  et~al.}{2016b}]{asassn16dm_spec_atel}
{Falco} E.,  {Challis} P.,  {Kirshner} R.,  {Berlind} P.,  {Calkins} M.,
  {Prieto} J.~L.,   {Stanek} K.~Z.,  2016b, The Astronomer's Telegram, \href
  {http://adsabs.harvard.edu/abs/2016ATel.8896....1F} {8896}

\bibitem[\protect\citeauthoryear{{Falco}, {Calkins}, {Challis}, {Kirshner},
  {Prieto}  \& {Stanek}}{{Falco} et~al.}{2016c}]{asassn16fn_spec_atel}
{Falco} E.,  {Calkins} M.,  {Challis} P.,  {Kirshner} R.,  {Prieto} J.~L.,
  {Stanek} K.~Z.,  2016c, The Astronomer's Telegram, \href
  {http://adsabs.harvard.edu/abs/2016ATel.9118....1F} {9118}

\bibitem[\protect\citeauthoryear{{Falco}, {Calkins}, {Challis}, {Kirshner},
  {Prieto}  \& {Stanek}}{{Falco} et~al.}{2016d}]{asassn16go_spec_atel}
{Falco} E.,  {Calkins} M.,  {Challis} P.,  {Kirshner} R.,  {Prieto} J.~L.,
  {Stanek} K.~Z.,  2016d, The Astronomer's Telegram, \href
  {http://adsabs.harvard.edu/abs/2016ATel.9219....1F} {9219}

\bibitem[\protect\citeauthoryear{{Falco}, {Calkins}, {Challis}, {Kirshner},
  {Prieto}  \& {Stanek}}{{Falco} et~al.}{2016e}]{asassn16gy_spec_atel}
{Falco} E.,  {Calkins} M.,  {Challis} P.,  {Kirshner} R.,  {Prieto} J.~L.,
  {Stanek} K.~Z.,  2016e, The Astronomer's Telegram, \href
  {http://adsabs.harvard.edu/abs/2016ATel.9237....1F} {9237}

\bibitem[\protect\citeauthoryear{{Faran} et~al.,}{{Faran}
  et~al.}{2016}]{asassn16bg_spec_atel}
{Faran} T.,  et~al., 2016, The Astronomer's Telegram, \href
  {http://adsabs.harvard.edu/abs/2016ATel.8724....1F} {8724}

\bibitem[\protect\citeauthoryear{{Fausnaugh} et~al.,}{{Fausnaugh}
  et~al.}{2016}]{asassn16gf_atel}
{Fausnaugh} M.,  et~al., 2016, The Astronomer's Telegram, \href
  {http://adsabs.harvard.edu/abs/2016ATel.9146....1F} {9146}

\bibitem[\protect\citeauthoryear{{Fernandez} et~al.,}{{Fernandez}
  et~al.}{2016a}]{asassn16bc_atel}
{Fernandez} J.~M.,  et~al., 2016a, The Astronomer's Telegram, \href
  {http://adsabs.harvard.edu/abs/2016ATel.8628....1F} {8628}

\bibitem[\protect\citeauthoryear{{Fernandez} et~al.,}{{Fernandez}
  et~al.}{2016b}]{asassn16ct_atel}
{Fernandez} J.~M.,  et~al., 2016b, The Astronomer's Telegram, \href
  {http://adsabs.harvard.edu/abs/2016ATel.8796....1F} {8796}

\bibitem[\protect\citeauthoryear{{Frank}, {Prieto}  \& {Stanek}}{{Frank}
  et~al.}{2016}]{asassn16bx_spec_atel}
{Frank} S.,  {Prieto} J.~L.,   {Stanek} K.~Z.,  2016, The Astronomer's
  Telegram, \href {http://adsabs.harvard.edu/abs/2016ATel.8830....1F} {8830}

\bibitem[\protect\citeauthoryear{{Fraser} et~al.,}{{Fraser}
  et~al.}{2016}]{asassn16gn_spec_atel}
{Fraser} M.,  et~al., 2016, The Astronomer's Telegram, \href
  {http://adsabs.harvard.edu/abs/2016ATel.9208....1F} {9208}

\bibitem[\protect\citeauthoryear{{Frieman} et~al.,}{{Frieman}
  et~al.}{2008}]{frieman08}
{Frieman} J.~A.,  et~al., 2008, \mn@doi [\aj] {10.1088/0004-6256/135/1/338},
  \href {http://adsabs.harvard.edu/abs/2008AJ....135..338F} {135, 338}

\bibitem[\protect\citeauthoryear{{Gal-Yam}, {Mazzali}, {Manulis}  \&
  {Bishop}}{{Gal-Yam} et~al.}{2013}]{galyam13}
{Gal-Yam} A.,  {Mazzali} P.~A.,  {Manulis} I.,   {Bishop} D.,  2013, \mn@doi
  [\pasp] {10.1086/671483}, \href
  {http://adsabs.harvard.edu/abs/2013PASP..125..749G} {125, 749}

\bibitem[\protect\citeauthoryear{{Gall} et~al.,}{{Gall}
  et~al.}{2016}]{asassn16ip_spec_atel}
{Gall} C.,  et~al., 2016, The Astronomer's Telegram, \href
  {http://adsabs.harvard.edu/abs/2016ATel.9368....1G} {9368}

\bibitem[\protect\citeauthoryear{{Godoy-Rivera} et~al.,}{{Godoy-Rivera}
  et~al.}{2017}]{godoy-rivera17}
{Godoy-Rivera} D.,  et~al., 2017, \mn@doi [\mnras] {10.1093/mnras/stw3237},
  \href {http://adsabs.harvard.edu/abs/2017MNRAS.466.1428G} {466, 1428}

\bibitem[\protect\citeauthoryear{{Gorbovskoy} et~al.,}{{Gorbovskoy}
  et~al.}{2013}]{gorbovskoy13}
{Gorbovskoy} E.~S.,  et~al., 2013, \mn@doi [Astronomy Reports]
  {10.1134/S1063772913040033}, \href
  {http://adsabs.harvard.edu/abs/2013ARep...57..233G} {57, 233}

\bibitem[\protect\citeauthoryear{{Harmanen} et~al.,}{{Harmanen}
  et~al.}{2016}]{asassn16hy_spec_atel}
{Harmanen} J.,  et~al., 2016, The Astronomer's Telegram, \href
  {http://adsabs.harvard.edu/abs/2016ATel.9326....1H} {9326}

\bibitem[\protect\citeauthoryear{{Harutyunyan} et~al.,}{{Harutyunyan}
  et~al.}{2008}]{harutyunyan08}
{Harutyunyan} A.~H.,  et~al., 2008, \mn@doi [\aap]
  {10.1051/0004-6361:20078859}, \href
  {http://adsabs.harvard.edu/abs/2008A%26A...488..383H} {488, 383}

\bibitem[\protect\citeauthoryear{{Herczeg} et~al.,}{{Herczeg}
  et~al.}{2016}]{herczeg16}
{Herczeg} G.~J.,  et~al., 2016, preprint, \href
  {http://adsabs.harvard.edu/abs/2016arXiv160706368H} {} (\mn@eprint {arXiv}
  {1607.06368})

\bibitem[\protect\citeauthoryear{{Hodgkin}, {Wyrzykowski}, {Blagorodnova}  \&
  {Koposov}}{{Hodgkin} et~al.}{2013}]{hodgkin13}
{Hodgkin} S.~T.,  {Wyrzykowski} L.,  {Blagorodnova} N.,   {Koposov} S.,  2013,
  \mn@doi [Philosophical Transactions of the Royal Society of London Series A]
  {10.1098/rsta.2012.0239}, \href
  {http://adsabs.harvard.edu/abs/2013RSPTA.37120239H} {371, 20120239}

\bibitem[\protect\citeauthoryear{{Holoien} et~al.,}{{Holoien}
  et~al.}{2014a}]{holoien14b}
{Holoien} T.~W.-S.,  et~al., 2014a, \mn@doi [\mnras] {10.1093/mnras/stu1922},
  \href {http://adsabs.harvard.edu/abs/2014MNRAS.445.3263H} {445, 3263}

\bibitem[\protect\citeauthoryear{{Holoien} et~al.,}{{Holoien}
  et~al.}{2014b}]{holoien14a}
{Holoien} T.~W.-S.,  et~al., 2014b, \mn@doi [\apjl]
  {10.1088/2041-8205/785/2/L35}, \href
  {http://adsabs.harvard.edu/abs/2014ApJ...785L..35H} {785, L35}

\bibitem[\protect\citeauthoryear{{Holoien} et~al.,}{{Holoien}
  et~al.}{2016a}]{holoien16c}
{Holoien} T.~W.-S.,  et~al., 2016a, \actaa, \href
  {http://adsabs.harvard.edu/abs/2016AcA....66..219H} {66, 219}

\bibitem[\protect\citeauthoryear{{Holoien} et~al.,}{{Holoien}
  et~al.}{2016b}]{holoien16a}
{Holoien} T.~W.-S.,  et~al., 2016b, \mn@doi [\mnras] {10.1093/mnras/stv2486},
  \href {http://adsabs.harvard.edu/abs/2016MNRAS.455.2918H} {455, 2918}

\bibitem[\protect\citeauthoryear{{Holoien} et~al.,}{{Holoien}
  et~al.}{2016c}]{holoien16b}
{Holoien} T.~W.-S.,  et~al., 2016c, \mn@doi [\mnras] {10.1093/mnras/stw2272},
  \href {http://adsabs.harvard.edu/abs/2016MNRAS.463.3813H} {463, 3813}

\bibitem[\protect\citeauthoryear{{Holoien} et~al.,}{{Holoien}
  et~al.}{2016d}]{asassn16av_atel}
{Holoien} T.~W.-S.,  et~al., 2016d, The Astronomer's Telegram, \href
  {http://adsabs.harvard.edu/abs/2016ATel.8569....1H} {8569}

\bibitem[\protect\citeauthoryear{{Holoien} et~al.,}{{Holoien}
  et~al.}{2016e}]{asassn16az_atel}
{Holoien} T.~W.-S.,  et~al., 2016e, The Astronomer's Telegram, \href
  {http://adsabs.harvard.edu/abs/2016ATel.8614....1H} {8614}

\bibitem[\protect\citeauthoryear{{Holoien}, {Godoy-Rivera}  \&
  {Prieto}}{{Holoien} et~al.}{2016f}]{asassn16ej_spec_atel}
{Holoien} T.~W.-S.,  {Godoy-Rivera} D.,   {Prieto} J.~L.,  2016f, The
  Astronomer's Telegram, \href
  {http://adsabs.harvard.edu/abs/2016ATel.8955....1H} {8955}

\bibitem[\protect\citeauthoryear{{Holoien} et~al.,}{{Holoien}
  et~al.}{2016g}]{asassn16eu_atel}
{Holoien} T.~W.-S.,  et~al., 2016g, The Astronomer's Telegram, \href
  {http://adsabs.harvard.edu/abs/2016ATel.9011....1H} {9011}

\bibitem[\protect\citeauthoryear{{Holoien} et~al.,}{{Holoien}
  et~al.}{2016h}]{asassn16fp_atel}
{Holoien} T.~W.-S.,  et~al., 2016h, The Astronomer's Telegram, \href
  {http://adsabs.harvard.edu/abs/2016ATel.9086....1H} {9086}

\bibitem[\protect\citeauthoryear{{Holoien} et~al.,}{{Holoien}
  et~al.}{2016i}]{asassn16go_atel}
{Holoien} T.~W.-S.,  et~al., 2016i, The Astronomer's Telegram, \href
  {http://adsabs.harvard.edu/abs/2016ATel.9195....1H} {9195}

\bibitem[\protect\citeauthoryear{{Holoien} et~al.,}{{Holoien}
  et~al.}{2016j}]{asassn16ic_atel}
{Holoien} T.~W.-S.,  et~al., 2016j, The Astronomer's Telegram, \href
  {http://adsabs.harvard.edu/abs/2016ATel.9336....1H} {9336}

\bibitem[\protect\citeauthoryear{{Holoien} et~al.,}{{Holoien}
  et~al.}{2016k}]{asassn16kw_atel}
{Holoien} T.~W.-S.,  et~al., 2016k, The Astronomer's Telegram, \href
  {http://adsabs.harvard.edu/abs/2016ATel.9555....1H} {9555}

\bibitem[\protect\citeauthoryear{{Holoien} et~al.,}{{Holoien}
  et~al.}{2016l}]{asassn16kz_atel}
{Holoien} T.~W.-S.,  et~al., 2016l, The Astronomer's Telegram, \href
  {http://adsabs.harvard.edu/abs/2016ATel.9568....1H} {9568}

\bibitem[\protect\citeauthoryear{{Holoien} et~al.,}{{Holoien}
  et~al.}{2016m}]{asassn16oj_atel}
{Holoien} T.~W.-S.,  et~al., 2016m, The Astronomer's Telegram, \href
  {http://adsabs.harvard.edu/abs/2016ATel.9825....1H} {9825}

\bibitem[\protect\citeauthoryear{{Holoien} et~al.,}{{Holoien}
  et~al.}{2016n}]{asassn16oq_atel}
{Holoien} T.~W.-S.,  et~al., 2016n, The Astronomer's Telegram, \href
  {http://adsabs.harvard.edu/abs/2016ATel.9838....1H} {9838}

\bibitem[\protect\citeauthoryear{{Holoien} et~al.,}{{Holoien}
  et~al.}{2017a}]{holoien17a}
{Holoien} T.~W.-S.,  et~al., 2017a, \mn@doi [\mnras] {10.1093/mnras/stx057},
  \href {http://adsabs.harvard.edu/abs/2017MNRAS.tmp...96H} {}

\bibitem[\protect\citeauthoryear{{Holoien} et~al.,}{{Holoien}
  et~al.}{2017b}]{holoien16d}
{Holoien} T.~W.-S.,  et~al., 2017b, \mn@doi [\mnras] {10.1093/mnras/stw2273},
  \href {http://adsabs.harvard.edu/abs/2017MNRAS.464.2672H} {464, 2672}

\bibitem[\protect\citeauthoryear{{Horiuchi} \& {Beacom}}{{Horiuchi} \&
  {Beacom}}{2010}]{horiuchi10}
{Horiuchi} S.,  {Beacom} J.~F.,  2010, \mn@doi [\apj]
  {10.1088/0004-637X/723/1/329}, \href
  {http://adsabs.harvard.edu/abs/2010ApJ...723..329H} {723, 329}

\bibitem[\protect\citeauthoryear{{Horiuchi}, {Beacom}, {Kochanek}, {Prieto},
  {Stanek}  \& {Thompson}}{{Horiuchi} et~al.}{2011}]{horiuchi11}
{Horiuchi} S.,  {Beacom} J.~F.,  {Kochanek} C.~S.,  {Prieto} J.~L.,  {Stanek}
  K.~Z.,   {Thompson} T.~A.,  2011, \mn@doi [\apj]
  {10.1088/0004-637X/738/2/154}, \href
  {http://adsabs.harvard.edu/abs/2011ApJ...738..154H} {738, 154}

\bibitem[\protect\citeauthoryear{{Horiuchi}, {Beacom}, {Bothwell}  \&
  {Thompson}}{{Horiuchi} et~al.}{2013}]{horiuchi13}
{Horiuchi} S.,  {Beacom} J.~F.,  {Bothwell} M.~S.,   {Thompson} T.~A.,  2013,
  \mn@doi [\apj] {10.1088/0004-637X/769/2/113}, \href
  {http://adsabs.harvard.edu/abs/2013ApJ...769..113H} {769, 113}

\bibitem[\protect\citeauthoryear{{Hosseinzadeh}, {Arcavi}, {McCully}, {Howell},
  {Sand}  \& {Valenti}}{{Hosseinzadeh} et~al.}{2016a}]{asassn16at_spec_atel}
{Hosseinzadeh} G.,  {Arcavi} I.,  {McCully} C.,  {Howell} D.~A.,  {Sand} D.,
  {Valenti} S.,  2016a, The Astronomer's Telegram, \href
  {http://adsabs.harvard.edu/abs/2016ATel.8567....1H} {8567}

\bibitem[\protect\citeauthoryear{{Hosseinzadeh}, {Yang}, {Arcavi}, {Howell},
  {McCully}  \& {Valenti}}{{Hosseinzadeh} et~al.}{2016b}]{asassn16ba_spec_atel}
{Hosseinzadeh} G.,  {Yang} Y.,  {Arcavi} I.,  {Howell} D.~A.,  {McCully} C.,
  {Valenti} S.,  2016b, The Astronomer's Telegram, \href
  {http://adsabs.harvard.edu/abs/2016ATel.8623....1H} {8623}

\bibitem[\protect\citeauthoryear{{Hosseinzadeh}, {Yang}, {Valenti}, {Arcavi},
  {Howell}  \& {McCully}}{{Hosseinzadeh} et~al.}{2016c}]{asassn16bm_spec_atel}
{Hosseinzadeh} G.,  {Yang} Y.,  {Valenti} S.,  {Arcavi} I.,  {Howell} D.~A.,
  {McCully} C.,  2016c, The Astronomer's Telegram, \href
  {http://adsabs.harvard.edu/abs/2016ATel.8679....1H} {8679}

\bibitem[\protect\citeauthoryear{{Hosseinzadeh}, {Yang}, {McCully}, {Arcavi},
  {Howell}  \& {Valenti}}{{Hosseinzadeh} et~al.}{2016d}]{asassn16cc_spec_atel}
{Hosseinzadeh} G.,  {Yang} Y.,  {McCully} C.,  {Arcavi} I.,  {Howell} D.~A.,
  {Valenti} S.,  2016d, The Astronomer's Telegram, \href
  {http://adsabs.harvard.edu/abs/2016ATel.8748....1H} {8748}

\bibitem[\protect\citeauthoryear{{Hosseinzadeh}, {Arcavi}, {Yang}, {Howell},
  {Valenti}  \& {McCully}}{{Hosseinzadeh} et~al.}{2016e}]{asassn16cr_spec_atel}
{Hosseinzadeh} G.,  {Arcavi} I.,  {Yang} Y.,  {Howell} D.~A.,  {Valenti} S.,
  {McCully} C.,  2016e, The Astronomer's Telegram, \href
  {http://adsabs.harvard.edu/abs/2016ATel.8807....1H} {8807}

\bibitem[\protect\citeauthoryear{{Hosseinzadeh}, {Arcavi}, {Howell}, {McCully}
  \& {Valenti}}{{Hosseinzadeh} et~al.}{2016f}]{asassn16ff_spec_atel}
{Hosseinzadeh} G.,  {Arcavi} I.,  {Howell} D.~A.,  {McCully} C.,   {Valenti}
  S.,  2016f, The Astronomer's Telegram, 9065

\bibitem[\protect\citeauthoryear{{Hosseinzadeh}, {Arcavi}, {Howell}, {McCully}
  \& {Valenti}}{{Hosseinzadeh} et~al.}{2016g}]{asassn16kv_spec_atel}
{Hosseinzadeh} G.,  {Arcavi} I.,  {Howell} D.~A.,  {McCully} C.,   {Valenti}
  S.,  2016g, The Astronomer's Telegram, \href
  {http://adsabs.harvard.edu/abs/2016ATel.9589....1H} {9589}

\bibitem[\protect\citeauthoryear{{Hosseinzadeh}, {Arcavi}, {Howell}, {McCully}
  \& {Valenti}}{{Hosseinzadeh} et~al.}{2016h}]{asassn16mc_spec_atel}
{Hosseinzadeh} G.,  {Arcavi} I.,  {Howell} D.~A.,  {McCully} C.,   {Valenti}
  S.,  2016h, The Astronomer's Telegram, \href
  {http://adsabs.harvard.edu/abs/2016ATel.9698....1H} {9698}

\bibitem[\protect\citeauthoryear{{James} et~al.,}{{James}
  et~al.}{2016}]{asassn16kz_spec_atel}
{James} P.,  et~al., 2016, The Astronomer's Telegram, \href
  {http://adsabs.harvard.edu/abs/2016ATel.9581....1J} {9581}

\bibitem[\protect\citeauthoryear{{Kaiser} et~al.,}{{Kaiser}
  et~al.}{2002}]{kaiser02}
{Kaiser} N.,  et~al., 2002, in {Tyson} J.~A.,  {Wolff} S.,  eds,  Proceedings
  of the SPIE Vol. 4836, Survey and Other Telescope Technologies and
  Discoveries. pp 154--164, \mn@doi{10.1117/12.457365}

\bibitem[\protect\citeauthoryear{{Kangas} et~al.,}{{Kangas}
  et~al.}{2016a}]{asassn16fc_spec_atel}
{Kangas} T.,  et~al., 2016a, The Astronomer's Telegram, \href
  {http://adsabs.harvard.edu/abs/2016ATel.9060....1K} {9060}

\bibitem[\protect\citeauthoryear{{Kangas} et~al.,}{{Kangas}
  et~al.}{2016b}]{asassn16oo_spec_atel}
{Kangas} T.,  et~al., 2016b, The Astronomer's Telegram, \href
  {http://adsabs.harvard.edu/abs/2016ATel.9836....1K} {9836}

\bibitem[\protect\citeauthoryear{{Kato} et~al.,}{{Kato} et~al.}{2014a}]{kato13}
{Kato} T.,  et~al., 2014a, \mn@doi [\pasj] {10.1093/pasj/psu014}, \href
  {http://adsabs.harvard.edu/abs/2014PASJ...66...30K} {66, 30}

\bibitem[\protect\citeauthoryear{{Kato} et~al.,}{{Kato} et~al.}{2014b}]{kato14}
{Kato} T.,  et~al., 2014b, \mn@doi [\pasj] {10.1093/pasj/psu072}, \href
  {http://adsabs.harvard.edu/abs/2014PASJ...66...90K} {66, 90}

\bibitem[\protect\citeauthoryear{{Kato} et~al.,}{{Kato} et~al.}{2015}]{kato15}
{Kato} T.,  et~al., 2015, \mn@doi [\pasj] {10.1093/pasj/psv072}, \href
  {http://adsabs.harvard.edu/abs/2015PASJ...67..105K} {67, 105}

\bibitem[\protect\citeauthoryear{{Kato} et~al.,}{{Kato} et~al.}{2016}]{kato16}
{Kato} T.,  et~al., 2016, \mn@doi [\pasj] {10.1093/pasj/psw064}, \href
  {http://adsabs.harvard.edu/abs/2016PASJ...68...65K} {68, 65}

\bibitem[\protect\citeauthoryear{{Kilpatrick}, {Siebert}, {Foley}, {Pan},
  {Jha}, {Rest}  \& {Scolnic}}{{Kilpatrick}
  et~al.}{2016a}]{asassn16iq_spec_atel}
{Kilpatrick} C.~D.,  {Siebert} M.~R.,  {Foley} R.~J.,  {Pan} Y.-C.,  {Jha}
  S.~W.,  {Rest} A.,   {Scolnic} D.,  2016a, The Astronomer's Telegram, \href
  {http://adsabs.harvard.edu/abs/2016ATel.9361....1K} {9361}

\bibitem[\protect\citeauthoryear{{Kilpatrick}, {Siebert}, {Coulter}, {Foley},
  {Pan}, {Jha}, {Rest}  \& {Scolnic}}{{Kilpatrick}
  et~al.}{2016b}]{asassn16io_spec_atel}
{Kilpatrick} C.~D.,  {Siebert} M.~R.,  {Coulter} D.~A.,  {Foley} R.~J.,  {Pan}
  Y.-C.,  {Jha} S.~W.,  {Rest} A.,   {Scolnic} D.,  2016b, The Astronomer's
  Telegram, \href {http://adsabs.harvard.edu/abs/2016ATel.9367....1K} {9367}

\bibitem[\protect\citeauthoryear{{Kiyota} et~al.,}{{Kiyota}
  et~al.}{2016a}]{asassn16al_atel}
{Kiyota} S.,  et~al., 2016a, The Astronomer's Telegram, \href
  {http://adsabs.harvard.edu/abs/2016ATel.8549....1K} {8549}

\bibitem[\protect\citeauthoryear{{Kiyota} et~al.,}{{Kiyota}
  et~al.}{2016b}]{asassn16aw_atel}
{Kiyota} S.,  et~al., 2016b, The Astronomer's Telegram, \href
  {http://adsabs.harvard.edu/abs/2016ATel.8607....1K} {8607}

\bibitem[\protect\citeauthoryear{{Kiyota} et~al.,}{{Kiyota}
  et~al.}{2016c}]{asassn16bb_atel}
{Kiyota} S.,  et~al., 2016c, The Astronomer's Telegram, \href
  {http://adsabs.harvard.edu/abs/2016ATel.8621....1K} {8621}

\bibitem[\protect\citeauthoryear{{Kiyota} et~al.,}{{Kiyota}
  et~al.}{2016d}]{asassn16dm_atel}
{Kiyota} S.,  et~al., 2016d, The Astronomer's Telegram, \href
  {http://adsabs.harvard.edu/abs/2016ATel.8882....1K} {8882}

\bibitem[\protect\citeauthoryear{{Kiyota} et~al.,}{{Kiyota}
  et~al.}{2016e}]{asassn16ej_atel}
{Kiyota} S.,  et~al., 2016e, The Astronomer's Telegram, \href
  {http://adsabs.harvard.edu/abs/2016ATel.8939....1K} {8939}

\bibitem[\protect\citeauthoryear{{Kiyota} et~al.,}{{Kiyota}
  et~al.}{2016f}]{asassn16ex_atel}
{Kiyota} S.,  et~al., 2016f, The Astronomer's Telegram, \href
  {http://adsabs.harvard.edu/abs/2016ATel.9020....1K} {9020}

\bibitem[\protect\citeauthoryear{{Kiyota} et~al.,}{{Kiyota}
  et~al.}{2016g}]{asassn16fa_atel}
{Kiyota} S.,  et~al., 2016g, The Astronomer's Telegram, \href
  {http://adsabs.harvard.edu/abs/2016ATel.9045....1K} {9045}

\bibitem[\protect\citeauthoryear{{Kiyota} et~al.,}{{Kiyota}
  et~al.}{2016h}]{asassn16of_atel}
{Kiyota} S.,  et~al., 2016h, The Astronomer's Telegram, \href
  {http://adsabs.harvard.edu/abs/2016ATel.9815....1K} {9815}

\bibitem[\protect\citeauthoryear{{Kochanek} et~al.,}{{Kochanek}
  et~al.}{2001}]{kochanek01}
{Kochanek} C.~S.,  et~al., 2001, \mn@doi [\apj] {10.1086/322488}, \href
  {http://adsabs.harvard.edu/abs/2001ApJ...560..566K} {560, 566}

\bibitem[\protect\citeauthoryear{{Koff} et~al.,}{{Koff}
  et~al.}{2016a}]{asassn16ay_atel}
{Koff} R.~A.,  et~al., 2016a, The Astronomer's Telegram, \href
  {http://adsabs.harvard.edu/abs/2016ATel.8609....1K} {8609}

\bibitem[\protect\citeauthoryear{{Koff} et~al.,}{{Koff}
  et~al.}{2016b}]{asassn16gn_atel}
{Koff} R.~A.,  et~al., 2016b, The Astronomer's Telegram, \href
  {http://adsabs.harvard.edu/abs/2016ATel.9194....1K} {9194}

\bibitem[\protect\citeauthoryear{{Lang}, {Hogg}, {Mierle}, {Blanton}  \&
  {Roweis}}{{Lang} et~al.}{2010}]{lang10}
{Lang} D.,  {Hogg} D.~W.,  {Mierle} K.,  {Blanton} M.,   {Roweis} S.,  2010,
  \mn@doi [\aj] {10.1088/0004-6256/139/5/1782}, \href
  {http://adsabs.harvard.edu/abs/2010AJ....139.1782L} {139, 1782}

\bibitem[\protect\citeauthoryear{{Law} et~al.,}{{Law} et~al.}{2009}]{law09}
{Law} N.~M.,  et~al., 2009, \mn@doi [\pasp] {10.1086/648598}, \href
  {http://adsabs.harvard.edu/abs/2009PASP..121.1395L} {121, 1395}

\bibitem[\protect\citeauthoryear{{Leloudas} et~al.,}{{Leloudas}
  et~al.}{2016}]{leloudas16}
{Leloudas} G.,  et~al., 2016, \mn@doi [Nature Astronomy]
  {10.1038/s41550-016-0002}, \href
  {http://adsabs.harvard.edu/abs/2016NatAs...1E...2L} {1, 0002}

\bibitem[\protect\citeauthoryear{{Li} et~al.,}{{Li} et~al.}{2000}]{li00}
{Li} W.~D.,  et~al., 2000, in {Holt} S.~S.,  {Zhang} W.~W.,  eds,  American
  Institute of Physics Conference Series Vol. 522, American Institute of
  Physics Conference Series. pp 103--106 (\mn@eprint {} {astro-ph/9912336}),
  \mn@doi{10.1063/1.1291702}

\bibitem[\protect\citeauthoryear{{Li} et~al.,}{{Li} et~al.}{2011}]{li11}
{Li} W.,  et~al., 2011, \mn@doi [\mnras] {10.1111/j.1365-2966.2011.18160.x},
  \href {http://adsabs.harvard.edu/abs/2011MNRAS.412.1441L} {412, 1441}

\bibitem[\protect\citeauthoryear{{Magee} et~al.,}{{Magee}
  et~al.}{2016}]{asassn16dw_spec_atel}
{Magee} M.,  et~al., 2016, The Astronomer's Telegram, \href
  {http://adsabs.harvard.edu/abs/2016ATel.8902....1M} {8902}

\bibitem[\protect\citeauthoryear{{Marples} et~al.,}{{Marples}
  et~al.}{2016a}]{asassn16jc_atel}
{Marples} P.,  et~al., 2016a, The Astronomer's Telegram, \href
  {http://adsabs.harvard.edu/abs/2016ATel.9393....1M} {9393}

\bibitem[\protect\citeauthoryear{{Marples} et~al.,}{{Marples}
  et~al.}{2016b}]{asassn16oy_atel}
{Marples} P.,  et~al., 2016b, The Astronomer's Telegram, \href
  {http://adsabs.harvard.edu/abs/2016ATel.9880....1M} {9880}

\bibitem[\protect\citeauthoryear{{Marples} et~al.,}{{Marples}
  et~al.}{2016c}]{asassn16pj_atel}
{Marples} P.,  et~al., 2016c, The Astronomer's Telegram, \href
  {http://adsabs.harvard.edu/abs/2016ATel.9914....1M} {9914}

\bibitem[\protect\citeauthoryear{{Masi} et~al.,}{{Masi}
  et~al.}{2016a}]{asassn16am_atel}
{Masi} G.,  et~al., 2016a, The Astronomer's Telegram, \href
  {http://adsabs.harvard.edu/abs/2016ATel.8556....1M} {8556}

\bibitem[\protect\citeauthoryear{{Masi} et~al.,}{{Masi}
  et~al.}{2016b}]{asassn16ar_atel}
{Masi} G.,  et~al., 2016b, The Astronomer's Telegram, \href
  {http://adsabs.harvard.edu/abs/2016ATel.8565....1M} {8565}

\bibitem[\protect\citeauthoryear{{Masi} et~al.,}{{Masi}
  et~al.}{2016c}]{asassn16ax_atel}
{Masi} G.,  et~al., 2016c, The Astronomer's Telegram, \href
  {http://adsabs.harvard.edu/abs/2016ATel.8594....1M} {8594}

\bibitem[\protect\citeauthoryear{{Masi} et~al.,}{{Masi}
  et~al.}{2016d}]{asassn16bn_atel}
{Masi} G.,  et~al., 2016d, The Astronomer's Telegram, \href
  {http://adsabs.harvard.edu/abs/2016ATel.8660....1M} {8660}

\bibitem[\protect\citeauthoryear{{Masi} et~al.,}{{Masi}
  et~al.}{2016e}]{asassn16cy_atel}
{Masi} G.,  et~al., 2016e, The Astronomer's Telegram, \href
  {http://adsabs.harvard.edu/abs/2016ATel.8801....1M} {8801}

\bibitem[\protect\citeauthoryear{{Masi} et~al.,}{{Masi}
  et~al.}{2016f}]{asassn16fs_atel}
{Masi} G.,  et~al., 2016f, The Astronomer's Telegram, \href
  {http://adsabs.harvard.edu/abs/2016ATel.9114....1M} {9114}

\bibitem[\protect\citeauthoryear{{Masi} et~al.,}{{Masi}
  et~al.}{2016g}]{asassn16gy_atel}
{Masi} G.,  et~al., 2016g, The Astronomer's Telegram, \href
  {http://adsabs.harvard.edu/abs/2016ATel.9217....1M} {9217}

\bibitem[\protect\citeauthoryear{{Masi} et~al.,}{{Masi}
  et~al.}{2016h}]{asassn16hc_atel}
{Masi} G.,  et~al., 2016h, The Astronomer's Telegram, \href
  {http://adsabs.harvard.edu/abs/2016ATel.9223....1M} {9223}

\bibitem[\protect\citeauthoryear{{Masi} et~al.,}{{Masi}
  et~al.}{2016i}]{asassn16pk_atel}
{Masi} G.,  et~al., 2016i, The Astronomer's Telegram, \href
  {http://adsabs.harvard.edu/abs/2016ATel.9918....1M} {9918}

\bibitem[\protect\citeauthoryear{{Mattila} et~al.,}{{Mattila}
  et~al.}{2016}]{asassn16ek_spec_atel}
{Mattila} S.,  et~al., 2016, The Astronomer's Telegram, \href
  {http://adsabs.harvard.edu/abs/2016ATel.8992....1M} {8992}

\bibitem[\protect\citeauthoryear{{Monard} et~al.,}{{Monard}
  et~al.}{2016a}]{asassn16cn_atel}
{Monard} L.~A.~G.,  et~al., 2016a, The Astronomer's Telegram, \href
  {http://adsabs.harvard.edu/abs/2016ATel.8776....1M} {8776}

\bibitem[\protect\citeauthoryear{{Monard} et~al.,}{{Monard}
  et~al.}{2016b}]{asassn16fg_atel}
{Monard} L.~A.~G.,  et~al., 2016b, The Astronomer's Telegram, \href
  {http://adsabs.harvard.edu/abs/2016ATel.9059....1M} {9059}

\bibitem[\protect\citeauthoryear{{Morrell} \& {Shappee}}{{Morrell} \&
  {Shappee}}{2016a}]{asassn16fh_spec_atel}
{Morrell} N.,  {Shappee} B.~J.,  2016a, The Astronomer's Telegram, \href
  {http://adsabs.harvard.edu/abs/2016ATel.9170....1M} {9170}

\bibitem[\protect\citeauthoryear{{Morrell} \& {Shappee}}{{Morrell} \&
  {Shappee}}{2016b}]{asassn16gm_spec_atel}
{Morrell} N.,  {Shappee} B.~J.,  2016b, The Astronomer's Telegram, \href
  {http://adsabs.harvard.edu/abs/2016ATel.9300....1M} {9300}

\bibitem[\protect\citeauthoryear{{Morrell}, {Phillips}, {Bose}, {Dong}  \&
  {Shappee}}{{Morrell} et~al.}{2016a}]{asassn16pd_spec_atel}
{Morrell} N.,  {Phillips} M.,  {Bose} S.,  {Dong} S.,   {Shappee} B.~J.,
  2016a, The Astronomer's Telegram, \href
  {http://adsabs.harvard.edu/abs/2016ATel.9899....1M} {9899}

\bibitem[\protect\citeauthoryear{{Morrell}, {Phillips}, {Shappee}  \&
  {Dong}}{{Morrell} et~al.}{2016b}]{asassn16pa_spec_atel}
{Morrell} N.,  {Phillips} M.,  {Shappee} B.~J.,   {Dong} S.,  2016b, The
  Astronomer's Telegram, \href
  {http://adsabs.harvard.edu/abs/2016ATel.9900....1M} {9900}

\bibitem[\protect\citeauthoryear{{Morrissey} et~al.,}{{Morrissey}
  et~al.}{2007}]{morrissey07}
{Morrissey} P.,  et~al., 2007, \mn@doi [\apjs] {10.1086/520512}, \href
  {http://adsabs.harvard.edu/abs/2007ApJS..173..682M} {173, 682}

\bibitem[\protect\citeauthoryear{{Murase}, {Thompson}, {Lacki}  \&
  {Beacom}}{{Murase} et~al.}{2011}]{murase11}
{Murase} K.,  {Thompson} T.~A.,  {Lacki} B.~C.,   {Beacom} J.~F.,  2011,
  \mn@doi [Physical Review D] {10.1103/PhysRevD.84.043003}, \href
  {http://adsabs.harvard.edu/abs/2011PhRvD..84d3003M} {84, 043003}

\bibitem[\protect\citeauthoryear{{Nakamura}, {Horiuchi}, {Tanaka}, {Hayama},
  {Takiwaki}  \& {Kotake}}{{Nakamura} et~al.}{2016}]{nakamura16}
{Nakamura} K.,  {Horiuchi} S.,  {Tanaka} M.,  {Hayama} K.,  {Takiwaki} T.,
  {Kotake} K.,  2016, \mn@doi [\mnras] {10.1093/mnras/stw1453}, \href
  {http://adsabs.harvard.edu/abs/2016MNRAS.461.3296N} {461, 3296}

\bibitem[\protect\citeauthoryear{{Nicholls} et~al.,}{{Nicholls}
  et~al.}{2016a}]{asassn16fl_atel}
{Nicholls} B.,  et~al., 2016a, The Astronomer's Telegram, \href
  {http://adsabs.harvard.edu/abs/2016ATel.9079....1N} {9079}

\bibitem[\protect\citeauthoryear{{Nicholls} et~al.,}{{Nicholls}
  et~al.}{2016b}]{asassn16fn_atel}
{Nicholls} B.,  et~al., 2016b, The Astronomer's Telegram, \href
  {http://adsabs.harvard.edu/abs/2016ATel.9081....1N} {9081}

\bibitem[\protect\citeauthoryear{{Nicholls} et~al.,}{{Nicholls}
  et~al.}{2016c}]{asassn16mc_atel}
{Nicholls} B.,  et~al., 2016c, The Astronomer's Telegram, \href
  {http://adsabs.harvard.edu/abs/2016ATel.9671....1N} {9671}

\bibitem[\protect\citeauthoryear{{Nicholls} et~al.,}{{Nicholls}
  et~al.}{2016d}]{asassn16mv_atel}
{Nicholls} B.,  et~al., 2016d, The Astronomer's Telegram, \href
  {http://adsabs.harvard.edu/abs/2016ATel.9713....1N} {9713}

\bibitem[\protect\citeauthoryear{{Nicholls} et~al.,}{{Nicholls}
  et~al.}{2016e}]{asassn16ok_atel}
{Nicholls} B.,  et~al., 2016e, The Astronomer's Telegram, \href
  {http://adsabs.harvard.edu/abs/2016ATel.9826....1N} {9826}

\bibitem[\protect\citeauthoryear{{Nicholls} et~al.,}{{Nicholls}
  et~al.}{2016f}]{asassn16om_atel}
{Nicholls} B.,  et~al., 2016f, The Astronomer's Telegram, \href
  {http://adsabs.harvard.edu/abs/2016ATel.9827....1N} {9827}

\bibitem[\protect\citeauthoryear{{Nicolas} et~al.,}{{Nicolas}
  et~al.}{2016a}]{asassn16hh_atel}
{Nicolas} J.,  et~al., 2016a, The Astronomer's Telegram, \href
  {http://adsabs.harvard.edu/abs/2016ATel.9254....1N} {9254}

\bibitem[\protect\citeauthoryear{{Nicolas} et~al.,}{{Nicolas}
  et~al.}{2016b}]{asassn16io_atel}
{Nicolas} J.,  et~al., 2016b, The Astronomer's Telegram, \href
  {http://adsabs.harvard.edu/abs/2016ATel.9349....1N} {9349}

\bibitem[\protect\citeauthoryear{{Nicolas} et~al.,}{{Nicolas}
  et~al.}{2016c}]{asassn16jj_atel}
{Nicolas} J.,  et~al., 2016c, The Astronomer's Telegram, \href
  {http://adsabs.harvard.edu/abs/2016ATel.9422....1N} {9422}

\bibitem[\protect\citeauthoryear{{Nicolas} et~al.,}{{Nicolas}
  et~al.}{2016d}]{asassn16lm_atel}
{Nicolas} J.,  et~al., 2016d, The Astronomer's Telegram, \href
  {http://adsabs.harvard.edu/abs/2016ATel.9611....1N} {9611}

\bibitem[\protect\citeauthoryear{{Nielsen} et~al.,}{{Nielsen}
  et~al.}{2016}]{asassn16lm_spec_atel}
{Nielsen} M.,  et~al., 2016, The Astronomer's Telegram, \href
  {http://adsabs.harvard.edu/abs/2016ATel.9630....1N} {9630}

\bibitem[\protect\citeauthoryear{{Pan}, {Downing}, {Foley}, {Jha}, {Rest}  \&
  {Scolnic}}{{Pan} et~al.}{2016a}]{asassn16aa_spec_atel}
{Pan} Y.-C.,  {Downing} S.,  {Foley} R.~J.,  {Jha} S.~W.,  {Rest} A.,
  {Scolnic} D.,  2016a, The Astronomer's Telegram, \href
  {http://adsabs.harvard.edu/abs/2016ATel.8506....1P} {8506}

\bibitem[\protect\citeauthoryear{{Pan}, {Kilpatrick}, {Siebert}, {Foley},
  {Jha}, {Rest}  \& {Scolnic}}{{Pan} et~al.}{2016b}]{asassn16hz_spec_atel}
{Pan} Y.-C.,  {Kilpatrick} C.~D.,  {Siebert} M.~R.,  {Foley} R.~J.,  {Jha}
  S.~W.,  {Rest} A.,   {Scolnic} D.,  2016b, The Astronomer's Telegram, \href
  {http://adsabs.harvard.edu/abs/2016ATel.9333....1P} {9333}

\bibitem[\protect\citeauthoryear{{Pan}, {Duarte}, {Foley}, {Jha}, {Rest}  \&
  {Scolnic}}{{Pan} et~al.}{2016c}]{asassn16lz_spec_atel}
{Pan} Y.-C.,  {Duarte} A.~S.,  {Foley} R.~J.,  {Jha} S.~W.,  {Rest} A.,
  {Scolnic} D.,  2016c, The Astronomer's Telegram, \href
  {http://adsabs.harvard.edu/abs/2016ATel.9696....1P} {9696}

\bibitem[\protect\citeauthoryear{{Pan}, {Foley}, {Jha}, {Rest}  \&
  {Scolnic}}{{Pan} et~al.}{2016d}]{asassn16pb_spec_atel}
{Pan} Y.-C.,  {Foley} R.~J.,  {Jha} S.~W.,  {Rest} A.,   {Scolnic} D.,  2016d,
  The Astronomer's Telegram, \href
  {http://adsabs.harvard.edu/abs/2016ATel.9904....1P} {9904}

\bibitem[\protect\citeauthoryear{{Piascik} \& {Steele}}{{Piascik} \&
  {Steele}}{2016a}]{asassn16ab_spec_atel}
{Piascik} A.~S.,  {Steele} I.~A.,  2016a, The Astronomer's Telegram, \href
  {http://adsabs.harvard.edu/abs/2016ATel.8505....1P} {8505}

\bibitem[\protect\citeauthoryear{{Piascik} \& {Steele}}{{Piascik} \&
  {Steele}}{2016b}]{asassn16am_spec_atel}
{Piascik} A.~S.,  {Steele} I.~A.,  2016b, The Astronomer's Telegram, \href
  {http://adsabs.harvard.edu/abs/2016ATel.8560....1P} {8560}

\bibitem[\protect\citeauthoryear{{Piascik} \& {Steele}}{{Piascik} \&
  {Steele}}{2016c}]{asassn16bb_spec_atel}
{Piascik} A.~S.,  {Steele} I.~A.,  2016c, The Astronomer's Telegram, \href
  {http://adsabs.harvard.edu/abs/2016ATel.8634....1P} {8634}

\bibitem[\protect\citeauthoryear{{Piascik} \& {Steele}}{{Piascik} \&
  {Steele}}{2016d}]{asassn16cm_spec_atel}
{Piascik} A.~S.,  {Steele} I.~A.,  2016d, The Astronomer's Telegram, \href
  {http://adsabs.harvard.edu/abs/2016ATel.8794....1P} {8794}

\bibitem[\protect\citeauthoryear{{Piascik} \& {Steele}}{{Piascik} \&
  {Steele}}{2016e}]{asassn16ct_spec_atel}
{Piascik} A.~S.,  {Steele} I.~A.,  2016e, The Astronomer's Telegram, \href
  {http://adsabs.harvard.edu/abs/2016ATel.8798....1P} {8798}

\bibitem[\protect\citeauthoryear{{Piascik} \& {Steele}}{{Piascik} \&
  {Steele}}{2016f}]{asassn16gv_spec_atel}
{Piascik} A.~S.,  {Steele} I.~A.,  2016f, The Astronomer's Telegram, \href
  {http://adsabs.harvard.edu/abs/2016ATel.9221....1P} {9221}

\bibitem[\protect\citeauthoryear{{Pignata} et~al.,}{{Pignata}
  et~al.}{2009}]{pignata09}
{Pignata} G.,  et~al., 2009, in {Giobbi} G.,  {Tornambe} A.,  {Raimondo} G.,
  {Limongi} M.,  {Antonelli} L.~A.,  {Menci} N.,   {Brocato} E.,  eds,
  American Institute of Physics Conference Series Vol. 1111, American Institute
  of Physics Conference Series. pp 551--554 (\mn@eprint {arXiv} {0812.4923}),
  \mn@doi{10.1063/1.3141608}

\bibitem[\protect\citeauthoryear{{Post} et~al.,}{{Post}
  et~al.}{2016a}]{asassn16oo_atel}
{Post} R.~S.,  et~al., 2016a, The Astronomer's Telegram, \href
  {http://adsabs.harvard.edu/abs/2016ATel.9837....1P} {9837}

\bibitem[\protect\citeauthoryear{{Post} et~al.,}{{Post}
  et~al.}{2016b}]{asassn16ov_atel}
{Post} R.~S.,  et~al., 2016b, The Astronomer's Telegram, \href
  {http://adsabs.harvard.edu/abs/2016ATel.9875....1P} {9875}

\bibitem[\protect\citeauthoryear{{Prieto}}{{Prieto}}{2016}]{asassn16cn_spec_atel}
{Prieto} J.~L.,  2016, The Astronomer's Telegram, \href
  {http://adsabs.harvard.edu/abs/2016ATel.8867....1P} {8867}

\bibitem[\protect\citeauthoryear{{Prieto} \& {Shappee}}{{Prieto} \&
  {Shappee}}{2016a}]{asassn16ec_spec_atel}
{Prieto} J.~L.,  {Shappee} B.~J.,  2016a, The Astronomer's Telegram, \href
  {http://adsabs.harvard.edu/abs/2016ATel.8924....1P} {8924}

\bibitem[\protect\citeauthoryear{{Prieto} \& {Shappee}}{{Prieto} \&
  {Shappee}}{2016b}]{asassn16dx_spec_atel}
{Prieto} J.~L.,  {Shappee} B.~J.,  2016b, The Astronomer's Telegram, \href
  {http://adsabs.harvard.edu/abs/2016ATel.8936....1P} {8936}

\bibitem[\protect\citeauthoryear{{Prieto} et~al.,}{{Prieto}
  et~al.}{2016a}]{prieto16}
{Prieto} J.~L.,  et~al., 2016a, preprint, \href
  {http://adsabs.harvard.edu/abs/2016arXiv160900013P} {} (\mn@eprint {arXiv}
  {1609.00013})

\bibitem[\protect\citeauthoryear{{Prieto}, {Rich}  \& {Shappee}}{{Prieto}
  et~al.}{2016b}]{asassn16az_spec_atel}
{Prieto} J.~L.,  {Rich} J.,   {Shappee} B.~J.,  2016b, The Astronomer's
  Telegram, \href {http://adsabs.harvard.edu/abs/2016ATel.8629....1P} {8629}

\bibitem[\protect\citeauthoryear{{Prieto}, {Morrell}, {Rudie}  \&
  {Shappee}}{{Prieto} et~al.}{2016c}]{asassn16fd_spec_atel}
{Prieto} J.~L.,  {Morrell} N.,  {Rudie} G.,   {Shappee} B.~J.,  2016c, The
  Astronomer's Telegram, \href
  {http://adsabs.harvard.edu/abs/2016ATel.9142....1P} {9142}

\bibitem[\protect\citeauthoryear{{Prieto}, {Seibert}, {Shappee}, {Dong}  \&
  {Stanek}}{{Prieto} et~al.}{2016d}]{asassn16ml_spec_atel}
{Prieto} J.~L.,  {Seibert} M.,  {Shappee} B.~J.,  {Dong} S.,   {Stanek} K.~Z.,
  2016d, The Astronomer's Telegram, \href
  {http://adsabs.harvard.edu/abs/2016ATel.9701....1P} {9701}

\bibitem[\protect\citeauthoryear{{Prieto}, {Madore}, {Shappee}, {Seidel},
  {Dong}  \& {Stanek}}{{Prieto} et~al.}{2016e}]{asassn16nm_spec_atel}
{Prieto} J.~L.,  {Madore} B.~F.,  {Shappee} B.~J.,  {Seidel} M.~K.,  {Dong} S.,
    {Stanek} K.~Z.,  2016e, The Astronomer's Telegram, \href
  {http://adsabs.harvard.edu/abs/2016ATel.9805....1P} {9805}

\bibitem[\protect\citeauthoryear{{Prieto}, {Seidel}, {Shappee}  \&
  {Stanek}}{{Prieto} et~al.}{2017}]{asassn16pp_spec_atel}
{Prieto} J.~L.,  {Seidel} M.~K.,  {Shappee} B.~J.,   {Stanek} K.~Z.,  2017, The
  Astronomer's Telegram, \href
  {http://adsabs.harvard.edu/abs/2017ATel.9932....1P} {9932}

\bibitem[\protect\citeauthoryear{{Pursiainen} et~al.,}{{Pursiainen}
  et~al.}{2016}]{asassn16mv_spec_atel}
{Pursiainen} M.,  et~al., 2016, The Astronomer's Telegram, \href
  {http://adsabs.harvard.edu/abs/2016ATel.9717....1P} {9717}

\bibitem[\protect\citeauthoryear{{Quimby}}{{Quimby}}{2006}]{quimby06}
{Quimby} R.~M.,  2006, PhD thesis, The University of Texas at Austin

\bibitem[\protect\citeauthoryear{{Reynolds} et~al.,}{{Reynolds}
  et~al.}{2016}]{asassn16hp_spec_atel}
{Reynolds} T.,  et~al., 2016, The Astronomer's Telegram, \href
  {http://adsabs.harvard.edu/abs/2016ATel.9272....1R} {9272}

\bibitem[\protect\citeauthoryear{{Romero-Ca{\~n}izales}, {Prieto}, {Chen},
  {Kochanek}, {Dong}, {Holoien}, {Stanek}  \& {Liu}}{{Romero-Ca{\~n}izales}
  et~al.}{2016}]{romero16}
{Romero-Ca{\~n}izales} C.,  {Prieto} J.~L.,  {Chen} X.,  {Kochanek} C.~S.,
  {Dong} S.,  {Holoien} T.~W.-S.,  {Stanek} K.~Z.,   {Liu} F.,  2016, preprint,
  \href {http://adsabs.harvard.edu/abs/2016arXiv160900010R} {} (\mn@eprint
  {arXiv} {1609.00010})

\bibitem[\protect\citeauthoryear{{Rui}, {Wang}, {Huang}, {Zhang}  \&
  {Zhai}}{{Rui} et~al.}{2016a}]{asassn16ad_spec_atel}
{Rui} L.,  {Wang} X.,  {Huang} F.,  {Zhang} T.,   {Zhai} M.,  2016a, The
  Astronomer's Telegram, \href
  {http://adsabs.harvard.edu/abs/2016ATel.8532....1R} {8532}

\bibitem[\protect\citeauthoryear{{Rui}, {Wang}, {Huang}, {Zhai}  \&
  {Zhang}}{{Rui} et~al.}{2016b}]{asassn16ch_spec_atel}
{Rui} L.,  {Wang} X.,  {Huang} F.,  {Zhai} M.,   {Zhang} T.,  2016b, The
  Astronomer's Telegram, \href
  {http://adsabs.harvard.edu/abs/2016ATel.8771....1R} {8771}

\bibitem[\protect\citeauthoryear{{SDSS Collaboration} et~al.,}{{SDSS
  Collaboration} et~al.}{2016}]{albareti16}
{SDSS Collaboration} et~al., 2016, preprint, \href
  {http://adsabs.harvard.edu/abs/2016arXiv160802013S} {} (\mn@eprint {arXiv}
  {1608.02013})

\bibitem[\protect\citeauthoryear{{Schlafly} \& {Finkbeiner}}{{Schlafly} \&
  {Finkbeiner}}{2011}]{schlafly11}
{Schlafly} E.~F.,  {Finkbeiner} D.~P.,  2011, \mn@doi [\apj]
  {10.1088/0004-637X/737/2/103}, \href
  {http://adsabs.harvard.edu/abs/2011ApJ...737..103S} {737, 103}

\bibitem[\protect\citeauthoryear{{Schmidt} et~al.,}{{Schmidt}
  et~al.}{2014}]{schmidt14}
{Schmidt} S.~J.,  et~al., 2014, \mn@doi [\apjl] {10.1088/2041-8205/781/2/L24},
  \href {http://adsabs.harvard.edu/abs/2014ApJ...781L..24S} {781, L24}

\bibitem[\protect\citeauthoryear{{Schmidt} et~al.,}{{Schmidt}
  et~al.}{2016}]{schmidt16}
{Schmidt} S.~J.,  et~al., 2016, \mn@doi [\apjl] {10.3847/2041-8205/828/2/L22},
  \href {http://adsabs.harvard.edu/abs/2016ApJ...828L..22S} {828, L22}

\bibitem[\protect\citeauthoryear{{Seidel}, {Seibert}, {Morrell}  \&
  {Shappee}}{{Seidel} et~al.}{2016}]{asassn16pj_spec_atel}
{Seidel} M.~K.,  {Seibert} M.,  {Morrell} N.,   {Shappee} B.~J.,  2016, The
  Astronomer's Telegram, \href
  {http://adsabs.harvard.edu/abs/2016ATel.9916....1S} {9916}

\bibitem[\protect\citeauthoryear{{Seidel}, {Morrell}  \& {Shappee}}{{Seidel}
  et~al.}{2017}]{asassn16pl_spec_atel}
{Seidel} M.~K.,  {Morrell} N.,   {Shappee} B.~J.,  2017, The Astronomer's
  Telegram, \href {http://adsabs.harvard.edu/abs/2017ATel.9922....1S} {9922}

\bibitem[\protect\citeauthoryear{{Shappee} et~al.,}{{Shappee}
  et~al.}{2014}]{shappee14}
{Shappee} B.~J.,  et~al., 2014, \mn@doi [\apj] {10.1088/0004-637X/788/1/48},
  \href {http://adsabs.harvard.edu/abs/2014ApJ...788...48S} {788, 48}

\bibitem[\protect\citeauthoryear{{Shappee} et~al.,}{{Shappee}
  et~al.}{2016a}]{shappee16}
{Shappee} B.~J.,  et~al., 2016a, \mn@doi [\apj] {10.3847/0004-637X/826/2/144},
  \href {http://adsabs.harvard.edu/abs/2016ApJ...826..144S} {826, 144}

\bibitem[\protect\citeauthoryear{{Shappee} et~al.,}{{Shappee}
  et~al.}{2016b}]{asassn16ab_atel}
{Shappee} B.~J.,  et~al., 2016b, The Astronomer's Telegram, \href
  {http://adsabs.harvard.edu/abs/2016ATel.8502....1S} {8502}

\bibitem[\protect\citeauthoryear{{Shappee} et~al.,}{{Shappee}
  et~al.}{2016c}]{asassn16hp_atel}
{Shappee} B.~J.,  et~al., 2016c, The Astronomer's Telegram, \href
  {http://adsabs.harvard.edu/abs/2016ATel.9268....1S} {9268}

\bibitem[\protect\citeauthoryear{{Shappee}, {Prieto}, {Rich}, {Seibert},
  {Madore}, {Poetrodjojo}  \& {D'Agostino}}{{Shappee}
  et~al.}{2016d}]{asassn16jc_spec_atel}
{Shappee} B.~J.,  {Prieto} J.~L.,  {Rich} J.,  {Seibert} M.,  {Madore} B.,
  {Poetrodjojo} H.,   {D'Agostino} J.,  2016d, The Astronomer's Telegram, \href
  {http://adsabs.harvard.edu/abs/2016ATel.9409....1S} {9409}

\bibitem[\protect\citeauthoryear{{Shappee}, {Prieto}, {Rich}, {Madore},
  {Poetrodjojo}  \& {D'Agostino}}{{Shappee}
  et~al.}{2016e}]{asassn16jh_spec_atel}
{Shappee} B.~J.,  {Prieto} J.~L.,  {Rich} J.,  {Madore} B.,  {Poetrodjojo} H.,
   {D'Agostino} J.,  2016e, The Astronomer's Telegram, \href
  {http://adsabs.harvard.edu/abs/2016ATel.9461....1S} {9461}

\bibitem[\protect\citeauthoryear{{Shields} et~al.,}{{Shields}
  et~al.}{2016}]{asassn16ip_atel}
{Shields} J.,  et~al., 2016, The Astronomer's Telegram, \href
  {http://adsabs.harvard.edu/abs/2016ATel.9353....1S} {9353}

\bibitem[\protect\citeauthoryear{{Short} et~al.,}{{Short}
  et~al.}{2016a}]{asassn16kk_spec_atel}
{Short} L.,  et~al., 2016a, The Astronomer's Telegram, \href
  {http://adsabs.harvard.edu/abs/2016ATel.9483....1S} {9483}

\bibitem[\protect\citeauthoryear{{Short} et~al.,}{{Short}
  et~al.}{2016b}]{asassn16kw_spec_atel}
{Short} L.,  et~al., 2016b, The Astronomer's Telegram, \href
  {http://adsabs.harvard.edu/abs/2016ATel.9566....1S} {9566}

\bibitem[\protect\citeauthoryear{{Skrutskie} et~al.,}{{Skrutskie}
  et~al.}{2006}]{skrutskie06}
{Skrutskie} M.~F.,  et~al., 2006, \mn@doi [\aj] {10.1086/498708}, \href
  {http://adsabs.harvard.edu/abs/2006AJ....131.1163S} {131, 1163}

\bibitem[\protect\citeauthoryear{{Smith} et~al.,}{{Smith}
  et~al.}{2016}]{asassn16oy_spec_atel}
{Smith} K.~W.,  et~al., 2016, The Astronomer's Telegram, \href
  {http://adsabs.harvard.edu/abs/2016ATel.9886....1S} {9886}

\bibitem[\protect\citeauthoryear{{Somero} et~al.,}{{Somero}
  et~al.}{2016}]{asassn16na_spec_atel}
{Somero} A.,  et~al., 2016, The Astronomer's Telegram, \href
  {http://adsabs.harvard.edu/abs/2016ATel.9734....1S} {9734}

\bibitem[\protect\citeauthoryear{{Stone} et~al.,}{{Stone}
  et~al.}{2016}]{asassn16oz_atel}
{Stone} G.,  et~al., 2016, The Astronomer's Telegram, \href
  {http://adsabs.harvard.edu/abs/2016ATel.9887....1S} {9887}

\bibitem[\protect\citeauthoryear{{Strader} \& {Prieto}}{{Strader} \&
  {Prieto}}{2017}]{asassn16po_spec_atel}
{Strader} J.,  {Prieto} J.~L.,  2017, The Astronomer's Telegram, \href
  {http://adsabs.harvard.edu/abs/2017ATel.9940....1S} {9940}

\bibitem[\protect\citeauthoryear{{Strader}, {Chomiuk}  \&
  {Shishkovsky}}{{Strader} et~al.}{2016a}]{asassn16cu_spec_atel}
{Strader} J.,  {Chomiuk} L.,   {Shishkovsky} L.,  2016a, The Astronomer's
  Telegram, \href {http://adsabs.harvard.edu/abs/2016ATel.8880....1S} {8880}

\bibitem[\protect\citeauthoryear{{Strader} et~al.,}{{Strader}
  et~al.}{2016b}]{asassn16eo_atel}
{Strader} J.,  et~al., 2016b, The Astronomer's Telegram, \href
  {http://adsabs.harvard.edu/abs/2016ATel.8965....1S} {8965}

\bibitem[\protect\citeauthoryear{{Strader}, {Chomiuk}  \& {Prieto}}{{Strader}
  et~al.}{2016c}]{asassn16gp_spec_atel}
{Strader} J.,  {Chomiuk} L.,   {Prieto} J.~L.,  2016c, The Astronomer's
  Telegram, \href {http://adsabs.harvard.edu/abs/2016ATel.9233....1S} {9233}

\bibitem[\protect\citeauthoryear{{Strader}, {Chomiuk}  \&
  {Shishkovsky}}{{Strader} et~al.}{2016d}]{asassn16ol_spec_atel}
{Strader} J.,  {Chomiuk} L.,   {Shishkovsky} L.,  2016d, The Astronomer's
  Telegram, \href {http://adsabs.harvard.edu/abs/2016ATel.9850....1S} {9850}

\bibitem[\protect\citeauthoryear{{Strader}, {Chomiuk}  \& {Salinas}}{{Strader}
  et~al.}{2016e}]{asassn16of_spec_atel}
{Strader} J.,  {Chomiuk} L.,   {Salinas} R.,  2016e, The Astronomer's Telegram,
  \href {http://adsabs.harvard.edu/abs/2016ATel.9897....1S} {9897}

\bibitem[\protect\citeauthoryear{{Taubenberger} et~al.,}{{Taubenberger}
  et~al.}{2016}]{asassn16bv_spec_atel}
{Taubenberger} S.,  et~al., 2016, The Astronomer's Telegram, \href
  {http://adsabs.harvard.edu/abs/2016ATel.8708....1T} {8708}

\bibitem[\protect\citeauthoryear{{Terndrup}, {Calhoun}, {Cannata}, {Schulze},
  {Dong}  \& {Prieto}}{{Terndrup} et~al.}{2016}]{asassn16fj_spec_atel}
{Terndrup} D.~M.,  {Calhoun} G.,  {Cannata} R.,  {Schulze} J.,  {Dong} S.,
  {Prieto} J.~L.,  2016, The Astronomer's Telegram, \href
  {http://adsabs.harvard.edu/abs/2016ATel.9075....1T} {9075}

\bibitem[\protect\citeauthoryear{{Terreran} et~al.,}{{Terreran}
  et~al.}{2016a}]{asassn16bl_spec_atel}
{Terreran} G.,  et~al., 2016a, The Astronomer's Telegram, \href
  {http://adsabs.harvard.edu/abs/2016ATel.8694....1T} {8694}

\bibitem[\protect\citeauthoryear{{Terreran} et~al.,}{{Terreran}
  et~al.}{2016b}]{ps16dtm_atel1}
{Terreran} G.,  et~al., 2016b, The Astronomer's Telegram, \href
  {http://adsabs.harvard.edu/abs/2016ATel.9417....1T} {9417}

\bibitem[\protect\citeauthoryear{{Tomasella} et~al.,}{{Tomasella}
  et~al.}{2016a}]{asassn16bp_spec_atel}
{Tomasella} L.,  et~al., 2016a, The Astronomer's Telegram, \href
  {http://adsabs.harvard.edu/abs/2016ATel.8672....1T} {8672}

\bibitem[\protect\citeauthoryear{{Tomasella} et~al.,}{{Tomasella}
  et~al.}{2016b}]{asassn16ex_spec_atel}
{Tomasella} L.,  et~al., 2016b, The Astronomer's Telegram, \href
  {http://adsabs.harvard.edu/abs/2016ATel.9024....1T} {9024}

\bibitem[\protect\citeauthoryear{{Tomasella}, {Benetti}, {Cappellaro},
  {Elias-Rosa}, {Ochner}, {Pastorello}, {Terreran}  \& {Turatto}}{{Tomasella}
  et~al.}{2016c}]{asassn16jj_spec_atel}
{Tomasella} L.,  {Benetti} S.,  {Cappellaro} E.,  {Elias-Rosa} N.,  {Ochner}
  P.,  {Pastorello} A.,  {Terreran} G.,   {Turatto} M.,  2016c, The
  Astronomer's Telegram, \href
  {http://adsabs.harvard.edu/abs/2016ATel.9420....1T} {9420}

\bibitem[\protect\citeauthoryear{{Tomasella} et~al.,}{{Tomasella}
  et~al.}{2016d}]{asassn16lg_spec_atel}
{Tomasella} L.,  et~al., 2016d, The Astronomer's Telegram, \href
  {http://adsabs.harvard.edu/abs/2016ATel.9610....1T} {9610}

\bibitem[\protect\citeauthoryear{{Tonry}}{{Tonry}}{2011}]{tonry11}
{Tonry} J.~L.,  2011, \mn@doi [\pasp] {10.1086/657997}, \href
  {http://adsabs.harvard.edu/abs/2011PASP..123...58T} {123, 58}

\bibitem[\protect\citeauthoryear{{Tonry}, {Denneau}, {Stalder}, {Heinze},
  {Sherstyuk}, {Rest}, {Smith}  \& {Smartt}}{{Tonry}
  et~al.}{2016}]{asassn16bn_spec_atel}
{Tonry} J.,  {Denneau} L.,  {Stalder} B.,  {Heinze} A.,  {Sherstyuk} A.,
  {Rest} A.,  {Smith} K.~W.,   {Smartt} S.~J.,  2016, The Astronomer's
  Telegram, \href {http://adsabs.harvard.edu/abs/2016ATel.8680....1T} {8680}

\bibitem[\protect\citeauthoryear{{Turatto}, {Benetti}, {Tomasella},
  {Cappellaro}, {Elias-Rosa}, {Ochner}  \& {Terreran}}{{Turatto}
  et~al.}{2016}]{asassn16ns_spec_atel}
{Turatto} M.,  {Benetti} S.,  {Tomasella} L.,  {Cappellaro} E.,  {Elias-Rosa}
  N.,  {Ochner} P.,   {Terreran} G.,  2016, The Astronomer's Telegram, \href
  {http://adsabs.harvard.edu/abs/2016ATel.9829....1T} {9829}

\bibitem[\protect\citeauthoryear{{Wiethoff} et~al.,}{{Wiethoff}
  et~al.}{2016}]{asassn16ch_atel}
{Wiethoff} W.,  et~al., 2016, The Astronomer's Telegram, \href
  {http://adsabs.harvard.edu/abs/2016ATel.8763....1W} {8763}

\bibitem[\protect\citeauthoryear{{Wright} et~al.,}{{Wright}
  et~al.}{2010}]{wright10}
{Wright} E.~L.,  et~al., 2010, \mn@doi [\aj] {10.1088/0004-6256/140/6/1868},
  \href {http://adsabs.harvard.edu/abs/2010AJ....140.1868W} {140, 1868}

\bibitem[\protect\citeauthoryear{{Wyrzykowski} et~al.,}{{Wyrzykowski}
  et~al.}{2014}]{wyrzykowski14}
{Wyrzykowski} {\L}.,  et~al., 2014, \actaa, \href
  {http://adsabs.harvard.edu/abs/2014AcA....64..197W} {64, 197}

\bibitem[\protect\citeauthoryear{{Xiang}, {Rui}, {Wang}, {Yang}, {Wu}, {Xiao},
  {Fan}  \& {Zhang}}{{Xiang} et~al.}{2017}]{asassn16pk_spec_atel}
{Xiang} D.,  {Rui} L.,  {Wang} X.,  {Yang} Q.,  {Wu} X.,  {Xiao} F.,  {Fan} Z.,
    {Zhang} T.,  2017, The Astronomer's Telegram, \href
  {http://adsabs.harvard.edu/abs/2017ATel.9926....1X} {9926}

\bibitem[\protect\citeauthoryear{{Xin} \& {Zhang}}{{Xin} \&
  {Zhang}}{2016}]{asassn16ai_spec_atel}
{Xin} Y.-X.,  {Zhang} J.-J.,  2016, The Astronomer's Telegram, \href
  {http://adsabs.harvard.edu/abs/2016ATel.8540....1X} {8540}

\bibitem[\protect\citeauthoryear{{Yamanaka} et~al.,}{{Yamanaka}
  et~al.}{2017}]{yamanaka17}
{Yamanaka} M.,  et~al., 2017, \mn@doi [\apj] {10.3847/1538-4357/aa5f57}, \href
  {http://adsabs.harvard.edu/abs/2017ApJ...837....1Y} {837, 1}

\bibitem[\protect\citeauthoryear{{Yaron} \& {Gal-Yam}}{{Yaron} \&
  {Gal-Yam}}{2012}]{yaron12}
{Yaron} O.,  {Gal-Yam} A.,  2012, \mn@doi [\pasp] {10.1086/666656}, \href
  {http://adsabs.harvard.edu/abs/2012PASP..124..668Y} {124, 668}

\bibitem[\protect\citeauthoryear{{Zhang}}{{Zhang}}{2016}]{asassn16aj_spec_atel}
{Zhang} J.-J.,  2016, The Astronomer's Telegram, \href
  {http://adsabs.harvard.edu/abs/2016ATel.8550....1Z} {8550}

\bibitem[\protect\citeauthoryear{{Zhang}, {Zheng}, {Wang}  \& {Rui}}{{Zhang}
  et~al.}{2016}]{asassn16fq_spec_atel}
{Zhang} J.,  {Zheng} X.,  {Wang} X.,   {Rui} L.,  2016, The Astronomer's
  Telegram, \href {http://adsabs.harvard.edu/abs/2016ATel.9093....1Z} {9093}

\makeatother
\end{thebibliography}

\newpage

%%%%%%%%%%%%%%%%%
% Table: ASAS-SN Supernovae
%%%%%%%%%%%%%%%%%
\begin{landscape}
\begin{table}
\begin{minipage}{\textwidth}
\centering
\fontsize{6}{7.2}\selectfont
\caption{ASAS-SN Supernovae}
\label{table:asassn_sne}
\begin{tabular}{@{}l@{\hspace{0.15cm}}l@{\hspace{0.15cm}}c@{\hspace{0.15cm}}c@{\hspace{0.15cm}}c@{\hspace{0.15cm}}l@{\hspace{0.15cm}}c@{\hspace{0.15cm}}c@{\hspace{0.15cm}}c@{\hspace{0.15cm}}c@{\hspace{0.15cm}}c@{\hspace{0.15cm}}l@{\hspace{0.15cm}}l@{\hspace{0.15cm}}l@{\hspace{-0.05cm}}} 
\hline
\vspace{-0.14cm}
 & & & & & & & & & & & & & \\
 & IAU & Discovery & & & & & & Offset & & Age & & & \\
SN Name & Name & Date & RA$^a$ & Dec.$^a$ & Redshift & $V_{disc}$$^b$ & $V_{peak}$$^b$ & (arcsec)$^c$ & Type & at Disc.$^d$ & Host Name & Discovery ATel & Classification ATel \\
\vspace{-0.23cm} \\
\hline
\vspace{-0.17cm}
 & & & & & & & & & & & & &\\
ASASSN-16aa  &  2016A  &  2016-01-02.42  &  08:09:14.55  &  $+$00:16:50.6  &  0.01738  &  16.8  &  16.4  &  7.99  &  Ia  &  -4  &  UGC 04251  & \citet{asassn16aa_atel} & \citet{asassn16aa_spec_atel} \\ 
ASASSN-16ab  &  2016B  &  2016-01-03.62  &  11:55:04.20  &  $+$01:43:07.2  &  0.00429  &  14.7  &  14.7  &  11.36  &  II  &  -4  &  CGCG 012-116  & \citet{asassn16ab_atel} & \citet{asassn16ab_spec_atel} \\ 
ASASSN-16ad  &  2016F  &  2016-01-09.28  &  01:39:32.05  &  $+$33:49:36.3  &  0.01614  &  16.2  &  15.1  &  13.99  &  Ia  &  ---  &  KUG 0136+335  & \citet{asassn16ad_atel} & \citet{asassn16ad_spec_atel} \\ 
ASASSN-16ah  &  2016J  &  2016-01-11.34  &  05:47:45.38  &  $+$53:36:32.3  &  0.02850  &  16.8  &  16.7  &  9.19  &  Ia  &  -8  &  Uncatalogued  & \citet{asassn16ah_atel} & \citet{asassn16ah_atel} \\ 
ASASSN-16ai  &  2016I  &  2016-01-12.58  &  14:39:44.77  &  $+$23:23:42.5  &  0.01490  &  17.0  &  16.7  &  8.99  &  IIP  &  ---  &  UGC 09450  & \citet{asassn16ai_atel} & \citet{asassn16ai_spec_atel} \\ 
ASASSN-16aj  &  2016K  &  2016-01-14.21  &  04:21:48.21  &  $-$15:45:19.3  &  0.03075  &  17.0  &  14.1  &  8.95  &  Ia  &  -2  &  NGC 1562  & \citet{asassn16aj_atel} & \citet{asassn16aj_spec_atel} \\ 
ASASSN-16al  &  2016L  &  2016-01-15.36  &  15:00:27.47  &  $-$13:33:09.0  &  0.00930  &  17.0  &  16.5  &  49.94  &  IIP  &  ---  &  UGCA 397  & \citet{asassn16al_atel} & \citet{asassn16al_spec_atel} \\ 
ASASSN-16am  &  2016N  &  2016-01-15.36  &  04:45:21.28  &  $+$73:23:41.1  &  0.01502  &  17.4  &  17.1  &  14.33  &  II  &  2  &  CGCG 328-018  & \citet{asassn16am_atel} & \citet{asassn16am_spec_atel} \\ 
ASASSN-16ar  &  2016Z  &  2016-01-17.21  &  04:28:30.83  &  $-$17:39:23.4  &  0.03108  &  17.0  &  16.5  &  0.14  &  Ia  &  1  &  2MASX J04283087-1739233  & \citet{asassn16ar_atel} & \citet{asassn16ar_atel} \\ 
ASASSN-16at  &  2016X  &  2016-01-20.59  &  12:55:15.50  &  $+$00:05:59.7  &  0.00441  &  15.1  &  14.0  &  73.00  &  II  &  ---  &  UGC 08041  & \citet{asassn16at_atel} & \citet{asassn16at_spec_atel} \\ 
ASASSN-16av  &  2016ac  &  2016-01-18.44  &  11:51:28.24  &  $+$22:01:33.7  &  0.02567  &  16.5  &  16.0  &  0.26  &  Ia  &  -7  &  NGC 3926 NED02  & \citet{asassn16av_atel} & \citet{asassn16av_atel} \\ 
ASASSN-16aw  &  2016yr  &  2016-01-29.27  &  05:39:57.45  &  $-$40:30:57.9  &  0.03728  &  17.0  &  16.3  &  61.53  &  Ia  &  -1  &  ESO 306-G 016  & \citet{asassn16aw_atel} & \citet{asassn16aw_spec_atel} \\ 
ASASSN-16ax  &  2016ag  &  2016-01-26.23  &  01:31:23.18  &  $+$60:19:13.7  &  0.01870  &  16.0  &  15.7  &  2.24  &  Ia  &  -5  &  2MASX J01312331+6019128  & \citet{asassn16ax_atel} & \citet{asassn16ax_spec_atel} \\ 
ASASSN-16ay  &  2016ys  &  2016-01-28.41  &  07:12:14.35  &  $+$07:14:23.2  &  0.02834  &  16.7  &  16.3  &  14.02  &  Ia  &  -2  &  UGC 03738  & \citet{asassn16ay_atel} & \citet{asassn16aw_spec_atel} \\ 
ASASSN-16az  &  2016za  &  2016-01-29.38  &  11:30:33.73  &  $-$42:33:31.4  &  0.03407  &  16.9  &  16.4  &  4.69  &  Ia  &  9  &  2MASX J11303364-4233359  & \citet{asassn16az_atel} & \citet{asassn16az_spec_atel} \\ 
ASASSN-16ba  &  2016zb  &  2016-01-29.34  &  09:42:29.22  &  $-$16:58:26.9  &  0.01392  &  16.8  &  14.9  &  10.06  &  II  &  5  &  MCG -03-25-015  & \citet{asassn16az_atel} & \citet{asassn16ba_spec_atel} \\ 
ASASSN-16bb  &  2016zc  &  2016-01-31.60  &  14:05:57.10  &  $+$43:53:02.2  &  0.03375  &  16.6  &  16.2  &  5.78  &  Ia-91T  &  -7  &  SDSS J140557.36+435257.2  & \citet{asassn16bb_atel} & \citet{asassn16bb_spec_atel} \\ 
ASASSN-16bc  &  2016zd  &  2016-02-02.28  &  12:05:25.83  &  $-$21:24:01.2  &  0.03194  &  17.0  &  16.4  &  14.48  &  Ia  &  -10  &  2MASX J12052488-2123572  & \citet{asassn16bc_atel} & \citet{asassn16bb_spec_atel} \\ 
ASASSN-16bg  &  2016acx  &  2016-02-06.47  &  12:59:25.10  &  $+$27:44:25.0  &  0.02022  &  17.4  &  16.8  &  5.68  &  Ia  &  14  &  2MASX J12592491+2744198  & \citet{asassn16bg_atel} & \citet{asassn16bg_spec_atel} \\ 
ASASSN-16bl  &  2016adk  &  2016-02-08.22  &  11:42:26.68  &  $-$36:54:23.4  &  0.02955  &  17.3  &  16.8  &  2.28  &  Ia  &  0  &  2MASX J11422674-3654256  & \citet{asassn16bl_atel} & \citet{asassn16bl_spec_atel} \\
\vspace{-0.22cm}
 & & & & & & & & & & & & &\\
\hline
\end{tabular}
\smallskip
\\
\raggedright
\noindent This table is available in its entirety in a machine-readable form in the online journal. A portion is shown here for guidance regarding its form and content.\\
$^a$ Right ascension and declination are given in the J2000 epoch. \\
$^b$ All magnitudes are $V$-band magnitudes from ASAS-SN. \\
$^c$ Offset indicates the offset of the SN in arcseconds from the coordinates of the host nucleus, taken from NED. \\
$^d$ Discovery ages are given in days relative to peak. All ages are approximate and are only listed if a clear age was given in the classification telegram. \\
\vspace{-0.5cm}
\end{minipage}
\end{table}

%%%%%%%%%%%%%%%%%
% Table: Other Supernovae
%%%%%%%%%%%%%%%%%

\begin{table}
\begin{minipage}{\textwidth}
\bigskip\bigskip
\centering
\fontsize{6}{7.2}\selectfont
\caption{Non-ASAS-SN Supernovae}
\label{table:other_sne}
\begin{tabular}{@{}l@{\hspace{0.15cm}}l@{\hspace{0.15cm}}c@{\hspace{0.15cm}}c@{\hspace{0.15cm}}c@{\hspace{0.15cm}}l@{\hspace{0.15cm}}c@{\hspace{0.15cm}}c@{\hspace{0.15cm}}c@{\hspace{0.15cm}}l@{\hspace{0.15cm}}c@{\hspace{0.15cm}}c} 
\hline
\vspace{-0.14cm}
 & & & & & & & & & & & \\
 & IAU & Discovery &  & & & & Offset & & & & \\
 SN Name & Name$^{a}$ & Date & RA$^b$ & Dec.$^b$ & Redshift & $m_{peak}$$^c$ & (arcsec)$^d$ & Type & Host Name & Discovered By$^e$ & Recovered?$^f$ \\
\vspace{-0.23cm} \\
\hline
\vspace{-0.17cm}
 & & & & & & & & & & & \\
2016D & 2016D & 2016-01-01.55 & 04:01:25.99 & $-$54:45:30.8 & 0.04510 & 16.5 & 2.23 & Ia & 2MASX J04012613-5445295 & Amateurs & Yes \\ 
2016C & 2016C & 2016-01-03.84 & 13:38:05.30 & $-$17:51:15.3 & 0.00452 & 14.9 & 112.25 & IIP & NGC 5247 & Amateurs & Yes \\ 
MASTER OT J165420.77-615258 & --- & 2016-01-06.00 & 16:54:20.77 & $-$61:52:58.0 & 0.01505 & 16.9 & 8.04 & IIP & ESO 138-G006 & MASTER & No \\ 
2016G & 2016G & 2016-01-09.94 & 03:03:57.74 & $+$43:24:03.5 & 0.00915 & 16.3 & 16.68 & Ic-BL & NGC 1171 & Amateurs & Yes \\ 
2016P & 2016P & 2016-01-19.19 & 13:57:31.10 & $+$06:05:51.0 & 0.01462 & 16.0 & 21.88 & Ic-BL & NGC 5374 & Amateurs & Yes \\ 
2016W & 2016W & 2016-01-20.49 & 02:30:39.69 & $+$42:14:08.9 & 0.01925 & 16.0 & 18.44 & Ia & NGC 946 & Amateurs & Yes \\ 
ATLAS16aaf & 2016ado & 2016-01-20.78 & 02:03:05.79 & $-$03:50:28.0 & 0.04248 & 17.0 & 3.18 & Ia-07if & 2MASX J02030578-0350240 & ATLAS & No \\ 
ATLAS16aab & 2016adp & 2016-01-21.81 & 03:21:42.43 & $+$42:05:49.4 & 0.01800 & 16.5 & 6.36 & Ia-91bg & 2MASX J03214217+4205549 & ATLAS & No \\ 
MASTER OT J105908.57+103834.8 & --- & 2016-01-31.06 & 10:59:08.57 & $+$10:38:34.8 & 0.03500 & 16.8 & 5.16 & Ia & SDSS J105908.63+103829.7 & MASTER & Yes \\ 
2016adi & 2016adi & 2016-02-03.54 & 13:47:43.11 & $-$30:55:57.0 & 0.01490 & 15.3 & 46.56 & Ia & NGC 5292 & Amateurs & No \\ 
2016adj & 2016adj & 2016-02-08.56 & 13:25:24.12 & $-$43:00:57.9 & 0.00183 & 14.3 & 39.56 & IIb & NGC 5128 & Amateurs & No \\ 
2016afa & 2016afa & 2016-02-12.92 & 15:36:32.47 & $+$16:36:36.7 & 0.00653 & 16.8 & 12.45 & II & NGC 5962 & Amateurs & Yes \\ 
2016ajf & 2016ajf & 2016-02-18.44 & 03:19:54.47 & $+$41:33:53.5 & 0.02031 & 16.9 & 6.42 & Ia & NGC 1278 & Amateurs & Yes \\ 
Gaia16aeu & 2016ajm & 2016-02-20.09 & 02:49:36.01 & $-$31:13:23.9 & 0.02136 & 16.9 & 14.70 & Ia-91bg & 2dFGRS S394Z183 & Gaia & No \\ 
Gaia16agf & 2016aqs & 2016-02-27.28 & 06:34:08.98 & $-$25:11:04.6 & 0.03000 & 17.0 & 20.09 & Ia & Uncatalogued & Gaia & Yes \\ 
2016aqt & 2016aqt & 2016-12-28.44 & 13:45:50.75 & $+$26:47:47.4 & 0.05046 & 16.5 & 2.16 & Ia-91T & SDSS J134550.90+264747.4 & Amateurs & Yes \\ 
PS16bdu & 2016bdu & 2016-02.28.62 & 13:10:13.95 & $+$32:31:14.1 & 0.01700 & 16.5 & 2.16 & IIn & SDSS J131014.04+323115.9 & Pan-STARRS & Yes \\ 
2016bam & 2016bam & 2016-03-07.43 & 07:46:52.72 & $+$39:01:21.8 & 0.01350 & 15.8 & 33.80 & II & NGC 2444 & Amateurs & Yes \\ 
ATLAS16ago & 2016bev & 2016-03-10.33 & 07:47:50.21 & $-$18:44:12.8 & 0.01080 & 16.7 & 50.82 & IIP & ESO 560-G013 & ATLAS & No \\ 
2016bas & 2016bas & 2016-03-12.51 & 07:38:05.53 & $-$55:11:47.0 & 0.00941 & 16.8 & 20.02 & IIb & ESO 163-G011 & Amateurs & No \\ 
\vspace{-0.22cm}
 & & & & & & & & & & & \\
\hline
\end{tabular}
\smallskip
\\
\raggedright
\noindent This table is available in its entirety in a machine-readable form in the online journal. A portion is shown here for guidance regarding its form and content.\\
$^a$ IAU name is not provided if one was not given to the supernova. In some cases the IAU name may also be the primary supernova name. \\
$^b$ Right ascension and declination are given in the J2000 epoch. \\
$^c$ All magnitudes are taken from D. W. Bishop's Bright Supernova website, as described in the text, and may be from different filters. \\
$^d$ Offset indicates the offset of the SN in arcseconds from the coordinates of the host nucleus, taken from NED. \\
$^e$ ``Amateurs'' indicates discovery by any number of non-professional astronomers, as described in the text. \\
$^f$ Indicates whether the supernova was independently recovered in ASAS-SN data or not.
\end{minipage}
\vspace{-0.5cm}
\end{table}

\end{landscape}
\pagebreak
\begin{landscape}

%%%%%%%%%%%%%%%%%
% Table: ASASSN Hosts
%%%%%%%%%%%%%%%%%

\begin{table}
\begin{minipage}{\textwidth}
\centering
\fontsize{6}{7.2}\selectfont
\caption{ASAS-SN Supernova Host Galaxies}
\label{table:asassn_hosts}
\begin{tabular}{@{}l@{\hspace{0.15cm}}l@{\hspace{0.15cm}}c@{\hspace{0.15cm}}c@{\hspace{0.15cm}}c@{\hspace{0.15cm}}c@{\hspace{0.15cm}}c@{\hspace{0.15cm}}c@{\hspace{0.15cm}}c@{\hspace{0.15cm}}c@{\hspace{0.15cm}}c@{\hspace{0.15cm}}c@{\hspace{0.15cm}}c@{\hspace{0.15cm}}c@{\hspace{0.15cm}}c@{\hspace{0.15cm}}c@{\hspace{0.15cm}}c} 
\hline
\vspace{-0.14cm}
 & & & & & & & & & & & & & & \\
 & & SN & SN & SN Offset & & & & & & & & & & \\
Galaxy Name & Redshift & Name & Type & (arcsec) & $A_V$$^a$ & $m_{NUV}$$^b$ & $m_u$$^c$ & $m_g$$^c$ & $m_r$$^c$ & $m_i$$^c$ & $m_z$$^c$ & $m_J$$^d$ & $m_H$$^d$ & $m_{K_S}$$^{d,e}$ & $m_{W1}$ & $m_{W2}$\\ 
\vspace{-0.23cm} \\
\hline
\vspace{-0.17cm}
 & & & & & & & & & & & & & & \\
UGC 04251 & 0.01738 & ASASSN-16aa & Ia & 7.99 & 0.124 & --- & 18.37 0.02 & 16.41 0.00 & 15.59 0.00 & 15.11 0.00 & 14.73 0.00 & 11.55 0.03 & 10.85 0.03 & 10.57 0.05 & 10.81 0.02 & 10.75 0.02 \\ 
PGC 037392 & 0.00429 & ASASSN-16ab & II & 11.36 & 0.058 & 16.56 0.02 & 16.05 0.01 & 15.21 0.00 & 14.92 0.00 & 14.78 0.00 & 14.67 0.01 & 13.91 0.09 & 13.34 0.10 & 13.02 0.16 & 13.69 0.03 & 13.57 0.04 \\ 
KUG 0136+335 & 0.01614 & ASASSN-16ad & Ia & 13.99 & 0.134 & 17.82 0.03 & 16.93 0.05 & 15.83 0.01 & 15.38 0.01 & 15.14 0.02 & 14.98 0.08 & 15.20 0.08 & 14.49 0.11 & 14.54 0.20 & 14.21 0.03 & 14.03 0.04 \\ 
Uncatalogued & 0.02850 & ASASSN-16ah & Ia & 9.19 & 0.719 & 20.34 0.25 & --- & --- & --- & --- & --- & $>$16.5 & $>$15.7 & 14.09 0.06* & 14.63 0.04 & 14.39 0.06 \\ 
UGC 09450 & 0.01490 & ASASSN-16ai & IIP & 8.99 & 0.088 & 17.61 0.06 & 17.31 0.02 & 16.38 0.01 & 16.08 0.01 & 15.87 0.01 & 16.06 0.03 & 16.09 0.10 & 15.66 0.15 & 15.78 0.19 & 15.62 0.04 & 15.29 0.08 \\ 
NGC 1562 & 0.03075 & ASASSN-16aj & Ia & 8.95 & 0.090 & 20.02 0.17 & --- & --- & --- & --- & --- & 12.08 0.03 & 11.37 0.04 & 11.08 0.06 & 11.37 0.02 & 11.44 0.02 \\ 
UGCA 397 & 0.00930 & ASASSN-16al & IIP & 49.94 & 0.276 & 16.89 0.04 & --- & --- & --- & --- & --- & $>$16.5 & $>$15.7 & 14.44 0.06* & 14.98 0.04 & 14.86 0.09 \\ 
CGCG 328-018 & 0.01502 & ASASSN-16am & II & 14.33 & 0.470 & 20.49 0.18 & --- & --- & --- & --- & --- & 11.63 0.03 & 10.91 0.04 & 10.67 0.05 & 11.21 0.02 & 11.28 0.02 \\ 
2MASX J04283087-1739233 & 0.03108 & ASASSN-16ar & Ia & 0.14 & 0.105 & --- & --- & --- & --- & --- & --- & 12.62 0.03 & 11.97 0.05 & 11.66 0.07 & 11.89 0.02 & 11.93 0.02 \\ 
UGC 08041 & 0.00441 & ASASSN-16at & II & 73.00 & 0.061 & 14.57 0.01 & 15.72 0.01 & 14.10 0.00 & 13.48 0.00 & 13.23 0.00 & 13.25 0.01 & 12.92 0.05 & 12.32 0.07 & 12.04 0.11 & 13.01 0.03 & 13.03 0.03 \\ 
NGC 3926 NED02 & 0.02567 & ASASSN-16av & Ia & 0.26 & 0.075 & 19.11 0.09 & 16.12 0.01 & 14.27 0.00 & 13.38 0.00 & 12.99 0.00 & 12.67 0.00 & 11.65 0.02 & 10.97 0.03 & 10.65 0.03 & 11.13 0.02 & 11.22 0.02 \\ 
ESO 306-G 016 & 0.03728 & ASASSN-16aw & Ia & 61.53 & 0.088 & --- & --- & --- & --- & --- & --- & 11.80 0.03 & 11.07 0.03 & 10.78 0.06 & 11.32 0.02 & 11.32 0.02 \\ 
2MASX J01312331+6019128 & 0.01870 & ASASSN-16ax & Ia & 2.24 & 1.512 & --- & --- & --- & --- & --- & --- & 12.47 0.03 & 11.65 0.03 & 11.36 0.05 & 11.44 0.02 & 11.46 0.02 \\ 
UGC 03738 & 0.02834 & ASASSN-16ay & Ia & 14.02 & 0.189 & --- & --- & --- & --- & --- & --- & 12.52 0.04 & 11.83 0.05 & 11.55 0.08 & 11.95 0.02 & 11.99 0.02 \\ 
2MASX J11303364-4233359 & 0.03407 & ASASSN-16az & Ia & 4.69 & 0.235 & --- & --- & --- & --- & --- & --- & 13.57 0.05 & 12.87 0.05 & 12.53 0.08 & 12.50 0.02 & 12.49 0.02 \\ 
MCG -03-25-015 & 0.01392 & ASASSN-16ba & II & 10.06 & 0.180 & 16.85 0.03 & --- & --- & --- & --- & --- & $>$16.5 & $>$15.7 & 14.12 0.06* & 14.66 0.03 & 14.58 0.06 \\ 
SDSS J140557.36+435257.2 & 0.03375 & ASASSN-16bb & Ia-91T & 5.78 & 0.020 & 18.89 0.08 & 18.31 0.04 & 17.29 0.01 & 16.96 0.01 & 16.76 0.01 & 16.71 0.08 & $>$16.5 & $>$15.7 & 15.06 0.06* & 15.60 0.04 & 15.40 0.08 \\ 
2MASX J12052488-2123572 & 0.03194 & ASASSN-16bc & Ia & 14.48 & 0.152 & 18.26 0.06 & --- & --- & --- & --- & --- & 12.78 0.04 & 12.11 0.04 & 11.76 0.07 & 11.99 0.02 & 11.58 0.02 \\ 
2MASX J12592491+2744198 & 0.02022 & ASASSN-16bg & Ia & 5.68 & 0.024 & 20.94 0.26 & 17.65 0.01 & 15.86 0.00 & 15.05 0.00 & 14.65 0.00 & 14.35 0.00 & 13.30 0.03 & 12.55 0.04 & 12.35 0.05 & 12.33 0.02 & 12.43 0.02 \\ 
2MASX J11422674-3654256 & 0.02955 & ASASSN-16bl & Ia & 2.28 & 0.318 & --- & --- & --- & --- & --- & --- & 12.72 0.04 & 11.92 0.04 & 11.63 0.07 & 11.81 0.02 & 11.83 0.02 \\ 
\vspace{-0.22cm}
 & & & & & & & & & & & & & & \\
\hline
\end{tabular}
\smallskip
\\
\raggedright
\noindent This table is available in its entirety in a machine-readable form in the online journal. A portion is shown here for guidance regarding its form and content. Uncertainty is given for all magnitudes, and in some cases is equal to zero.\\
$^a$ Galactic extinction taken from \citet{schlafly11}. \\
$^b$ No magnitude is listed for those galaxies not detected in GALEX survey data. \\
$^c$ No magnitude is listed for those galaxies not detected in SDSS data or those located outside of the SDSS footprint. \\
$^d$ For those galaxies not detected in 2MASS data, we assume an upper limit of the faintest galaxy detected in each band from our sample. \\
$^e$ $K_S$-band magnitudes marked with a ``*'' indicate those estimated from the WISE $W1$-band data, as described in the text. \\
\end{minipage}
\vspace{-0.5cm}
\end{table}

%%%%%%%%%%%%%%%%%
% Table: Non-ASASSN Hosts
%%%%%%%%%%%%%%%%%

\begin{table}
\begin{minipage}{\textwidth}
\bigskip\bigskip
\centering
\fontsize{6}{7.2}\selectfont
\caption{Non-ASAS-SN Supernova Host Galaxies}
\label{table:other_hosts}
\begin{tabular}{@{}l@{\hspace{0.15cm}}l@{\hspace{0.15cm}}c@{\hspace{0.15cm}}c@{\hspace{0.05cm}}c@{\hspace{0.15cm}}c@{\hspace{0.15cm}}c@{\hspace{0.15cm}}c@{\hspace{0.15cm}}c@{\hspace{0.15cm}}c@{\hspace{0.15cm}}c@{\hspace{0.15cm}}c@{\hspace{0.15cm}}c@{\hspace{0.15cm}}c@{\hspace{0.15cm}}c@{\hspace{0.15cm}}c@{\hspace{0.15cm}}c} 
\hline
\vspace{-0.14cm}
 & & & & & & & & & & & & & & \\
 & & SN & SN & SN Offset & & & & & & & & & & \\
Galaxy Name & Redshift & Name & Type & (arcsec) & $A_V$$^a$ & $m_{NUV}$$^b$ & $m_u$$^c$ & $m_g$$^c$ & $m_r$$^c$ & $m_i$$^c$ & $m_z$$^c$ & $m_J$$^d$ & $m_H$$^d$ & $m_{K_S}$$^{d,e}$ & $m_{W1}$ & $m_{W2}$ \\ 
\vspace{-0.23cm} \\
\hline
\vspace{-0.17cm}
 & & & & & & & & & & & & & & \\
2MASX J04012613-5445295 & 0.04510 & 2016D & Ia & 2.23 & 0.035 & 19.96 0.18 & --- & --- & --- & --- & --- & 12.34 0.03 & 11.63 0.04 & 11.26 0.06 & 11.49 0.02 & 11.55 0.02 \\ 
NGC 5247 & 0.00452 & 2016C & IIP & 112.25 & 0.244 & --- & --- & --- & --- & --- & --- & 8.80 0.02 & 8.17 0.02 & 7.93 0.03 & 10.46 0.02 & 10.34 0.02 \\ 
ESO 138-G006 & 0.01505 & MASTER J165420.77 & IIP & 8.04 & 0.425 & --- & --- & --- & --- & --- & --- & $>$16.5 & $>$15.7 & 13.57 0.07* & 14.21 0.04 & 14.00 0.04 \\ 
NGC 1171 & 0.00915 & 2016G & Ic-BL & 16.68 & 0.431 & 15.89 0.02 & 15.31 0.01 & 13.54 0.00 & 12.64 0.00 & 12.18 0.00 & 11.84 0.00 & 10.79 0.02 & 10.10 0.02 & 9.80 0.03 & 10.87 0.02 & 10.89 0.02 \\ 
NGC 5374 & 0.01462 & 2016P & Ic-BL & 21.88 & 0.074 & 15.12 0.01 & 14.52 0.01 & 13.31 0.00 & 12.53 0.00 & 12.13 0.00 & 11.82 0.00 & 10.81 0.03 & 10.16 0.04 & 9.88 0.06 & 10.86 0.02 & 10.67 0.02 \\ 
NGC 946 & 0.01925 & 2016W & Ia & 18.44 & 0.191 & 18.83 0.12 & --- & --- & --- & --- & --- & 10.83 0.02 & 10.11 0.02 & 9.82 0.02 & 10.25 0.02 & 10.25 0.02 \\ 
2MASX J02030578-0350240 & 0.04248 & ATLAS16aaf & Ia-07if & 3.18 & 0.069 & 20.06 0.11 & 17.59 0.02 & 15.92 0.00 & 15.26 0.00 & 14.93 0.00 & 14.71 0.00 & 13.76 0.04 & 13.07 0.05 & 12.80 0.07 & 12.85 0.04 & 12.83 0.04 \\ 
2MASX J03214217+4205549 & 0.01800 & ATLAS16aab & Ia-91bg & 6.36 & 0.513 & --- & 18.14 0.03 & 16.30 0.00 & 15.37 0.00 & 14.90 0.00 & 14.53 0.01 & 13.62 0.05 & 12.87 0.07 & 12.58 0.09 & 13.16 0.04 & 13.20 0.04 \\ 
SDSS J105908.63+103829.7 & 0.03500 & MASTER J105908.57 & Ia & 5.16 & 0.069 & 21.29 0.22 & 21.31 0.32 & 20.18 0.05 & 19.77 0.06 & 19.57 0.06 & 19.93 0.32 & $>$16.5 & $>$15.7 & 16.92 0.22* & 17.56 0.21 & 16.49 0.29 \\ 
NGC 5292 & 0.01490 & 2016adi & Ia & 46.56 & 0.164 & 15.76 0.02 & --- & --- & --- & --- & --- & 9.78 0.02 & 9.13 0.03 & 8.84 0.03 & 10.07 0.02 & 10.12 0.02 \\ 
NGC 5128 & 0.00183 & 2016adj & IIb & 39.56 & 0.315 & --- & --- & --- & --- & --- & --- & 5.03 0.02 & 4.31 0.02 & 3.99 0.02 & 5.26 0.05 & 4.76 0.04 \\ 
NGC 5962 & 0.00653 & 2016afa & II & 12.45 & 0.149 & 14.11 0.01 & 13.89 0.00 & 12.27 0.00 & 11.51 0.00 & 11.14 0.00 & 10.69 0.00 & 9.58 0.01 & 8.89 0.01 & 8.61 0.02 & 9.67 0.02 & 9.53 0.02 \\ 
NGC 1278 & 0.02031 & 2016ajf & Ia & 6.42 & 0.451 & 19.11 0.05 & --- & --- & --- & --- & --- & 10.30 0.02 & 9.57 0.02 & 9.25 0.03 & 10.10 0.02 & 10.14 0.02 \\ 
2dFGRS S394Z183 & 0.02136 & Gaia16aeu & Ia-91bg & 14.70 & 0.069 & 20.43 0.13 & --- & --- & --- & --- & --- & $>$16.5 & $>$15.7 & 14.77 0.07* & 15.41 0.04 & 15.80 0.11 \\ 
Uncatalogued & 0.03000 & Gaia16agf & Ia & 20.09 & 0.182 & --- & --- & --- & --- & --- & --- & $>$16.5 & $>$15.7 & $>$15.6 & --- & --- \\ 
SDSS J134550.90+264747.4 & 0.05046 & 2016aqt & Ia-91T & 2.16 & 0.046 & 19.14 0.11 & 18.79 0.07 & 17.75 0.01 & 17.45 0.01 & 17.24 0.02 & 17.10 0.05 & $>$16.5 & $>$15.7 & 15.51 0.08* & 16.15 0.05 & 16.09 0.15 \\ 
SDSS J131014.04+323115.9 & 0.01700 & PS16bdu & IIn & 2.16 & 0.041 & --- & 22.08 0.28 & 21.19 0.06 & 20.94 0.07 & 20.99 0.11 & 21.07 0.39 & $>$16.5 & $>$15.7 & $>$15.6 & --- & --- \\ 
NGC 2444 & 0.01350 & 2016bam & II & 33.80 & 0.140 & --- & 15.57 0.01 & 13.73 0.00 & 12.84 0.00 & 12.41 0.00 & 12.10 0.00 & 11.17 0.02 & 10.47 0.02 & 10.21 0.03 & 10.39 0.02 & 10.48 0.02 \\ 
ESO 560-G013 & 0.01080 & ATLAS16ago & IIP & 50.82 & 1.050 & --- & --- & --- & --- & --- & --- & 10.09 0.02 & 9.26 0.02 & 8.90 0.02 & 10.45 0.02 & 10.27 0.02 \\ 
ESO 163-G011 & 0.00941 & 2016bas & IIb & 20.02 & 0.407 & --- & --- & --- & --- & --- & --- & 10.81 0.02 & 10.01 0.02 & 9.64 0.03 & 10.32 0.02 & 10.08 0.02 \\ 
\vspace{-0.22cm}
 & & & & & & & & & & & & & & \\
\hline
\end{tabular}
\smallskip
\\
\raggedright
\noindent This table is available in its entirety in a machine-readable form in the online journal. A portion is shown here for guidance regarding its form and content. Uncertainty is given for all magnitudes, and in some cases is equal to zero. ``MASTER'' supernova names have been abbreviated for space reasons.\\
$^a$ Galactic extinction taken from \citet{schlafly11}. \\
$^b$ No magnitude is listed for those galaxies not detected in GALEX survey data. \\
$^c$ No magnitude is listed for those galaxies not detected in SDSS data or those located outside of the SDSS footprint. \\
$^d$ For those galaxies not detected in 2MASS data, we assume an upper limit of the faintest galaxy detected in each band from our sample. \\
$^e$ $K_S$-band magnitudes marked with a ``*'' indicate those estimated from the WISE $W1$-band data, as described in the text. \\
\end{minipage}
\vspace{-0.5cm}
\end{table}

\end{landscape}

\end{document}